\documentclass[twocolumn]{aastex701}

\begin{document}

\title{Through the Veil: Ly$\alpha$ Illuminates the Host Galaxies of Little Red Dots}

\correspondingauthor{Zhiyuan Ji}
\email{zhiyuanji@arizona.edu}

\author[0000-0001-7673-2257]{Zhiyuan Ji}
\affiliation{Steward Observatory, University of Arizona, 933 N. Cherry Avenue, Tucson, AZ 85721, USA}
\email{zhiyuanji@arizona.edu}

\author[0000-0001-6561-9443]{Yang Sun}
\affiliation{Steward Observatory, University of Arizona, 933 N. Cherry Avenue, Tucson, AZ 85721, USA}
\email{sunyang@arizona.edu}

\author[0000-0002-7831-8751]{Mauro Giavalisco}
\affiliation{University of Massachusetts Amherst, 710 North Pleasant Street, Amherst, MA 01003-9305, USA}
\email{mauro@umass.edu}

\author[0000-0002-2380-9801]{Anna de Graaff}
\thanks{Clay Fellow}
\affiliation{Center for Astrophysics, Harvard \& Smithsonian, 60 Garden St, Cambridge, MA 02138, USA}
\affiliation{Max-Planck-Institut f\"ur Astronomie, K\"onigstuhl 17, D-69117 Heidelberg, Germany}
\email{anna.de_graaff@cfa.harvard.edu}

\author[0000-0003-2919-7495]{Christina C. Williams}
\affiliation{NSF–DOE Vera C. Rubin Observatory/NSF NOIRLab, 950 N. Cherry Ave., Tucson, AZ 85719, USA}
\affiliation{Steward Observatory, University of Arizona, 933 N. Cherry Avenue, Tucson, AZ 85721, USA}
\email{christina.williams@noirlab.edu}

\author[0000-0003-3307-7525]{Yongda Zhu}
\affiliation{Steward Observatory, University of Arizona, 933 N. Cherry Avenue, Tucson, AZ 85721, USA}
\email{yongdaz@arizona.edu}

\author[0000-0003-2303-6519]{George H. Rieke}
\affiliation{Steward Observatory, University of Arizona, 933 N. Cherry Avenue, Tucson, AZ 85721, USA}
\email{ghrieke@gmail.com}

\author[0000-0002-7893-6170]{Marcia Rieke}
\affiliation{Steward Observatory, University of Arizona, 933 N. Cherry Avenue, Tucson, AZ 85721, USA}
\email{mrieke@gmail.com}

\begin{abstract}

Little Red Dots (LRDs) are enigmatic, compact red sources ubiquitous in JWST deep fields whose physical nature remains elusive. As  one of the most sensitive tracers of neutral hydrogen in galaxy environments, Ly$\alpha$ is uniquely positioned to probe the gaseous structures proposed to explain LRDs' unusual properties. We present a systematic study of Ly$\alpha$ emission in LRDs, using a sample of 110 spectroscopically confirmed LRDs at $z \geq 4$ from the \citet{deGraaff2026} catalog, all with NIRSpec/PRISM coverage of the Ly$\alpha$ line. We detect Ly$\alpha$ at signal-to-noise S/N $\geq$ 3 in 32 LRDs, finding Ly$\alpha$ luminosities and the distribution of rest-frame equivalent widths consistent with normal star-forming galaxies at comparable redshifts. Yet the Ly$\alpha$/H$\alpha$ ratios fall systematically below those of star-forming galaxies, and the Ly$\alpha$ luminosity tracks [\ion{O}{3}] luminosity more closely than [\ion{O}{3}] equivalent width, together suggesting that Ly$\alpha$ is primarily associated with the host-scale component rather than the compact component responsible for the broad Balmer lines and red continuum. For 13 LRDs at $z \gtrsim 5.5$, we construct continuum-subtracted Ly$\alpha$ maps using broadband imaging from HST/ACS or JWST/NIRCam, revealing spatially extended, asymmetric, and often offset emission relative to the rest-optical light, consistent with resonant scattering through clumpy, anisotropic gas commonly observed in high-redshift Ly$\alpha$ emitters. These results support a two-component picture in which the compact rest-optical source is embedded within a more extended host-galaxy environment whose interstellar and circumgalactic gas shapes Ly$\alpha$ escape and spatial redistribution. Ly$\alpha$ opens a new window into the relation between the compact red component, the host galaxy, and the surrounding gas in LRDs.

\end{abstract}

%\keywords{\uat{Galaxies}{573} --- \uat{Cosmology}{343} --- \uat{High Energy astrophysics}{739} --- \uat{Interstellar medium}{847} --- \uat{Stellar astronomy}{1583} --- \uat{Solar physics}{1476}}


\section{Introduction}

The \textit{James Webb Space Telescope} (JWST) has uncovered a remarkable population of compact, red sources at high redshift, commonly referred to as little red dots (LRDs; e.g., \citealt{Matthee2024}). These systems occupy a region of parameter space not well represented by previously known galaxy or quasar populations, exhibiting V-shaped spectral energy distributions from rest-frame UV to near-infrared (NIR), extremely compact rest-frame optical/NIR morphologies, and, in many cases, broad Balmer emission lines \citep[e.g.,][]{Greene2024,Hviding2025}. This unusual combination of properties, together with their apparently high abundance at $z\gtrsim4$, has  made LRDs central to current debates on the emergence of early black holes and the nature of luminous compact red sources in the first $\sim$1--2~Gyr of cosmic history.

A central question is what powers LRDs and whether they represent a single physical class. One widely discussed interpretation is that many LRDs host obscured active galactic nuclei (AGN), motivated by their compact morphologies, broad Balmer lines, and luminous red optical continua \citep{Matthee2024,Greene2024,Kocevski2025,Taylor2025,Labbe2025}. At the same time, alternative interpretations have invoked massive post-starburst or dusty star-forming galaxies; and the weak X-ray emission, low dust masses, and lack of strong hot-dust signatures suggest that LRDs are not simply  scaled-down analogs of ordinary unobscured quasars \citep{Labbe2023,PerezGonzalez2024,Ananna2024,Yue2024,Williams2024b,Wang2025b,Setton2025,Casey2025,Xiao2025}. 

More recent work has proposed a picture in which at least some LRDs are powered by accreting black holes embedded in dense gaseous environments, where the observed continuum and line emission are strongly shaped by reprocessing in the surrounding medium \citep{InayoshiMaiolino2025,Ji2025a,Naidu2025a,cliff,Rusakov2026,Chang2025}. This interpretation is supported by \citet{deGraaff2026}, who argue from a large spectroscopic sample that many LRD continua are well described by modified blackbodies, and that the optical continuum, H$\alpha$, H$\beta$, and O\,{\sc i} emission are tightly linked, while [O\,{\sc iii}] is more naturally associated with the host galaxy. 

The picture of a dense nuclear medium is also consistent with the idea that Balmer-line widths need not be purely virial, but may be shaped in part by scattering in dense gas, with important implications for inferred black-hole masses and growth rates \citep{Rusakov2026,Chang2025,Naidu2025a,Torralba2025b,Greene2026}.  This possibility is additionally supported by the observations from \citet{Torralba2026} and \citet{PerezGonzalez2026}, who find that the Fe\,{\sc ii}/Mg\,{\sc ii} ratios of LRDs imply Eddington ratios of $\sim 0.5$. When combined with the bolometric luminosities of the nuclear sources, these Eddington ratios yield black-hole masses that are lower by orders of magnitude than estimates based purely on the Balmer-line widths. More broadly, several theoretical studies have connected such dense envelopes to quasistar-, black-hole-star-, or super-Eddington-accretion-like scenarios, in which high gas columns can naturally explain the red optical continua, strong Balmer features, and X-ray weakness of many LRDs \citep{Begelman2008,PacucciNarayan2024,Inayoshi2024,Lambrides2024,BegelmanDexter2025,Kido2025,Liu2025,Zhang2026}. Establishing how these different components are spatially related is therefore central to understanding the LRD phenomenon.

A key empirical insight from recent studies is that the rest-frame UV
and optical emission in LRDs likely trace primarily different
components. Spectral decompositions of large LRD
samples indicate that the central engine contributes a median of
$\sim$20\% to the rest-UV continuum of a typical LRD, with the
remainder arising from the host \citep{Sun2026}. Yet identifying the
dominant ionizing source from rest-UV emission lines alone has proven
challenging, both because single-component AGN or stellar models
cannot simultaneously reproduce the UV continuum, Balmer break, and
broad lines of individual LRDs \citep{Ma2025}, and because the
relevant UV emission-line diagnostics suffer from well-known
degeneracies between AGN- and star-formation-driven ionization at
high redshift \citep[e.g.,][]{Treiber2025}. On the imaging side,
while the redder light is frequently extremely compact, the UV can
be extended, asymmetric, and/or multi-component
\citep[e.g.,][]{Killi2024,Kocevski2025,Rinaldi2025,Iani2025,Cloonan2026},
independently suggesting that the UV emission often arises at least
partly in the putative host galaxy rather than in the compact source
that dominates the optical continuum. At the same time, however, UV
morphology alone cannot determine how the ionizing source is embedded
within, or obscured by, the surrounding neutral gas.

In this context, Ly$\alpha$ offers a uniquely powerful probe of the gaseous environment around LRDs. Because Ly$\alpha$ is resonantly scattered, its strength, line profile, and spatial distribution are highly sensitive to the neutral-gas column density, covering fraction, kinematics, dust content, and the topology of escape channels \citep[e.g.,][]{Dijkstra2014,Hayes2015}. Recent ultra-deep UV spectroscopy of the prototypical LRD Abell2744-QSO1 at $z=7.04$, initially discovered by \citet{Furtak2023}, has already shown that the Ly$\alpha$ profile itself strongly constrains the gas geometry, disfavoring a uniform high-covering-fraction envelope and instead favoring a clumpy or porous medium with low-column-density escape channels \citep{Tang2026,Ji2026}. Whereas the Ly$\alpha$ spectral profile provides a global, galaxy-integrated view of how photons escape through the surrounding gas, its spatial distribution reveals where that escaping and scattered emission emerges. In particular, spatially resolved Ly$\alpha$ can distinguish whether the emission emerges directly from the nuclear region, diffuses through an extended scattering halo, or appears offset because it escapes anisotropically through the surrounding medium. So far, however, our knowledge of spatially resolved Ly$\alpha$ in LRDs remains very limited: in one case, the MUSE observation shows weak and modestly extended Ly$\alpha$ emission from an LRD at $z\sim4.5$, favoring a highly enshrouded central source rather than a luminous quasar-like Ly$\alpha$ halo \citep{Torralba2026}. Extending such analyses to larger samples is essential for understanding how Ly$\alpha$ properties connect to the nature of the LRD population.

In this paper, we investigate the Ly$\alpha$ properties of a large sample of spectroscopically confirmed LRDs from \citet{deGraaff2026}; full details of the sample and associated data are provided in Section~\ref{sec:sample}. We first characterize the integrated Ly$\alpha$ properties across the full sample at $z\ge4$ (Section~\ref{sec:lya_global}). While integral-field spectroscopy
would offer the most direct probe of spatially resolved Ly$\alpha$,
such observations remain scarce for the high-redshift LRD population.
We therefore leverage the wealth of existing broadband imaging from
JWST/NIRCam and \textit{Hubble Space Telescope} (HST) to construct continuum-subtracted Ly$\alpha$
maps for the subsample with robust Ly$\alpha$ detections at $z \gtrsim 5.5$
(Section~\ref{sec:lya_map}). By comparing the integrated and spatially resolved views of Ly$\alpha$ with the rest-frame UV and optical continuum, we constrain the relation between the compact red component, the UV host emission, and the surrounding neutral gas, thereby clarifying which physical components are traced by Ly$\alpha$ in LRDs (Section~\ref{sec:diss}). 

Throughout this work, we adopt the AB magnitude system and the $\Lambda$CDM cosmology with \citet{Planck2020} parameters, i.e., $\Omega_m = 0.315$ and $\rm{h = H_0/(100 \,km\,s^{-1}\,Mpc^{-1}) = 0.673}$.

\section{The Sample and Data}\label{sec:sample}

Our parent sample is drawn from \citet{deGraaff2026}\footnote{\href{https://doi.org/10.5281/zenodo.17665942}{https://doi.org/10.5281/zenodo.17665942}}, who used the DAWN JWST Archive (DJA) to construct a uniformly selected sample of 146 LRDs at $z = 2.0$--$9.3$. Their selection requires both a V-shaped UV--optical continuum in the NIRSpec/PRISM spectra and a compact morphology in NIRCam/F444W imaging, yielding an intentionally conservative, high-purity sample with a remarkably high broad-Balmer-line detection rate of 98\%. \citet{deGraaff2026} show that these objects form a coherent spectroscopic family that is likely distinct from ordinary obscured Type~1 AGNs. This makes their sample a large, high-confidence set of classical LRDs well suited for population studies. The major surveys included in this LRD selection are CANUCS \citep{Sarrouh2026}, CAPERS \citep{Dickinson2024}, CEERS \citep{Finkelstein2025}, JADES \citep{Eisenstein2026}, MoM \citep{Naidu2026}, NEXUS \citep{Shen2024}, NIRSpec GTO-Wide \citep{Maseda2024}, RUBIES \citep{rubies}, and UNCOVER \citep{Bezanson2024}. Additional programs were also considered; we refer readers to \citet[][and references therein]{deGraaff2026} for further details.

\begin{figure*}
    \centering
    \includegraphics[width=0.727\linewidth]{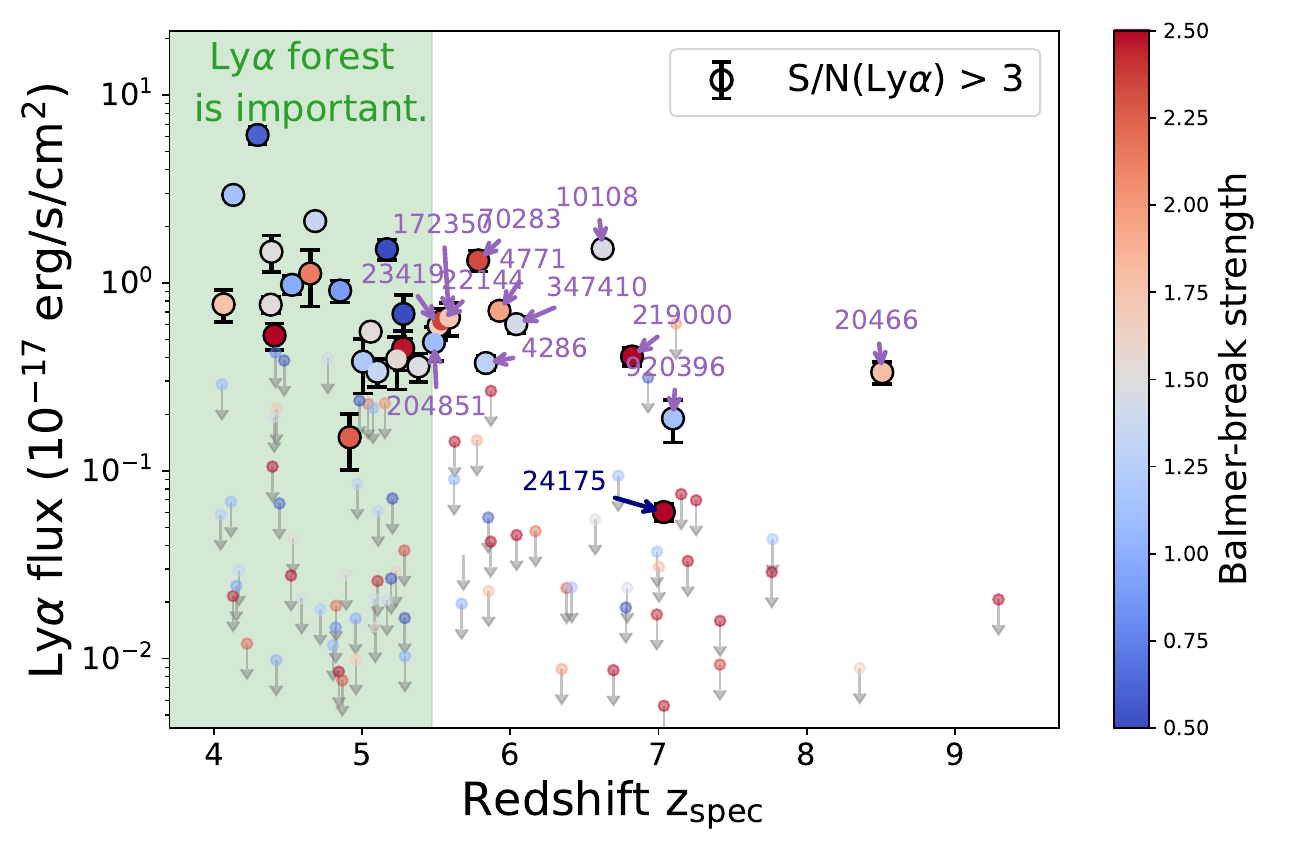}
    \caption{The final sample of 110 LRDs at $z\geq4$ with NIRSpec/PRISM coverage of Ly$\alpha$. Sources with Ly$\alpha$ detections at S/N $\geq3$ (see Section \ref{sec:int_lya_ana}) are shown as circles with error bars, while the remaining sources are shown as upper limits on the Ly$\alpha$ flux. Sources are color-coded according to their Balmer-break strengths reported in \citet{deGraaff2026}. The LRDs at $z\geq5.5$ labeled by their source IDs in this plot are those included in our spatially resolved Ly$\alpha$ analysis using broadband imaging (Section \ref{sec:lya_map}).}
    \label{fig:sample}
\end{figure*}

In this work, we restrict our analysis to $z\ge4$, where Ly$\alpha$ is covered by the nominal wavelength range of NIRSpec/PRISM ($0.6$--$5.3\,\mu$m). Excluding 3 sources lacking usable Ly$\alpha$ coverage because the corresponding spectral region is affected by detector gaps or masking during data reduction, we have a final sample of 110 LRDs. Figure \ref{fig:sample} shows the redshift distribution of the sample. We note that Abell2744-QSO1 (24175 in Figure \ref{fig:sample}) is multiply imaged by strong gravitational lensing, and its Ly$\alpha$ fluxes are fully consistent across the images within uncertainties (see Appendix \ref{app:lya_fit}), so we include only one image of this source in Figure \ref{fig:sample}.

We use the NIRSpec/PRISM spectroscopic data from the DJA. Specifically, we adopt the public DJA NIRSpec reduction products from version 4.4 \citep{Heintz2025,rubies}. This latest version improves on the previous one by applying an empirical wavelength calibration correction based on the source centroid in the shutter, updating the flux-calibration reference files to extend the usable spectral range, and using optimal 1D extraction \citep{Horne1986} with a wavelength-dependent slit-loss correction.

For the spatially resolved Ly$\alpha$ analysis, we use the corresponding HST/ACS and NIRCam imaging data from the DJA reductions \citep{Valentino2023}. We adopt the latest DJA imaging release, version 7, for all fields containing LRDs, including GOODS-S (DJA imaging prefix: \texttt{gds}), Abell 2744 (\texttt{abell2744clu}), EGS (\texttt{ceers-full}), UDS (\texttt{primer-uds}), and COSMOS (\texttt{primer-cosmos}). The HST and NIRCam/LW images are all on $0.04''$ pixel grids, while the NIRCam/SW data also have $0.04''$ pixels in most fields, except in Abell 2744 and GOODS-S, where the mosaics are provided on $0.02''$ pixel grids. All mosaics are astrometrically registered  to a common world coordinate system (WCS) tied to Gaia DR3 \citep{Gaia2021}.

\section{Integrated Ly-alpha Emission in LRDs}
\label{sec:lya_global}

In this section, we investigate the integrated Ly$\alpha$ emission properties of the LRD sample using the JWST/NIRSpec PRISM spectra ($R\sim100$). We first measure the Ly$\alpha$ line properties from the PRISM data through forward modeling of the continuum and line emission. We then compare the Ly$\alpha$ measurements with the UV-continuum, Balmer-line, and [\ion{O}{3}] properties of the sources. These comparisons allow us to assess how Ly$\alpha$ emission relates to the UV and optical nebular properties of LRDs, and thereby provide insight into the physical origin and nature of Ly$\alpha$ emission in this population.

\subsection{Analysis}\label{sec:int_lya_ana}

We measure Ly$\alpha$ fluxes from the PRISM spectra by fitting the rest-frame spectral region blueward of 2500\,\AA. For each source, we model the intrinsic spectrum as the sum of a power-law continuum and a single Gaussian component representing Ly$\alpha$,
\begin{equation}
f_{\lambda}^{\rm int}(\lambda)
=
A\left(\frac{\lambda}{\lambda_0}\right)^{\beta}
+
\frac{F_{\rm Ly\alpha}}{\sqrt{2\pi}\sigma_{\lambda}}
\exp\left[
-\frac{(\lambda-\lambda_{\rm Ly\alpha})^2}{2\sigma_{\lambda}^2}
\right],
\end{equation}
where $A$ is the continuum normalization, $\beta$ is the continuum slope, $\lambda_0$ is an arbitrary reference wavelength, $F_{\rm Ly\alpha}$ is the integrated Ly$\alpha$ flux and $\sigma_{\lambda}$ is the Gaussian width. $\beta$ is constrained by the continuum over the wavelength range 1300--2500\,\AA\ \citep{Calzetti1994}. Given the spectral resolution of the PRISM data, we fix $\lambda_{\rm Ly\alpha}$ based on the source redshift and use a Gaussian component as a simple empirical description of the Ly$\alpha$ emission profile.

To account for absorption by the  Intergalactic Medium (IGM), we multiply the intrinsic model by the IGM transmission curve appropriate for the source redshift following \citet{Inoue2014}. The transmitted model spectrum is therefore
\begin{equation}
f_{\lambda}^{\rm trans}(\lambda)
=
f_{\lambda}^{\rm int}(\lambda)\,T_{\rm IGM}(\lambda,z),
\end{equation}
where $T_{\rm IGM}(\lambda,z)$ is the IGM transmission function. Because the observed spectra are additionally broadened by the instrumental response, we convolve the transmitted model with the wavelength-dependent NIRSpec/PRISM line-spread function (LSF)\footnote{\href{https://jwst-docs.stsci.edu/jwst-near-infrared-spectrograph/nirspec-instrumentation/nirspec-dispersers-and-filters\#gsc.tab=0}{https://jwst-docs.stsci.edu/jwst-near-infrared-spectrograph/nirspec-instrumentation/nirspec-dispersers-and-filters\#gsc.tab=0}},
\begin{equation}
f_{\lambda}^{\rm model}(\lambda)
=
f_{\lambda}^{\rm trans}(\lambda)\otimes {\rm LSF}_{\rm PRISM}(\lambda).
\end{equation}

We determine the best-fit parameters using a Markov Chain Monte Carlo (MCMC) analysis with the package \textsc{emcee} \citep{emcee}. The likelihood is constructed assuming Gaussian-distributed flux uncertainties,
\begin{equation}
\ln \mathcal{L}
=
-\frac{1}{2}
\sum_i
\left[
\frac{\left(f_{\lambda,i}^{\rm obs}-f_{\lambda,i}^{\rm model}\right)^2}{\sigma_i^2}
+
\ln(2\pi\sigma_i^2)
\right],
\end{equation}
where $f_{\lambda,i}^{\rm obs}$ and $\sigma_i$ are the observed flux density and its uncertainty in the $i$th spectral pixel, respectively. The free parameters in the fit are the continuum normalization, continuum slope, Ly$\alpha$ width, and Ly$\alpha$ flux. We adopt the posterior distribution from the MCMC chains to estimate the best-fit Ly$\alpha$ flux and its associated uncertainty. Throughout this work, we define a Ly$\alpha$ detection as a source with ${\rm S/N} \equiv F_{\rm Ly\alpha}/\sigma(F_{\rm Ly\alpha}) \geq 3$. For sources with ${\rm S/N} < 3$, we instead report the $1\sigma$ upper limit, $F_{\rm Ly\alpha}^{\rm UL}$, derived from the posterior distribution such that $\int_{0}^{F_{\rm Ly\alpha}^{\rm UL}} P(F_{\rm Ly\alpha})\, dF_{\rm Ly\alpha} = 0.68$. In addition to the Ly$\alpha$ flux, the MCMC fitting also constrains the underlying UV continuum, from which we derive $M_{\rm UV}$ for each source. All fitting results of the Ly$\alpha$-detected LRDs, including the best-fit parameters and data--model comparison plots, are presented in Appendix \ref{app:lya_fit}.

In our fiducial Ly$\alpha$ modeling, we do not include a damped Ly$\alpha$ absorption (DLA) component, because DLA detectability is generally poor in PRISM $R\sim100$ spectra \citep{Huberty2025}. However, we test the impact of this assumption in Appendix \ref{app:dla} by repeating the Ly$\alpha$ flux measurements with an additional DLA component. Based on the Bayesian information criterion (BIC), the vast majority of the LRD PRISM spectra do not show evidence for strong DLA absorption. Including the DLA component changes the measured Ly$\alpha$ fluxes by only $\sim20\%$ and therefore does not affect any of the conclusions of this work.

Finally, to place the Ly$\alpha$ measurements in the broader context of the optical spectral properties of LRDs, we also make use of existing measurements of their rest-frame optical features. Specifically, in the subsequent analysis, we adopt H$\alpha$, the Balmer decrement, and [O~III] measurements from \citet{deGraaff2026}. Because these measurements are based on NIRSpec/PRISM spectra, we use the total H$\alpha$ luminosities rather than attempting a broad/narrow decomposition.

\subsection{Results}

\subsubsection{Ly$\alpha$ vs. UV continuum}
\label{sec:lya_uv}

We first compare the integrated Ly$\alpha$ properties of the LRD sample with their UV continuum. The analysis proceeds in two complementary
steps: (i)~a comparison of the Ly$\alpha$-detected LRDs to ordinary
Ly$\alpha$ emitters (LAEs) in the $L_{\rm Ly\alpha}$--$M_{\rm UV}$ plane; and
(ii)~a comparison of the Ly$\alpha$ detection fraction, $X_{\rm Ly\alpha}$, against UV-selected Lyman-break galaxies for the
full LRD sample. The two steps together cover the LRD populations with and without Ly$\alpha$ detections.

For the detected-vs-detected comparison, we use the LAEs of \citet{Kerutt2022} as a benchmark for ordinary high-redshift
galaxies. By construction this
LAE sample is restricted to Ly$\alpha$-detected sources, containing $\sim 2000$ LAEs at
$2.9 < z < 6.6$ with Ly$\alpha$ measurements from MUSE-Wide$+$Deep
\citep{Bacon2017,Inami2017,Herenz2017,Urrutia2019} and UV continuum
constraints from archival HST imaging
\citep{Giavalisco2004,Koekemoer2011,Grogin2011}. We focus on their subsample at $z \geq 4$.

\begin{figure*}
    \includegraphics[width=0.97\textwidth]{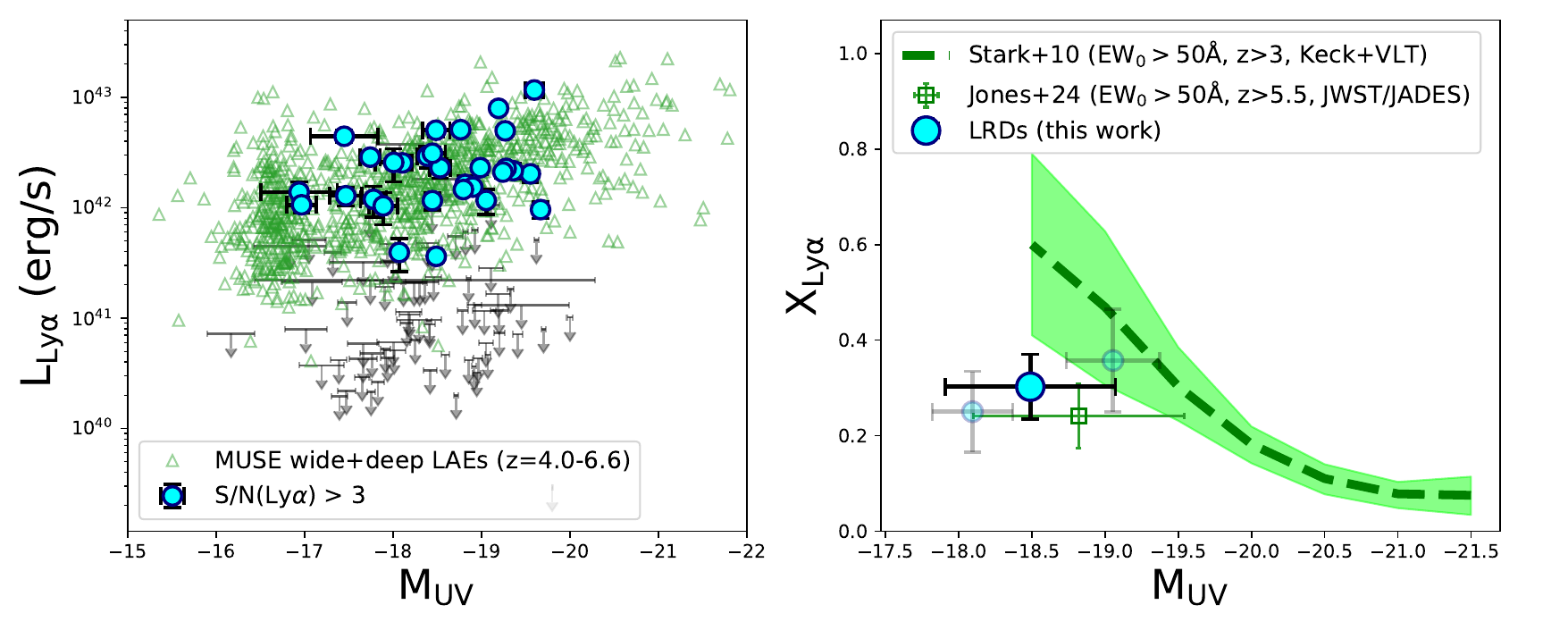}
    \caption{\textbf{Left:} Ly$\alpha$ luminosity as a function of UV absolute magnitude for the LRD sample. Sources with Ly$\alpha$ detections at S/N $\geq3$ are shown as circles with error bars, while non-detections are shown as upper limits. For comparison, the green triangles show MUSE-Wide+Deep Ly$\alpha$ emitters at $z=4.0$--6.6 from \citet{Kerutt2022}. \textbf{Right:} Ly$\alpha$ detection fraction, $X_{\rm Ly\alpha}$, as a function of $M_{\rm UV}$ for the full LRD sample, shown by the cyan circle with a black edge. We also divide the LRD sample into two subsamples using the median $M_{\rm UV}$, shown by the light-cyan circles with gray edges. For comparison, we show measurements for Lyman-break galaxies from \citet{Stark2010} and \citet{Jones2024} with rest-frame ${\rm EW}_0>50$\,\AA, broadly matching the observed Ly$\alpha$ equivalent widths of our Ly$\alpha$-detected LRDs (see Figure \ref{fig:lya_EW}).}
    \label{fig:lya_muv}
\end{figure*}

As shown in the left panel of Figure~\ref{fig:lya_muv}, the
Ly$\alpha$-detected LRDs do not appear exceptional compared with the
MUSE LAEs in the $L_{\rm Ly\alpha}$--$M_{\rm UV}$ plane: they overlap
with the locus occupied by the LAEs with $-20 \lesssim M_{\rm UV} \lesssim -17$,
with a $p$-value of $\sim 0.2$ from a two-dimensional K--S test
\citep{Peacock1983,Fasano1987}. We emphasize that this statement
applies to the Ly$\alpha$-detected subsamples; whether the
similarity extends to the full LRD population, which contains a
substantial fraction of Ly$\alpha$ upper limits visible in
Figure~\ref{fig:lya_muv}, is addressed separately by the
$X_{\rm Ly\alpha}$ analysis later presented in this section.

\begin{figure}
    \centering
    \includegraphics[width=0.97\linewidth]{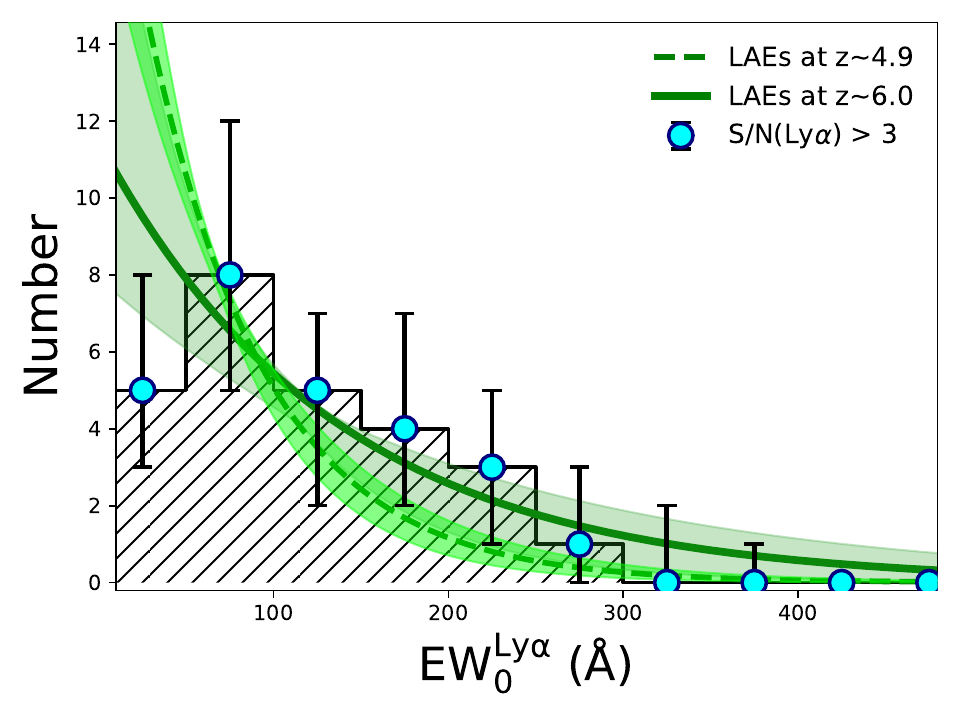}
    \caption{Rest-frame Ly$\alpha$ equivalent-width (EW$_{\rm Ly\alpha}^{0}$) distribution. The blue circles with error bars show the Ly$\alpha$-detected LRDs in our sample. The uncertainties are estimated by bootstrapping the sample and resampling each EW measurement using its corresponding fitting uncertainty derived in Section \ref{sec:int_lya_ana}. For comparison, the green shaded histograms and curves show the EW$_{\rm Ly\alpha}^{0}$ distributions of MUSE Hubble Ultra Deep Field LAEs at $z\sim4.9$ and $z\sim6.0$ from \citet{Hashimoto2017}.}
    \label{fig:lya_EW}
\end{figure}

We also examine the rest-frame Ly$\alpha$ equivalent-width distribution
shown in Figure~\ref{fig:lya_EW}. We do not attempt to fit the
intrinsic EW distribution of the LRD sample here, because such an
analysis would require a careful treatment of completeness and
selection effects. In particular, the LRD spectra used in this work are
drawn from a heterogeneous set of surveys and programs, whose combined
selection function is difficult to quantify in a uniform way.
Nevertheless, it is notable that the observed EW distribution of the
Ly$\alpha$-detected LRDs is broadly consistent with that of normal LAEs
at similar redshifts. Taken together, these indicate that, \emph{when Ly$\alpha$ is detected}, the UV and Ly$\alpha$ properties of LRDs are not unusual compared with those of ordinary high-redshift star-forming galaxies.

To extend the comparison to the full LRD population regardless of Ly$\alpha$ detections, we turn to the Ly$\alpha$ detection fraction at fixed UV luminosity, $X_{\rm Ly\alpha}$, defined as the number of LRDs with Ly$\alpha$ detections divided by the total number of LRDs in a given $M_{\rm UV}$ bin. This second-tier comparison is benchmarked against
UV-selected Lyman-break galaxies, which are
not pre-selected on Ly$\alpha$ and therefore include both
detections and non-detections.

The right panel of Figure \ref{fig:lya_muv} shows the $X_{\rm Ly\alpha}$ of the LRDs. For comparison, we show the measurements from \citet{Stark2010}, based on Keck and VLT spectroscopy for Lyman-break galaxies over $3<z<7$, and from \citet{Jones2024}, based on recent deep NIRSpec spectroscopy from the JADES program that provides updated constraints on the Ly$\alpha$ emitter fraction at $z\gtrsim5$. Many other studies have reached broadly similar conclusions for Lyman-break galaxies over this redshift range using a variety of surveys and spectroscopic data sets (e.g., \citealt{Ono2012,Pentericci2014,Pentericci2018}). The measured $X_{\rm Ly\alpha}$ values for LRDs are broadly consistent with these previous measurements for Lyman-break galaxies at similar redshifts and UV luminosities. This suggests that, from the perspective of UV-selected Ly$\alpha$ visibility, LRDs do not appear to form a distinct Ly$\alpha$ population. Instead, their Ly$\alpha$ detection fraction appears broadly consistent with that of the broader high-redshift galaxy population, although a fully homogeneous comparison would require a detailed treatment of the heterogeneous selection functions and spectroscopic depths.

\subsubsection{Ly$\alpha$ vs. Balmer lines}
\label{sec:lya_H}
\begin{figure*}
    \includegraphics[width=0.97\textwidth]{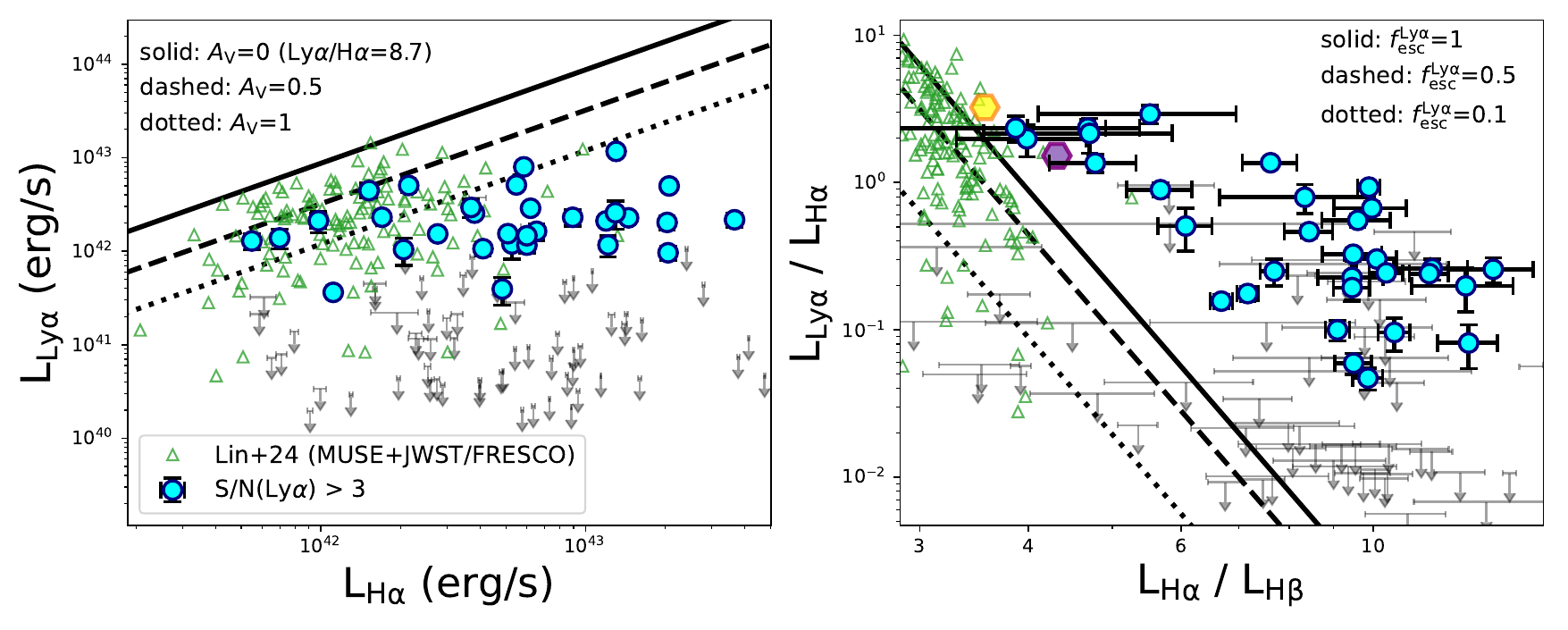}
    \caption{\textbf{Left:} Ly$\alpha$ luminosity as a function of H$\alpha$ luminosity for the LRD sample. Sources with Ly$\alpha$ detections at S/N $\geq 3$ are shown as circles with error bars, while non-detections are shown as upper limits. The green triangles show the comparison sample of normal star-forming galaxies from \citet{Lin2024}. The black curves show the expected relation assuming Case B recombination with $n_{\rm e}=10^2$~cm$^{-3}$ and $T_{\rm e}=10^4$~K \citep{Osterbrock2006}, attenuated with the \citet{Calzetti2000} dust law for $A_V=0$, 0.5, and 1.0 mag. \textbf{Right:} $L_{\mathrm{Ly}\alpha}/L_{\mathrm{H}\alpha}$ as a function of $L_{\mathrm{H}\alpha}/L_{\mathrm{H}\beta}$ for the same samples. The black curves are constructed by assuming the Case B recombination and adopting different Ly$\alpha$ escape fractions, $f_{\rm esc}^{\rm Ly\alpha}=1$, 0.5, and 0.1. The orange hexagon marks the line ratios of the SDSS Type I quasar composite \citep{VandenBerk2001}. The purple hexagon marks the median line ratios of the $z\gtrsim6$ red quasars from the SHELLQs survey \citep{Matsuoka2025}.}
    \label{fig:lya_balmer} 
\end{figure*}

We next compare the integrated Ly$\alpha$ emission of LRDs with their Balmer-line properties in Figure~\ref{fig:lya_balmer}. As a benchmark for normal high-redshift star-forming galaxies, we use the sample of \citet{Lin2024}, which combines H$\alpha$ measurements from NIRCam/WFSS spectroscopy acquired by the FRESCO program \citep{Oesch2023} with Ly$\alpha$ measurements from MUSE for star-forming galaxies at $z=4.9$--6.3 \citep{Bacon2017,Bacon2023}. In the left panel of Figure~\ref{fig:lya_balmer}, we compare $L_{\rm Ly\alpha}$ and $L_{\rm H\alpha}$, and also show the expected relation from Case B recombination for different levels of dust attenuation. Relative to both these expectations and the \citet{Lin2024} comparison sample, the LRDs lie systematically low in Ly$\alpha$ at fixed H$\alpha$. In other words, although many LRDs have luminous Balmer emission, their escaping Ly$\alpha$ is faint compared with that of ordinary star-forming galaxies with similar H$\alpha$ luminosities. This indicates that the Balmer emission in LRDs is not accompanied by proportionally strong Ly$\alpha$, in contrast to what would be expected if both lines traced the same gas under ordinary nebular conditions.

We further compare $L_{\rm Ly\alpha}/L_{\rm H\alpha}$ with $L_{\rm H\alpha}/L_{\rm H\beta}$ in the right panel of Figure~\ref{fig:lya_balmer}. Because \citet{Lin2024} do not have H$\beta$ measurements for their sample, we estimate $L_{\rm H\alpha}/L_{\rm H\beta}$ from their SED-based dust attenuation using the \citet{Calzetti2000} attenuation law. In this panel, we also show the expected relation from Case B recombination for different Ly$\alpha$ escape fractions. Normal galaxies from the \citet{Lin2024} sample broadly follow the expected trend in which dust attenuation and varying Ly$\alpha$ escape fraction move galaxies through this plane. By contrast, the LRDs occupy a clearly different region, with relatively low ${\rm Ly}\alpha/{\rm H}\alpha$ yet large ${\rm H}\alpha/{\rm H}\beta$. We also show the typical line ratios of Type~I quasars \citep{VandenBerk2001} and $z\gtrsim6$ red quasars \citep{Matsuoka2025}, shown as orange and purple hexagons, respectively. The observed hydrogen line ratios in LRDs are likewise distinct from those of quasars.

Taken together, these results indicate that the Ly$\alpha$ and Balmer emission
in LRDs are not simply scaled versions of the nebular emission seen in ordinary
galaxies, but instead point to a different effective line-emitting and/or Ly$\alpha$-escaping environment.
As discussed in Section~\ref{sec:lya_uv}, the escaping Ly$\alpha$ is consistent
with being dominated by host-galaxy star formation. If so, the Ly$\alpha$-emitting
gas should also produce H$\alpha$ and H$\beta$, but with a
Balmer ratio characteristic of normal \ion{H}{2} regions rather than that of the
dominant LRD component.

This picture makes a testable spatial prediction. If the Ly$\alpha$-emitting
gas is distributed differently from the compact central engine, as suggested by the spatially resolved analysis in 
Section~\ref{sec:lya_map}, then the Balmer decrement should vary with
position in LRDs where Ly$\alpha$ is detected. To test this, we stack the 2D
NIRSpec/PRISM spectra of the Ly$\alpha$-detected and Ly$\alpha$-undetected
subsamples separately and measure H$\alpha$/H$\beta$ as a function of spatial
position, restricting the analysis to $4\leq z\leq7$, where H$\alpha$ falls
within the PRISM range. For each source we subtract a per-row linear continuum
fitted to the rest-frame sidebands of each line, resample onto a common
rest-frame grid, and normalize by the continuum at rest-frame $5100$~\AA{}
before stacking into mean, median, and inverse-variance-weighted (IVW) 2D
spectra. For each spatial row we then measure H$\alpha$ by integrating over
$6400$--$6750$~\AA{} (wide enough to enclose both the narrow and broad
components) and H$\beta$ over $4800$--$4925$~\AA{}, after removing the per-row
best-fit [\ion{O}{3}]$\lambda\lambda4959,5007$ Gaussians with the doublet ratio
fixed at $2.98$ \citep{Storey2000}. Gaussian fitting yields ratios consistent
with this box integration to within $\sim5\%$, and uncertainties are derived
from a bootstrap-plus-Monte~Carlo procedure (200 resamplings per subsample,
each spectrum perturbed by its error map before re-stacking and re-fitting).

\begin{figure*}
\centering
    \includegraphics[width=0.87\textwidth]{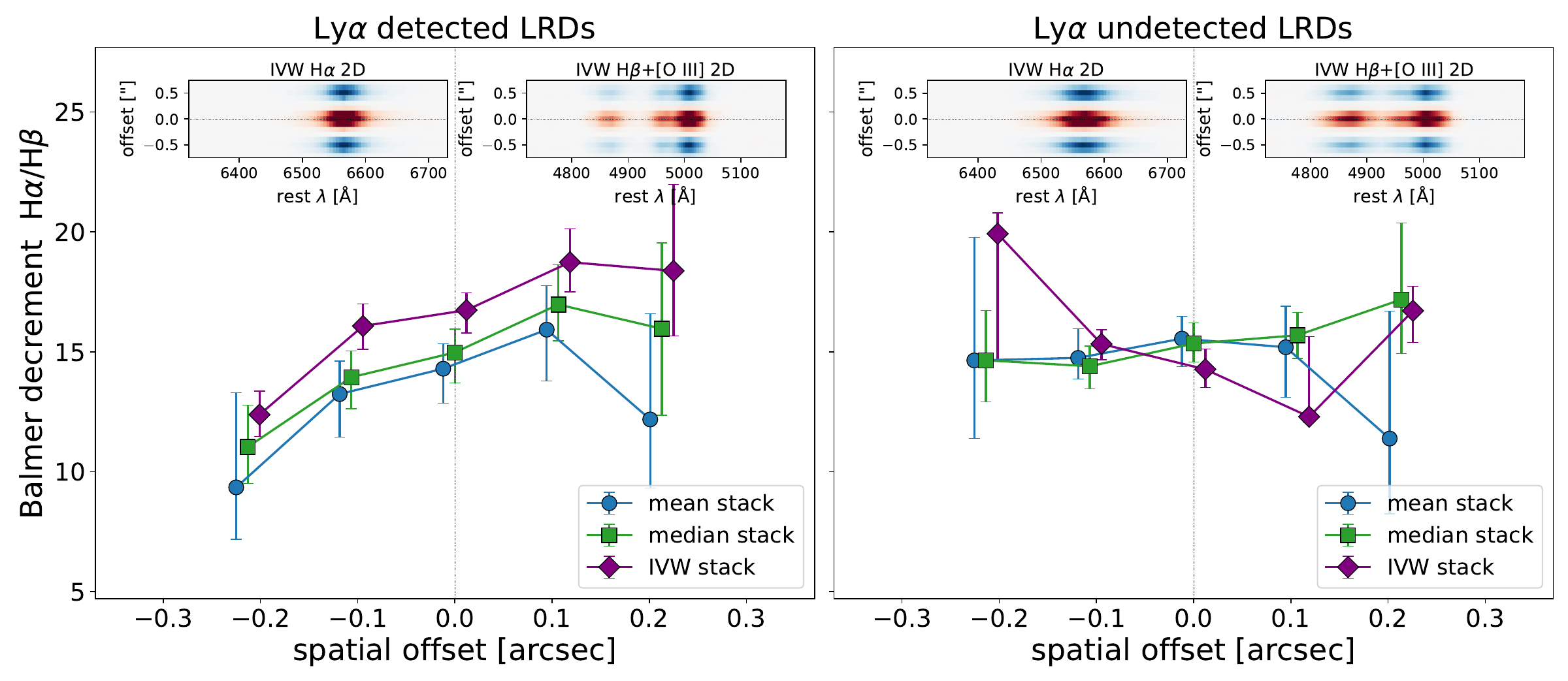}
    \caption{Balmer decrement, H$\alpha$/H$\beta$, as a function of spatial position for the Ly$\alpha$-detected (left) and Ly$\alpha$-undetected (right) LRD subsamples, measured from stacked 2D NIRSpec/PRISM spectra. Results from three different stacking weights are shown: mean (blue circles), median (green squares), and inverse-variance-weighted (IVW; purple diamonds). Error bars are estimated from bootstrap resamplings, as described in Section~\ref{sec:lya_H}. The insets at the top of each panel show the IVW H$\alpha$ 2D stack and the IVW H$\beta$+[O~{\sc iii}] 2D stack.}
    \label{fig:stack_decrement}
\end{figure*}

Figure~\ref{fig:stack_decrement} shows the resulting H$\alpha$/H$\beta$ as a
function of position. While the absolute decrement depends on the stacking
weights, the trends are qualitatively consistent across the three stacks. The
Ly$\alpha$-undetected stack shows no clear spatial variation, remaining
consistent with a constant decrement across the inner $\pm0.2''$. The
Ly$\alpha$-detected stack, however, shows evidence that H$\alpha$/H$\beta$ decreases from the
trace center toward off-trace positions.

This off-trace trend must be interpreted with care. At the sample median
$z\approx5$, the observed H$\alpha$ and H$\beta$ fall at $\approx4$ and
$\approx3~\mu$m, where the NIRSpec/PRISM cross-dispersion PSF FWHM ($\approx0.2''$)
is comparable to the outermost bin in which the decrement is still measurable,
so the off-trace spectra remain significantly contaminated by the central
source through the PSF wings. Even so, the fact that H$\alpha$/H$\beta$ declines
with offset in the Ly$\alpha$-detected stack but not in the Ly$\alpha$-undetected
one suggests that, when Ly$\alpha$ is detectable, an additional emission
component contributes that is both more spatially extended and has a lower
Balmer ratio than the dominant central component. Together with
Figure~\ref{fig:lya_balmer}, these measurements suggest that the dominant Balmer-line emission and the escaping Ly$\alpha$ are not tracing the same gas under ordinary nebular conditions.

\subsubsection{Ly$\alpha$ vs. [O III]$\lambda$5007 emission}
\label{sec:lya_O}

\begin{figure*}
    \includegraphics[width=0.97\textwidth]{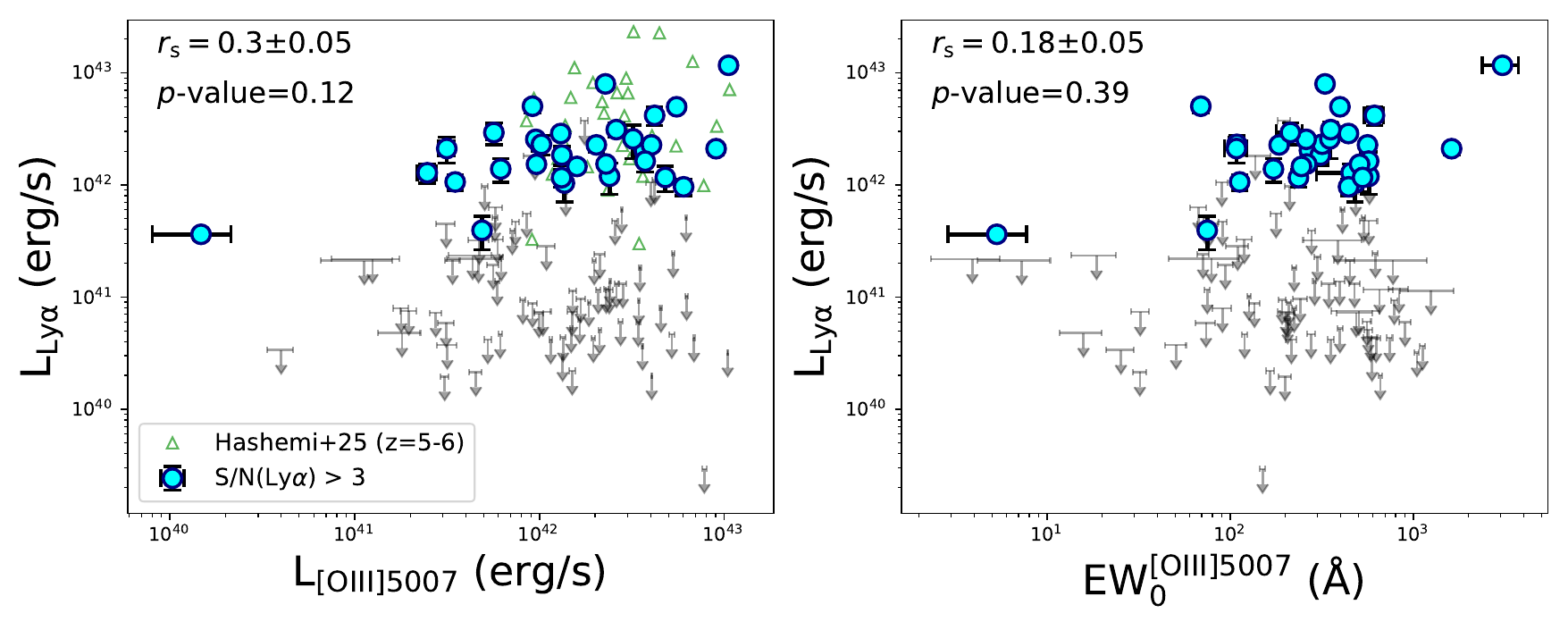}
    \caption{\textbf{Left}: Ly$\alpha$ luminosity as a function of [O III]$\lambda5007$ luminosity. The green triangles show the comparison sample of normal $z\sim5$--6 galaxies from \citet{Hashemi2025}.  \textbf{Right}: Ly$\alpha$ luminosity as a function of rest-frame [O III]$\lambda5007$ equivalent width. In each panel, the quoted Spearman rank correlation coefficient $r_{\rm s}$ is estimated from bootstrap resampling; the quoted uncertainty corresponds to the dispersion of the resulting $r_{\rm s}$ distribution. }
    \label{fig:lya_oiii}
\end{figure*}

In Figure~\ref{fig:lya_oiii} we compare the integrated Ly$\alpha$ emission of LRDs with their [\ion{O}{3}] properties. In the left panel we plot $L_{\rm Ly\alpha}$ against $L_{\rm [O\,III]}$ and compare the LRDs with the $z\sim5$--$6$ galaxy sample of \citet{Hashemi2025}. The LRDs broadly overlap with the locus occupied by these normal high-redshift galaxies, suggesting that, at fixed [\ion{O}{3}] luminosity, the escaping Ly$\alpha$ emission in LRDs is consistent with that of the broader galaxy population.

We quantify the trends in Figure~\ref{fig:lya_oiii} using the Spearman rank correlation coefficient, $r_s$. We estimate the uncertainty in $r_s$ with a bootstrap procedure, in which we resample the Ly$\alpha$-detected sources and perturb their measurements according to the observational uncertainties. The Ly$\alpha$ luminosity is correlated with $L_{\rm [O\,III]}$, with $r_s=0.30\pm0.05$. Its correlation with EW$_{\rm [O\,III]}$ is weaker, with $r_s=0.18\pm0.05$, and is driven almost entirely by a single source with EW$_{\rm [O\,III]}>2000$~\AA{}; removing this object reduces the coefficient to $r_s<0.1$ ($p>0.9$), consistent with no correlation. We further assess whether the difference between the two correlation strengths is meaningful by computing
$\Delta r_s = r_s(L_{\rm Ly\alpha},L_{\rm [O\,III]}) -
r_s(L_{\rm Ly\alpha},{\rm EW}_{\rm [O\,III]})$
within each bootstrap realization. Among the $10^4$ bootstrap realizations, $\approx95\%$ yield $\Delta r_s>0$, suggesting, at modest significance, that the luminosity–luminosity relation is stronger than the relation with equivalent width.

Although the modest size of the Ly$\alpha$-detected LRD sample means that larger samples will be needed to confirm it, this difference in correlation strength is already physically suggestive. If the rest-frame optical continuum near
[\ion{O}{3}] arose from the same physical component as the [\ion{O}{3}]-emitting gas, then normalizing the line luminosity by that continuum, i.e.\ using EW$_{\rm [O\,III]}$ rather than $L_{\rm [O\,III]}$, should not substantially weaken its correlation with Ly$\alpha$. Instead, we observe the opposite: the luminosity--luminosity relation is stronger, suggesting that the optical continuum introduces additional scatter that is less closely linked to the gas regulating Ly$\alpha$ escape. This interpretation is consistent with recent work arguing that the red optical continuum in LRDs is closely tied to the compact
central component, whereas [\ion{O}{3}] emission arises predominantly in the host galaxy \citep[e.g.,][]{deGraaff2026,Sun2026}. If Ly$\alpha$ likewise traces
gas on host-galaxy scales, as suggested by the preceding sections, then a tighter relation with $L_{\rm [O\,III]}$ than with EW$_{\rm [O\,III]}$ follows naturally: the continuum entering the equivalent-width denominator originates, at
least in part, from a distinct physical component. Our results therefore provide supporting evidence for a picture in which the optical continuum and the [\ion{O}{3}]-emitting gas are at least partly associated with distinct physical components in LRDs.

\section{Spatially Resolved Ly-alpha Emission in LRDs at $z\ge5.5$} \label{sec:lya_map}

The HST and NIRCam imaging provide a powerful way to constrain the two-dimensional Ly$\alpha$ emission in LRDs at sub-kpc resolution by combining a Ly$\alpha$-sensitive broadband filter with adjacent continuum filters. Such an analysis, however, requires a statistically robust treatment of IGM transmission. At lower redshifts, the Ly$\alpha$ forest transmission affects the flux blueward of Ly$\alpha$, introducing redshift-dependent and sightline-dependent contamination into broadband Ly$\alpha$ measurements \citep[e.g.,][]{Becker2015,Bosman2018,Eilers2018}. Once the Gunn--Peterson trough \citep{Gunn1965} becomes effective, the IGM is essentially opaque blueward of Ly$\alpha$, allowing the broadband flux to be more cleanly separated into Ly$\alpha$ emission and redward UV continuum. We therefore restrict this analysis to $z\gtrsim5.5$. This redshift cut is intentionally conservative relative to the nominal transition at $z\approx5.3$, often associated with the end of reionization \citep[e.g.,][]{Zhu2021,Bosman2022}, because even modest residual uncertainties in IGM transmission can bias spatially resolved Ly$\alpha$ maps more severely than integrated flux measurements.

There are a total of 47 galaxies at $z \gtrsim 5.5$ in the \citet{deGraaff2026} sample with usable NIRSpec/PRISM Ly$\alpha$ coverage. Of these, 13 show Ly$\alpha$ emission detected at S/N $\geq 3$ in NIRSpec and have the required HST and/or NIRCam Ly$\alpha$ imaging coverage. These 13 LRDs are the focus of the subsequent spatially resolved Ly$\alpha$ analysis. Their detailed information can be found in Table \ref{tab:sample_map}.

\begin{deluxetable*}{lcccccc}
\tablecaption{Ly$\alpha$-detected LRD sample at $z \ge 5.5$ for the spatially resolved analysis \label{tab:sample_map}}
\tablewidth{0pt}
\tablehead{
  \colhead{Source ID} &
  \colhead{Field} &
  \colhead{$z_{\rm spec}$} &
  \colhead{$F_{\rm Ly\alpha, DJA}$} &
  \colhead{S/N$_{\rm Ly\alpha}$} &
  \colhead{Ly$\alpha$ Filter} & 
  \colhead{UV Continuum Filter}\\
  \colhead{} &
  \colhead{} &
  \colhead{} &
  \colhead{($10^{-18}$ erg/s/cm$^2$)} &
  \colhead{} &
  \colhead{} &
  \colhead{} 
}
\startdata
204851 & GOODS-S & 5.49 & 4.83 & 4.9 & HST/F775W & (F850LP\tablenotemark{$^\dagger$}, F090W, F115W, F150W) \\
23419 & UDS & 5.52 & 5.93 & 6.2 & HST/F814W\tablenotemark{$^\dagger$} & (F090W, F115W, F150W) \\
22144 & COSMOS & 5.54 & 6.27 & 6.1 & HST/F814W\tablenotemark{$^\dagger$} & (F090W, F115W, F150W) \\
172350 & UDS & 5.59 & 6.53 & 5.0 & F090W & (F115W, F150W, F200W\tablenotemark{$^\dagger$}) \\
70283 & COSMOS & 5.78 & 13.20 & 7.8 & F090W & (F115W, F150W, F200W\tablenotemark{$^\dagger$}) \\
4286 & A2744 & 5.84 & 6.02 & 11.3 & F090W & (F115W, F150W, F200W\tablenotemark{$^\dagger$}) \\
4771 & COSMOS & 5.93 & 7.10 & 11.3 & F090W & (F115W, F150W, F200W\tablenotemark{$^\dagger$}) \\
347410 & COSMOS & 6.04 & 6.03 & 10.0 & F090W & (F115W, F150W, F200W\tablenotemark{$^\dagger$}) \\
10108 & EGS & 6.62 & 15.21 & 26.9 & F090W & (F115W, F150W, F200W\tablenotemark{$^\dagger$}) \\
219000 & GOODS-S & 6.82 & 4.06 & 9.3 & F090W & (F115W, F150W, F200W\tablenotemark{$^\dagger$}) \\
24175 & A2744 & 7.04 & 4.14 & 9.6 & F090W & (F115W, F150W, F200W\tablenotemark{$^\dagger$}) \\
920396 & UDS & 7.10 & 1.90 & 3.9 & F090W & (F115W, F150W, F200W\tablenotemark{$^\dagger$}) \\
20466 & A2744 & 8.51 & 4.51 & 7.5 & F115W & (F150W, F200W\tablenotemark{$^\dagger$}) \\
\enddata
\tablenotetext{$$^\dagger$$}{The filter to which all other filters are PSF-matched.}
\tablecomments{Columns: (1) source ID in \citet{deGraaff2026}; (2) survey field; (3) spectroscopic redshift; (4) apparent DJA spectroscopic Ly$\alpha$ flux; (5) DJA Ly$\alpha$ signal-to-noise ratio; (6) broadband filter containing Ly$\alpha$, with HST filters indicated; and (7) broadband filters used to derive the continuum at the Ly$\alpha$ wavelength.}
\end{deluxetable*}

\subsection{Analysis}
\label{sec:map_analysis}

To produce the spatially resolved Ly$\alpha$ maps, we first identify, for each source, the broadband filter that contains redshifted Ly$\alpha$ and a set of adjacent redward UV filters, covering rest-frame $\approx1500-3000$~\AA, that can be used to constrain the underlying continuum (Table \ref{tab:sample_map}). 

We then homogenize the imaging to a common PSF. In most legacy fields, we use the NIRCam empirical point spread functions (ePSFs) released by the DJA \citep{Genin2025}\footnote{\href{https://dawn-cph.github.io/dja/blog/2024/08/16/morphological-data/}{https://dawn-cph.github.io/dja/blog/2024/08/16/morphological-data/}}. Abell 2744 is treated separately because the DJA does not provide NIRCam PSFs for that field. Although the UNCOVER team provides PSFs for Abell 2744 \citep{Weaver2024}, these are defined on a $0\farcs04$ pixel grid, whereas we retain the native DJA $0\farcs02$ pixel scale for the NIRCam/SW mosaics in order to preserve the best possible spatial information. We therefore construct our own NIRCam/SW ePSFs for Abell 2744 following \citet{Anderson2000}, and verify that they are highly consistent with the UNCOVER ePSFs (see Appendix~\ref{app:psf}). Similarly, we construct ePSFs for the HST/ACS bands relevant to this study. We then compute the PSF-matching kernels using \textsc{pypher} \citep{Boucaud2016} and convolve all relevant images to match the broadest PSF among the filters used for that source (see Table \ref{tab:sample_map}).

After PSF matching, we predict the UV continuum contribution in the Ly$\alpha$ band on a pixel-by-pixel basis. We model the UV continuum as a power law, $f_\lambda = A\lambda^\beta$, and compute the expected continuum flux in the Ly$\alpha$-sensitive filter by integrating this spectrum through the filter transmission curve, $R(\lambda)$, while also including the wavelength-dependent IGM transmission, $T_{\rm IGM}$, at the source redshift:
\begin{equation}
    f_{\rm cont} = \frac{\int A\lambda^\beta \, T_{\rm IGM}(z,\lambda)\, R(\lambda)\,\lambda\, d\lambda}{\int R(\lambda)\,\lambda\, d\lambda}.
\end{equation}
For the IGM model, we adopt that of \citet{Inoue2014}. We have verified that our results do not change if we instead use the \citet{Madau1995} model. We consider two approaches for the continuum modeling. In our fiducial method, we allow $\beta$ to vary from pixel to pixel and determine the best-fitting value directly from the imaging, thereby accounting for possible spatial variations in the UV continuum slope across the source. As a consistency check, we also perform an alternative fit in which $\beta$ is fixed to the global UV slope inferred from the integrated photometry and only the normalization is solved for in each pixel. These two methods provide complementary estimates of the continuum and allow us to assess the robustness of the inferred Ly$\alpha$ morphology.

With the PSF-matched imaging and the predicted UV continuum map at the Ly$\alpha$ wavelength in hand, we construct the Ly$\alpha$ map by subtracting the continuum image from the observed image of the Ly$\alpha$-sensitive band. We then convert the residual broadband excess into a Ly$\alpha$ line-flux map. In practice, we note that noise and continuum-model uncertainties can lead to formally negative Ly$\alpha$ fluxes in some pixels, which do not represent physically meaningful emission. We therefore define the displayed Ly$\alpha$ map as $\max(F_{\mathrm{Ly}\alpha}, 0)$ on a pixel-by-pixel basis, while retaining the full signed residuals for uncertainty estimation and other quantitative checks.

The spatially resolved Ly$\alpha$ maps presented in this work rely on accurate WCS alignment across different filters. As described in Appendix \ref{app:wcs}, we perform a detailed astrometric analysis for each source and find that the relative WCS alignment reaches sub-pixel precision in all cases, sufficient for the subsequent analysis of the Ly$\alpha$ spatial distribution.

\begin{figure}
    \centering
    \includegraphics[width=0.97\linewidth]{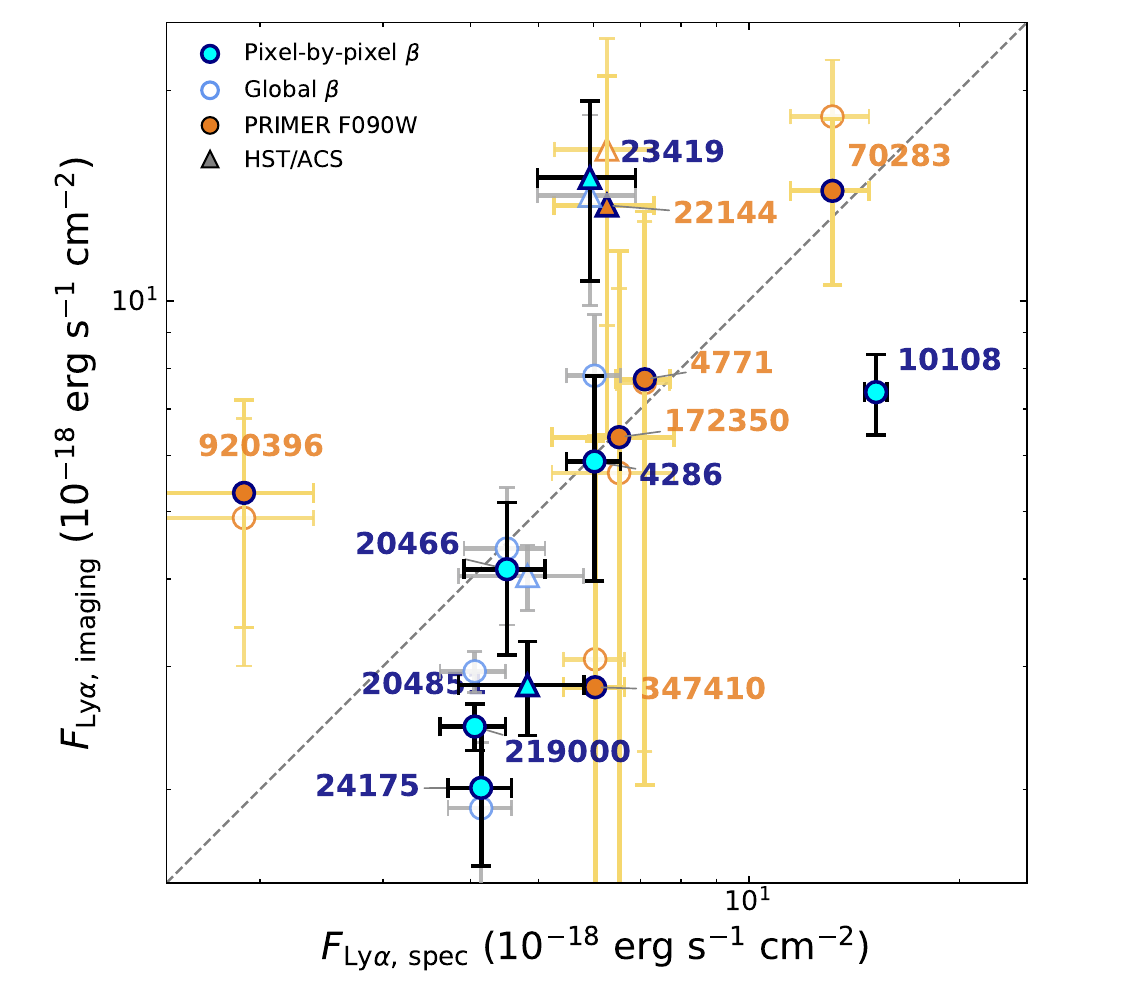}
    \caption{Comparison of Ly$\alpha$ fluxes measured from spectroscopy ($x$-axis) and from our broadband imaging continuum-subtraction method ($y$-axis), where the latter are computed by summing the Ly$\alpha$ maps within a circular aperture of radius $r = 0\farcs16$ centered on the F444W centroid. Filled symbols show the results obtained with pixel-by-pixel UV $\beta$ fitting, while open symbols show those obtained using a global $\beta$. Circles denote Ly$\alpha$ measurements derived from NIRCam imaging, while triangles denote those derived from HST/ACS imaging. The dashed line marks the 1:1 relation.}
    \label{fig:spec_vs_img}
\end{figure}

We note that the fidelity of the resulting Ly$\alpha$ maps depends on the depth of the Ly$\alpha$-sensitive band, which varies across fields. For the majority of the sources, Ly$\alpha$ falls in F090W, and the PRIMER COSMOS and UDS F090W mosaics are appreciably shallower than those in the other fields. The Ly$\alpha$ maps for sources in these two fields are therefore noticeably noisier, and we present them separately in Appendix~\ref{app:map_cosmos_uds}.

Finally, in Figure~\ref{fig:spec_vs_img}, we compare the Ly$\alpha$ fluxes derived from NIRSpec spectroscopy with those measured from our Ly$\alpha$ maps. We find overall good agreement within the uncertainties. This consistency is seen for both the pixel-by-pixel $\beta$ fitting and the global $\beta$ fitting methods, demonstrating that our results are not highly sensitive to the specific treatment of the UV continuum slope. Consistent with the depth differences noted above, the COSMOS and UDS sources (orange symbols in Figure~\ref{fig:spec_vs_img}), whose Ly$\alpha$-sensitive imaging is drawn from the shallower PRIMER mosaics, exhibit significantly larger uncertainties than the sources in the deeper fields.

\subsection{Results}

\begin{figure*}
    \includegraphics[width=1\textwidth]{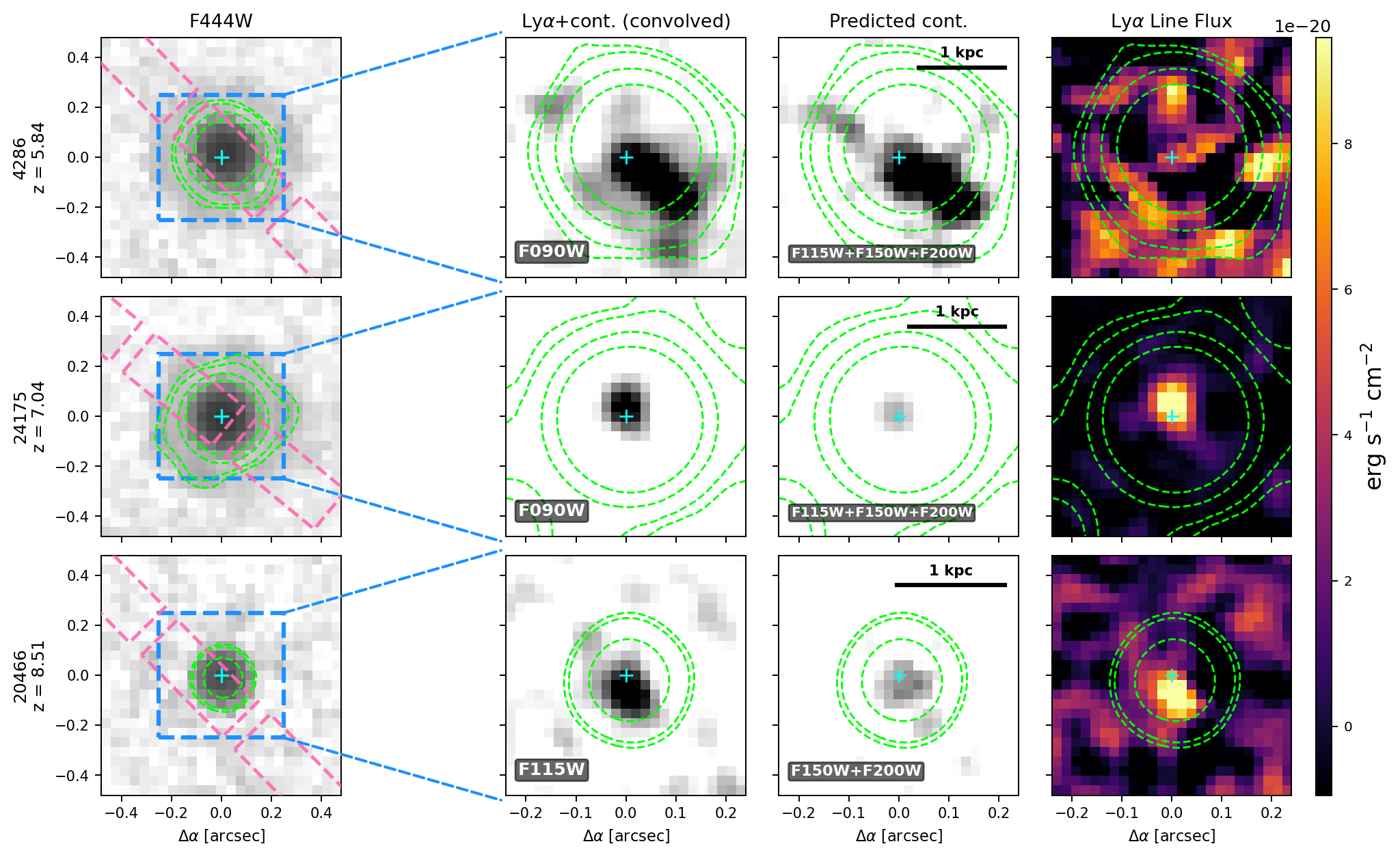}
    \caption{Spatially resolved Ly$\alpha$ maps for three LRDs in Abell~2744. The sources are identified by the IDs shown on the far left. For each source (row), the panels show: (1) the NIRCam/F444W image displayed with a logarithmic color scale, with the MSA slit overplotted in dark pink; (2) the PSF-matched broadband image containing Ly$\alpha$ + continuum; (3) the predicted UV continuum in the Ly$\alpha$ filter from the pixel-by-pixel $\beta$ fit; and (4) the continuum-subtracted Ly$\alpha$ line-flux map. Green dashed contours show the F444W surface brightness at $4\sigma$, $5\sigma$, $10\sigma$, and $20\sigma$, while cyan crosses mark the F444W centroid determined using \texttt{photutils.centroids}. The Ly$\alpha$-filter and continuum panels are PSF-matched and share a common grayscale; therefore, a darker appearance in the Ly$\alpha$ + continuum image than in the continuum-only image implies a flux excess due to Ly$\alpha$. Filter names are annotated in the lower-left corner of each panel.}
    \label{fig:map_1}
\end{figure*}

\begin{figure*}
    \centering
    \includegraphics[width=0.87\textwidth]{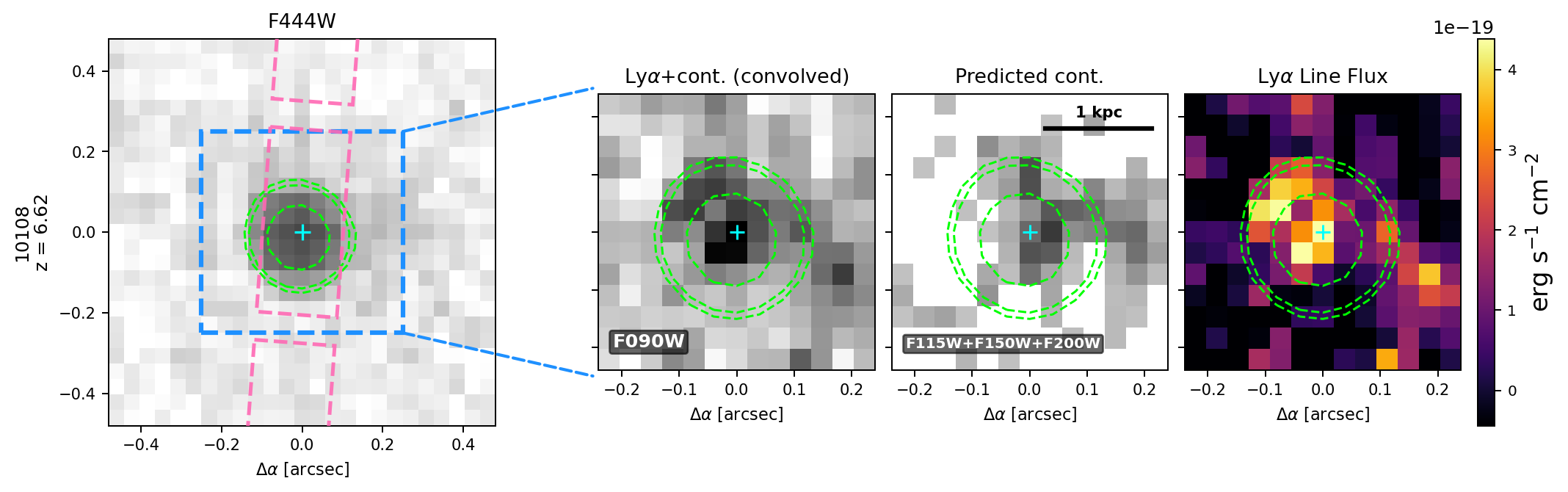}
    \includegraphics[width=0.87\textwidth]{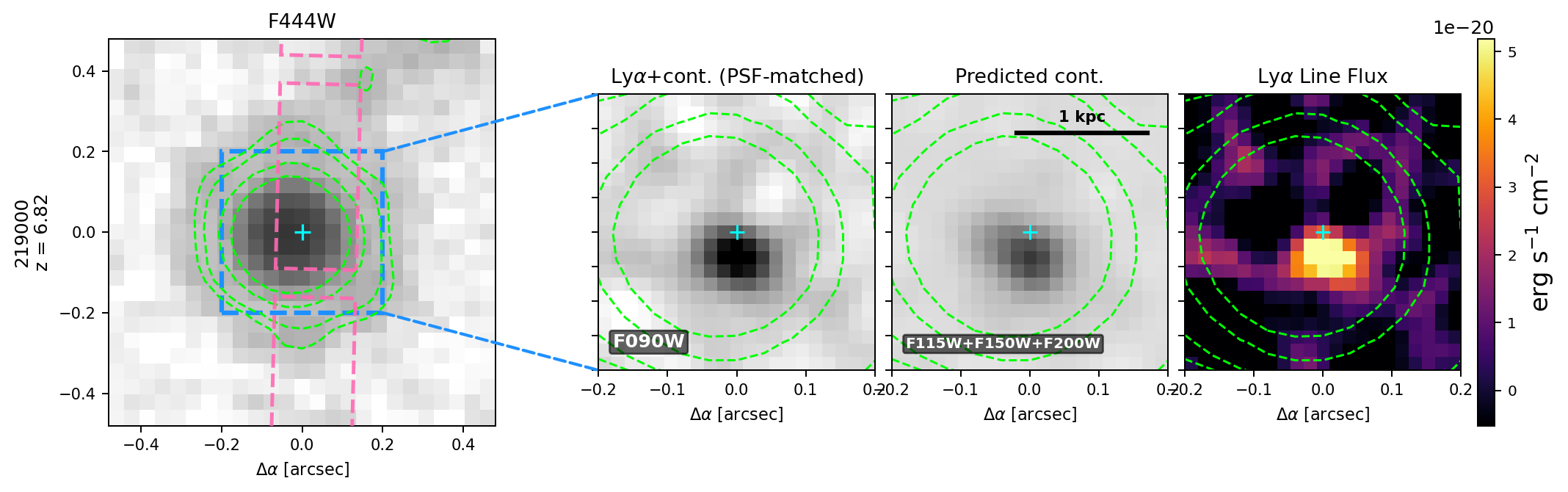}
    \includegraphics[width=0.87\textwidth]{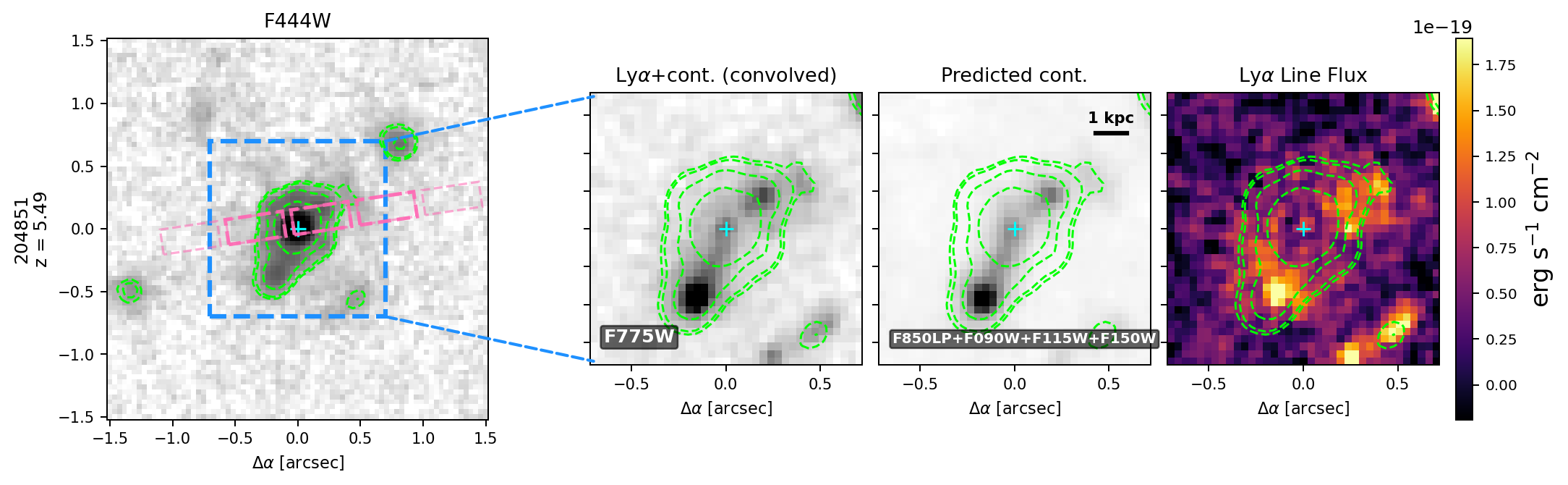}
    \caption{Similar to Figure~\ref{fig:map_1}, but for the LRDs in EGS (first row) and GOODS-S (second and third rows).}
    \label{fig:map_2}
\end{figure*}

Figures~\ref{fig:map_1} and \ref{fig:map_2} present the spatially resolved Ly$\alpha$ maps for the LRDs at $z \gtrsim 5.5$ in Abell~2744, GOODS-S, and EGS; the maps for the COSMOS and UDS sources are shown in Appendix~\ref{app:map_cosmos_uds}.

In the resolved maps, the Ly$\alpha$-sensitive image shows a clear flux excess over the predicted continuum, and in many cases the inferred Ly$\alpha$ emission is not centered exactly on the compact rest-optical component. Instead, across most of the sample, the inferred Ly$\alpha$ morphology is asymmetric, patchy, and/or displaced relative to the F444W centroid. In some objects, the emission is clearly offset from the compact F444W centroid (24175, 20466, 10108 and 219000), while in others it is distributed in an irregular or clumpy pattern around the rest-optical centroid (4286 and 204851). A particularly remarkable case is the GOODS-S source 204851 (Figure~\ref{fig:map_2}), which, in addition to a diffuse component, exhibits two distinct clump-like structures in the Ly$\alpha$-sensitive imaging rather than a single centrally concentrated component. The lower/southern clump is independently confirmed to belong to the same system and also shows strong Ly$\alpha$ emission (see Appendix \ref{app:204851}), demonstrating that escaping Ly$\alpha$ can be associated with a spatially complex, multi-component structure rather than being confined to the compact F444W core. A more detailed study of this source will be presented by Z. Ji et al. (2026, in prep.).

\begin{figure}
    \centering
    \includegraphics[width=1\linewidth]{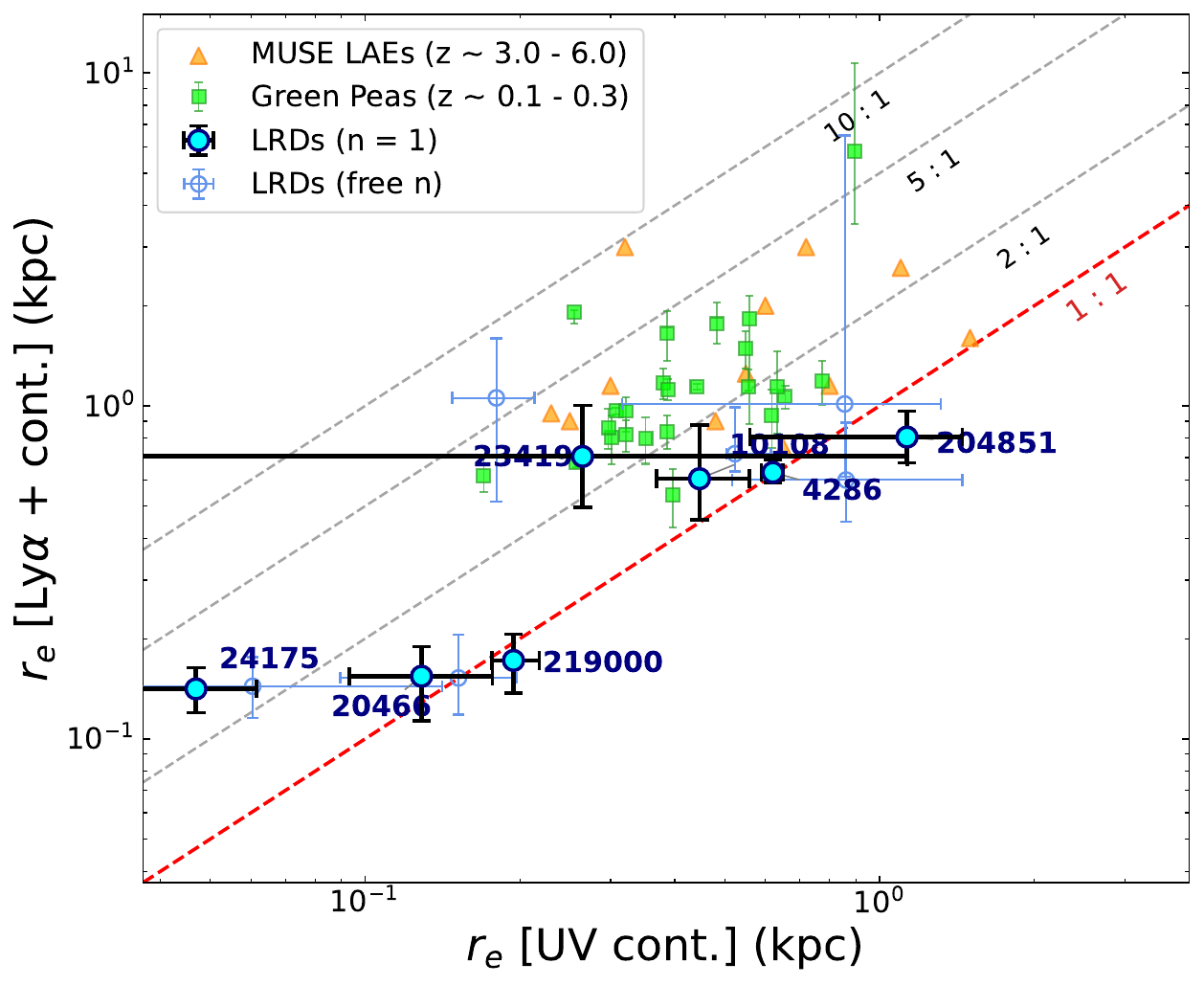}
    \caption{Effective radius measured from the Ly$\alpha$-sensitive image ($y$-axis) versus that measured from the adjacent UV-continuum image ($x$-axis) for the $z \geq 5.5$ LRDs. Orange and green points show comparison samples of MUSE LAEs at $z \sim 3$--6 and Green Peas at $z \sim 0.1$--0.3, respectively, measured by \citet{Yang2017}. We note that these comparison-sample measurements trace the Ly$\alpha$ size directly, as they are derived from either MUSE IFU  or HST/COS 2D spectroscopy, whereas our measurements are based on the Ly$\alpha$-sensitive broadband image and therefore correspond to ${\rm Ly}\alpha + {\rm continuum}$. Error bars (1$\sigma$) are estimated from resampling of the image pixel values using the sigma maps, followed by 100 repeated \texttt{galfit} fits. Only the non-PRIMER sources are shown, because the shallower COSMOS and UDS images do not yield reliable size measurements.}
    \label{fig:size}
\end{figure}

The Ly$\alpha$ emission also appears to be more extended than the adjacent UV continuum. We explore this difference by performing single-S\'{e}rsic fitting with \textsc{galfit} \citep{Peng2010}. We restrict this morphological analysis to the non-PRIMER sources, as the shallower COSMOS and UDS images (Section~\ref{sec:map_analysis}) do not yield reliable size measurements. Because the continuum-subtracted Ly$\alpha$ maps do not have sufficient S/N for a meaningful morphological analysis, we compare the morphology measured from the Ly$\alpha$-sensitive image, which contains Ly$\alpha$+continuum, with that measured from the adjacent continuum-only image. All \textsc{galfit} analyses presented here were performed on the corresponding images at their nominal resolutions. Given the generally faint UV emission of the LRDs, we perform two sets of fits: one with the S\'{e}rsic index $n$ fixed at 1 (i.e., an exponential disk) and one with $n$ left free.  As shown in Figure~\ref{fig:size}, the Ly$\alpha$-sensitive image is typically more extended than the adjacent continuum-only images. We note that this result does not provide a direct measurement of the Ly$\alpha$ effective radius, but it is consistent with an additional Ly$\alpha$ component contributing at larger radii than the UV continuum.

Such offset, asymmetric, and spatially extended Ly$\alpha$ morphologies are not unique to LRDs. Similar phenomenology is commonly observed in ordinary high-redshift Ly$\alpha$ emitters and in local compact starbursts, where resonant scattering redistributes Ly$\alpha$ photons into a more extended and irregular component than the stellar UV light \citep[e.g.,][]{Hayes2005,Atek2008,Ostlin2009,Steidel2011,Wisotzki2016,Leclercq2017}. These studies show that extended Ly$\alpha$ halos and spatial offsets between Ly$\alpha$ and the stellar continuum are common outcomes of radiative transfer through an inhomogeneous neutral medium rather than signatures unique to any one galaxy class. Therefore, at least at the phenomenological level, the Ly$\alpha$ morphologies of LRDs seem to be broadly consistent with this picture.

\section{Discussion}
\label{sec:diss}

\subsection{The origin of Ly$\alpha$ in LRDs}

From the perspective of integrated Ly$\alpha$ properties alone, LRDs are
not obviously distinguishable from the broader population of star-forming
galaxies at similar redshifts.  Their Ly$\alpha$ luminosities, UV--Ly$\alpha$
relations, and rest-frame equivalent-width distributions are all broadly
consistent with those of ordinary high-redshift galaxies of comparable UV
brightness (Section~\ref{sec:lya_uv}).  Phenomenologically, therefore, the
Ly$\alpha$ emission emerging from LRDs appears fairly typical.

What makes LRDs distinctive is the contrast between these ordinary
Ly$\alpha$ properties and their extraordinary optical spectra.  Relative to
their strong Balmer emission, the escaping Ly$\alpha$ is systematically
weak: the LRDs in our sample are offset toward low
$L_\mathrm{Ly\alpha}/L_\mathrm{H\alpha}$ compared to star-forming galaxies
at similar redshifts (Section~\ref{sec:lya_H}).  Meanwhile, the Ly$\alpha$
luminosity correlates more closely with $L_\mathrm{[O\,III]}$ than with
[O\,\textsc{iii}] EW (Section~\ref{sec:lya_O}), and the
spatially resolved maps show Ly$\alpha$ morphologies that are extended, asymmetric,
patchy, and frequently offset from the compact rest-optical component traced
by F444W (Section~\ref{sec:lya_map}).
Taken together, these results suggest that the observed Ly$\alpha$ does not
primarily trace the compact component that dominates the red optical
continuum and broad Balmer lines, but instead behaves as a tracer of gas on
host-galaxy scales.

This connection is consistent with the emerging picture from optical
spectroscopy.  \citet{deGraaff2026} and \citet{Sun2026} argue that [O\,\textsc{iii}] is more
naturally associated with the host galaxy than with the compact component
that produces the red continuum and broad Balmer lines, and
\citet{Pang2026} find that [O\,\textsc{iii}] correlates with the UV
continuum rather than with the optical.  The UV continuum itself is
increasingly attributed to young stellar populations in the host galaxy
\citep{Asada2026, Killi2024, Rinaldi2025}, although the contribution from the central component may still be non-negligible \citep{Sun2026,Ando2026}. The fact that Ly$\alpha$ tracks
[O\,\textsc{iii}] and shares qualitatively similar spatial characteristics with the UV emission
is consistent with Ly$\alpha$ being primarily associated with the same host-scale component.

A natural question is why Ly$\alpha$ associated with the compact and much brighter (relative to UV) rest-optical
component, if produced there at all, does not dominate the observed
emission.  In the dense-envelope framework, several factors may contribute.
A growing body of theoretical work has shown that the red optical continuum,
strong Balmer breaks, and broad Balmer lines of LRDs can be reproduced by
models in which an accreting black hole is embedded in a dense gaseous
cocoon, whether framed as a dense neutral gas envelope
\citep{InayoshiMaiolino2025, Ji2025a}, a ``black hole star'' atmosphere
\citep{Naidu2025a, Kido2025}, a dense ionized cocoon with electron-scattering
line broadening \citep{Rusakov2026, Chang2025}, a late-stage quasi-star
\citep{PacucciNarayan2024, Begelman2026}, or an orientation-dependent
super-Eddington accretion flow \citep{Madau2026}.  A common feature of these
models is that the compact component is surrounded by gas of high column
density, which has direct implications for Ly$\alpha$.  Self-consistent
radiative-transfer calculations predict that the inner ionized region is
bounded by a cold gas reservoir with neutral column densities
$N_\mathrm{HI} \sim 10^{23}$~cm$^{-2}$, sufficient to produce strong
Ly$\alpha$ absorption, and that little UV flux is transmitted directly from
the accretion region \citep{Sneppen2026}.  In a complementary analysis,
\citet{Asada2026} argue that both the UV continuum and much of the
H$\alpha$ emission in LRDs are powered by young massive stars surrounding
the envelope rather than by the AGN itself, and find that Ly$\alpha$
occurrence rates in LRDs are comparable to those in ordinary
star-forming galaxies, which is consistent with Ly$\alpha$ originating in, or being reprocessed through, host-scale gas associated with the extended stellar component rather than in the cocoon. 

A further possible connection between Ly$\alpha$ and the dense-envelope picture
emerges from the anomalous Balmer decrements observed in many LRDs (Figure \ref{fig:lya_balmer}).  At the
gas densities inferred for the cocoon ($n_\mathrm{H} \sim
10^{9\text{--}11}$~cm$^{-3}$), collisional excitation populates the hydrogen
$n = 2$ level to the point where the gas becomes optically thick to the
Balmer transitions themselves \citep{InayoshiMaiolino2025, Ji2025a}.  In this regime,
H$\beta$ photons resonantly scatter off $n = 2$ hydrogen and can convert to
Pa$\alpha$ + H$\alpha$ via the $n = 4 \to 3$ cascade, systematically
destroying H$\beta$ and enhancing H$\alpha$
\citep{Chang2025, Naidu2025a}.  This process steepens the observed Balmer
decrement well beyond the Case~B value (Figure \ref{fig:lya_balmer}) and beyond what dust
reddening alone can account for \citep{Nikopoulos2025}. Our stacking analysis of 2D spectra (Figure~\ref{fig:stack_decrement}) offers additional, though currently tentative observational support for this association: among the Ly$\alpha$-detected LRDs, the Balmer decrement is steepest at the position of the compact rest-optical source and appears to decline toward larger spatial offsets, as expected if the anomalous decrement originates in the dense central component while the more extended host gas approaches a more nearly Case~B ratio.  The same
dense medium that drives this anomalous decrement would, by extension, be
even more opaque to Ly$\alpha$: if the $n = 2$ population is sufficient to
make H$\alpha$ a resonant line, the $n = 1$ population would imply that any Ly$\alpha$ produced within or passing through the
cocoon experiences extreme optical depths.  The steep Balmer decrements and
the weakness of Ly$\alpha$ from the compact component are therefore not
independent observations but rather two manifestations of the same
underlying condition: a dense gaseous envelope in which hydrogen radiative
transfer departs strongly from the optically thin (Case~B) limit.

The current data cannot uniquely determine whether the compact component fails to produce significant Ly$\alpha$, produces it but the surrounding neutral gas absorbs it before escape, or contributes only a subdominant fraction that is overwhelmed by host-galaxy emission.  What they do
strongly suggest is that the observed Ly$\alpha$ behaves as a host-galaxy
tracer.  The complex, extended morphologies seen in our maps are the expected
signatures of resonant scattering through a clumpy, anisotropic interstellar
and circumgalactic medium \citep[e.g.,][]{Laursen2009, Zheng2010,
Gronke2017}, and are commonly observed in ordinary high-redshift Ly$\alpha$
emitters and local compact starbursts \citep[e.g.,][]{Hayes2013,
Leclercq2017}. Detailed Ly$\alpha$-profile studies of individual LRDs likewise favor porous or non-uniform gas geometries rather than fully closed
configurations \citep{Tang2026, Ji2026}.

The picture that emerges is one in which the observed Ly$\alpha$ in LRDs is
governed by the structure and geometry of gas on host-galaxy scales, aligning
it with the UV continuum and [O\,\textsc{iii}] rather than with the red
optical continuum and broad Balmer lines.  This reinforces the broader
empirical picture in which the properties of LRDs reflect at least two
spatially and physically distinct components:
a compact inner region that dominates the red continuum and Balmer emission,
and a host-galaxy environment traced by the UV, [O\,\textsc{iii}], and now
Ly$\alpha$.

\subsection{Implications for the physical diversity of LRDs}

Whether LRDs constitute a single physical class remains an open question,
and the answer depends sensitively, even primarily, on how the sample is defined.
Photometric selections based on V-shaped continua, spectroscopic selections
requiring broad Balmer lines, and morphological selections based on
compactness in the rest-frame optical all yield overlapping but distinct
samples with different levels of contamination and completeness
\citep[e.g.,][]{PerezGonzalez2026,Rinaldi2026}.  Studies that
adopt broader photometric criteria tend to find a continuous distribution of
properties with no sharp boundary separating LRDs from the general galaxy
population \citep{Rinaldi2026}, while spectroscopically confirmed samples
with strict broad-line and continuum-shape requirements isolate a more
homogeneous subset \citep{deGraaff2026, Wang2025b}.  Much of the apparent
tension in the literature over whether LRDs are a distinct class likely
reflects these differences in sample definition rather than a genuine
disagreement in the underlying physics.

In this work, our sample is drawn from the spectroscopic catalog of
\citet{deGraaff2026}, i.e. a sample that
is by construction more uniform in its compact-component properties than
purely photometric selections (Section~\ref{sec:sample}).  Our Ly$\alpha$ results therefore pertain
specifically to this broad-line, V-shaped continuum, and compact population.

Within this sample, our central finding is that the observed Ly$\alpha$
more likely traces host-galaxy gas rather than the compact component.
Because Ly$\alpha$ is governed by a different set of physical
conditions (e.g., ISM covering fraction, dust/gas geometry,
and star-formation activity) than those that define the LRD selection
(continuum shape and compact morphology), the
Ly$\alpha$ properties provide an additional degree of freedom that is largely
independent of the compact-component diagnostics.  The $\approx$30\%
Ly$\alpha$ detection rate, the scatter in Ly$\alpha$ luminosity at fixed
H$\alpha$ or [O\,\textsc{iii}] luminosity, and the diversity of Ly$\alpha$
morphologies across our spatially resolved subsample likely reflect
variations in host-galaxy environment rather than variations in the central
engine or cocoon.

This interpretation naturally accommodates diversity even within a
spectroscopically homogeneous LRD sample.  Two LRDs with similar
compact-component properties (e.g., comparable Balmer-break strength, optical luminosity) could exhibit very different Ly$\alpha$
properties if their host galaxies differ in ISM structure or star-formation
activity.  Conversely, the fact that LRD Ly$\alpha$ properties overlap
broadly with those of ordinary star-forming galaxies suggests that the host-galaxy environments of these LRDs are not
exceptional: it is the compact component that makes them unusual, not the
galaxies in which they reside.

This two-component perspective, an unusual compact source embedded in a
relatively ordinary host galaxy, offers a framework for understanding why
even spectroscopically selected LRDs can appear diverse when examined across
multiple diagnostics.  Indicators sensitive to the compact component (broad
Balmer lines, red continuum, Balmer break) track one axis of variation,
while indicators sensitive to the host galaxy (UV morphology,
[O\,\textsc{iii}], and now Ly$\alpha$) track another, likely independent
axis.  The observed scatter in LRD properties may therefore reflect the
natural variation expected when two physically distinct components -- each
with its own range of properties -- contribute to the observed emission in
different proportions.  Future studies that jointly analyze Ly$\alpha$, UV
morphology, and optical spectral properties for larger samples will be able
to test this picture by examining whether Ly$\alpha$ detection and morphology
correlate with host-galaxy indicators (e.g., UV half-light radius,
[O\,\textsc{iii}] luminosity) independently of compact-component properties
(e.g., Balmer-break strength, broad H$\alpha$ luminosity).

\section{Caveats}

Our analysis relies on the spectroscopic catalog of \citet{deGraaff2026}, which selects LRDs based on V-shaped UV-to-optical continua and compact rest-optical morphology. These criteria yield a clean, spectroscopically coherent sample in which broad Balmer lines are found in $\sim$98\% of cases, but prioritize purity over completeness, a trade-off that, in our view, is necessary when the physical nature of the population itself remains uncertain. This choice, however, necessarily excludes sources that may share similar physical origins but fall outside these strict criteria. \citet{Hviding2025} demonstrate that photometric and spectroscopic LRD selections overlap by only $\sim$50–60\%, and broader photometric criteria reveal a continuous distribution of properties with no sharp boundary separating LRDs from the general galaxy population \citep[e.g.,][]{Rinaldi2026,PerezGonzalez2026}. The practical consequence is that our Ly$\alpha$ results, including detection rates, luminosity distributions, and spatial morphologies, may not be representative of the full population of objects commonly ``labeled'' as LRDs. Quantifying this effect is very challenging, if possible at all, given that no consensus yet exists on how to define an LRD selection that is simultaneously pure and complete. Our results thus apply specifically to the high-purity subsample identified by \citet{deGraaff2026}. Extending them to the broader LRD population will require Ly$\alpha$ analyses of complementary LRD samples selected using other criteria, together with a dedicated assessment of the systematics inherent to those selection methods.

\section{Summary}

We have presented a systematic study of Ly$\alpha$ emission in Little Red Dots, using a sample of 110 spectroscopically confirmed LRDs at $z \geq 4$ drawn from the \citet{deGraaff2026} catalog. All sources have NIRSpec/PRISM coverage of the Ly$\alpha$ line, and a subset of 13 at $z \gtrsim 5.5$ have the broadband imaging required for spatially resolved Ly$\alpha$ mapping.

We detect Ly$\alpha$ at S/N $\geq 3$ in 32 LRDs, with $L_{\mathrm{Ly\alpha}}$ luminosities of $\sim 10^{41}$--$10^{43}$~erg~s$^{-1}$, rest-frame equivalent widths, and detection fractions all broadly consistent with those of ordinary star-forming galaxies at comparable redshifts and UV luminosities. From the perspective of their UV and Ly$\alpha$ properties alone, LRDs do not appear to form a distinct population. Yet what sets LRDs apart is the contrast between these ordinary Ly$\alpha$ properties and their extraordinary rest-optical spectra. The Ly$\alpha$/H$\alpha$ luminosity ratio is systematically lower than expected under Case~B recombination, 
suggesting that the escaping Ly$\alpha$ emission is not simply proportional to the dominant Balmer-emitting component. Meanwhile, Ly$\alpha$ luminosity shows a stronger trend with [\ion{O}{3}]$\lambda5007$ luminosity than with [\ion{O}{3}] equivalent width. Although statistically modest, this behavior is consistent with Ly$\alpha$ being associated with host-scale gas rather than with the compact component that dominates the red continuum and broad Balmer lines.

The spatially resolved analysis reinforces this picture. Continuum-subtracted Ly$\alpha$ maps for 13 LRDs at $z \geq 5.5$ reveal extended, asymmetric, and often offset emission relative to the rest-optical light traced by F444W. GALFIT modeling of the Ly$\alpha$-sensitive broadband images yields larger effective radii than adjacent continuum-only images, consistent with Ly$\alpha$ contributing on larger spatial scales than the adjacent UV continuum. These morphologies are consistent with resonant scattering through clumpy, anisotropic interstellar and circumgalactic gas, as commonly observed in high-redshift Ly$\alpha$ emitters and local compact starbursts.

Together, our integrated and spatially resolved measurements support a two-component picture of LRDs: a compact inner region that dominates the red optical continuum and broad Balmer emission, embedded within a more 
extended host-galaxy environment whose gas governs Ly$\alpha$ production and radiative transfer. Within this framework, the steep Balmer decrements and the weakness of Ly$\alpha$ from the compact component may be two manifestations of the same underlying condition: a dense gaseous envelope in which collisional excitation of hydrogen populates the $n=2$ level, rendering both Balmer and Ly$\alpha$ photons susceptible to resonant scattering and destruction. The diversity of Ly$\alpha$ properties across the sample, spanning non-detections to luminous, spatially extended emission, more likely reflects variation in host-galaxy environments rather than in the compact component, establishing Ly$\alpha$ as a new and independent diagnostic for understanding the physical nature of LRDs.

\section*{Acknowledgments}
This work is based on observations made with the NASA/ESA/CSA
James Webb Space Telescope. The data were obtained from the
Mikulski Archive for Space Telescopes at the Space Telescope Science Institute, which is operated by the Association of Universities
for Research in Astronomy, Inc., under NASA contract NAS 5-03127
for JWST. The spectroscopic observations are associated with programs 1180,
1181, 1208, 1212, 1213, 1215, 1286, 1345, 1433, 2198, 2561, 2750,
2767, 4106, 4233, 5105, 5224, 6368, 6541 and 6585. The imaging observations are associated with programs 1180, 1181, 1264, 1324, 1345, 1727, 1810, 1837, 1840, 1895, 1963, 2079, 2234, 2279, 2561, 2738, 2750, 2756, 3215, 3516, 3577, 4111, 4762, 6434, 6585.

ZJ, YS, YZ, GHR, and MR acknowledge support from 
the NIRCam Science Team contract to the University of Arizona, NAS5-02015. AdG acknowledges support from a Clay Fellowship awarded by the Smithsonian Astrophysical Observatory. The work of CCW is supported by NOIRLab, which is managed by the Association of Universities for Research in Astronomy (AURA) under a cooperative agreement with the National Science Foundation.

\appendix

\section{UV continuum and Ly$\alpha$ spectral fitting}
\label{app:lya_fit}

\begin{figure}
\centering
\includegraphics[width=0.32\textwidth]{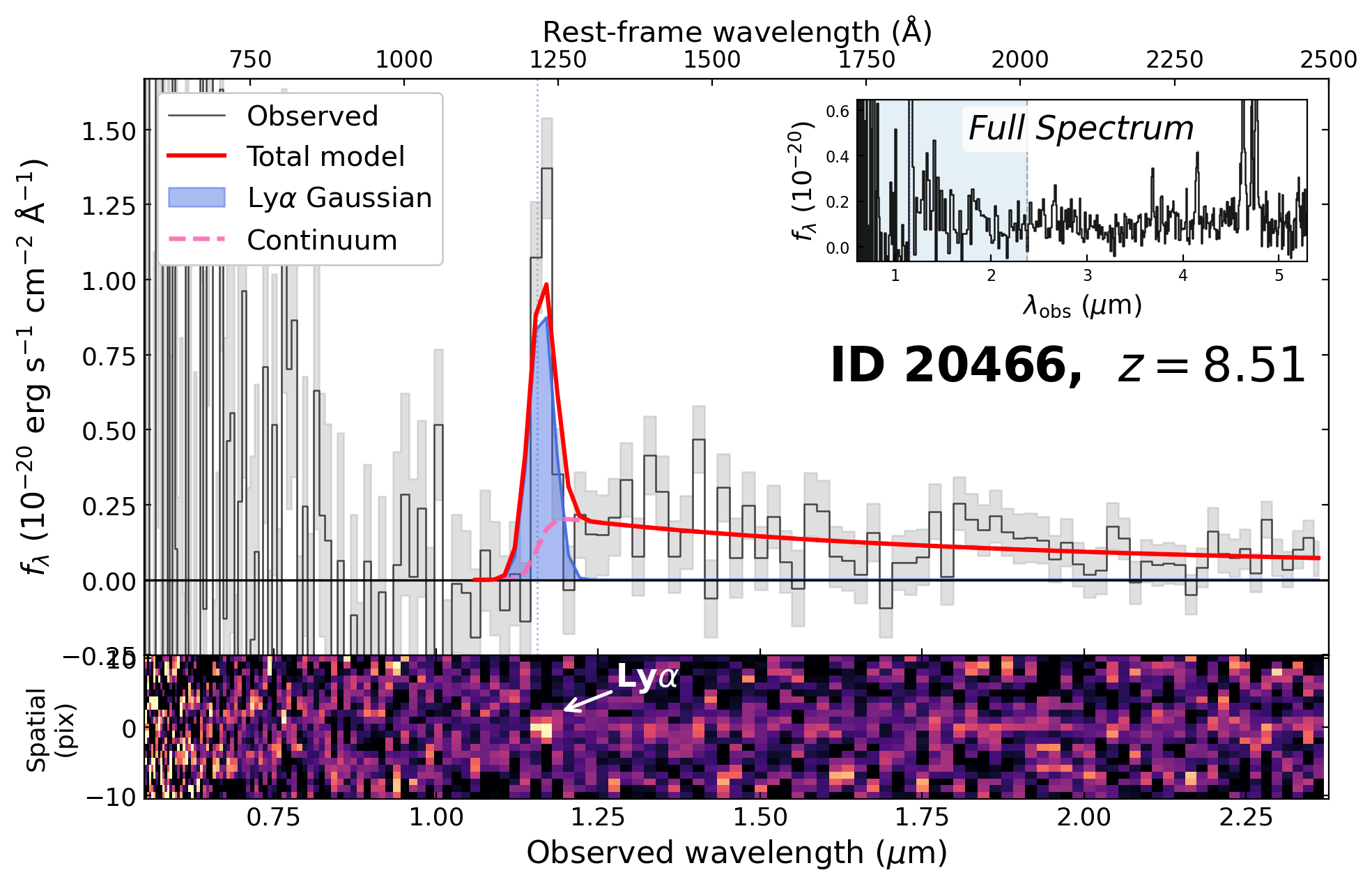}
\includegraphics[width=0.32\textwidth]{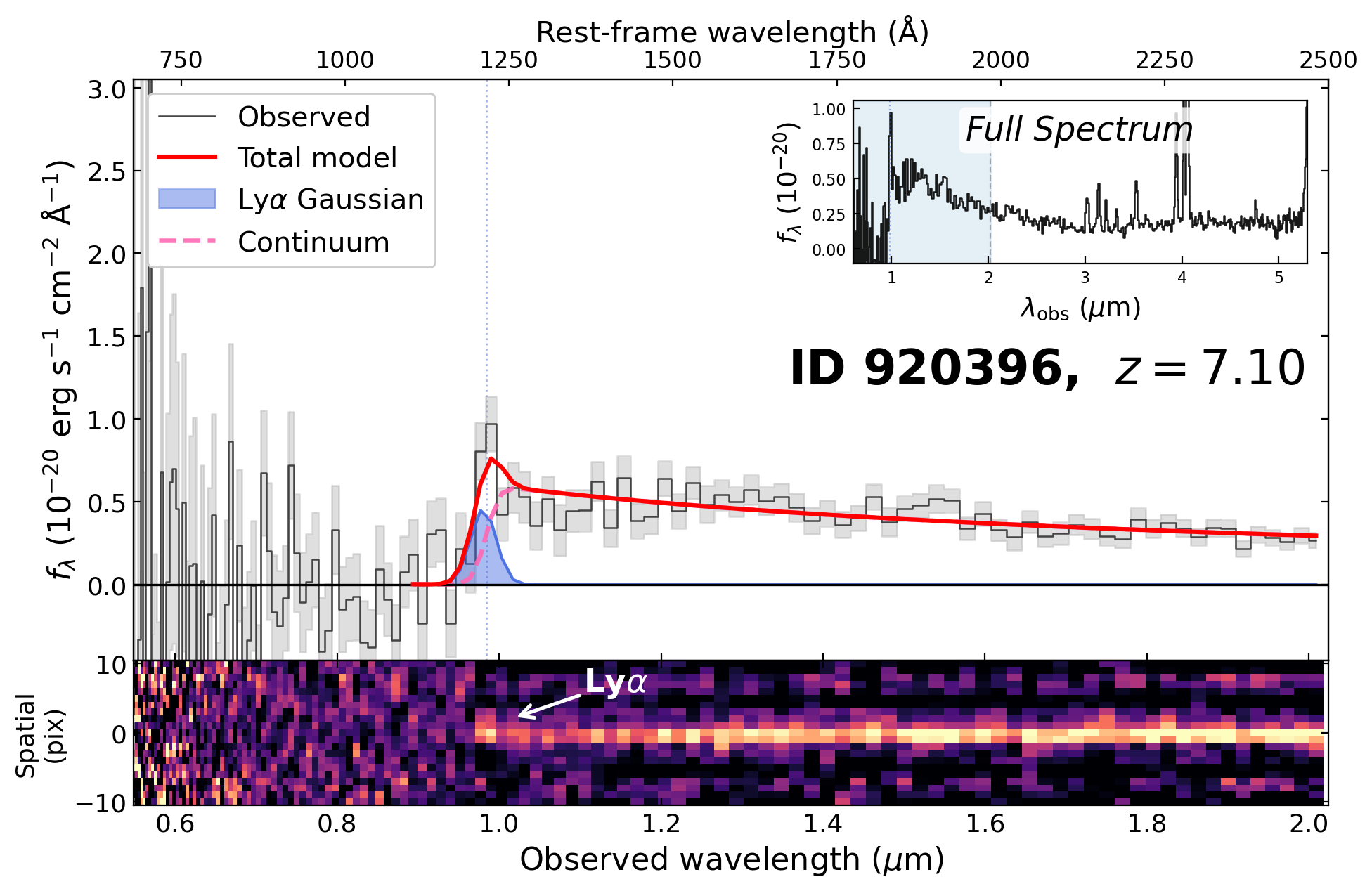}
\includegraphics[width=0.32\textwidth]{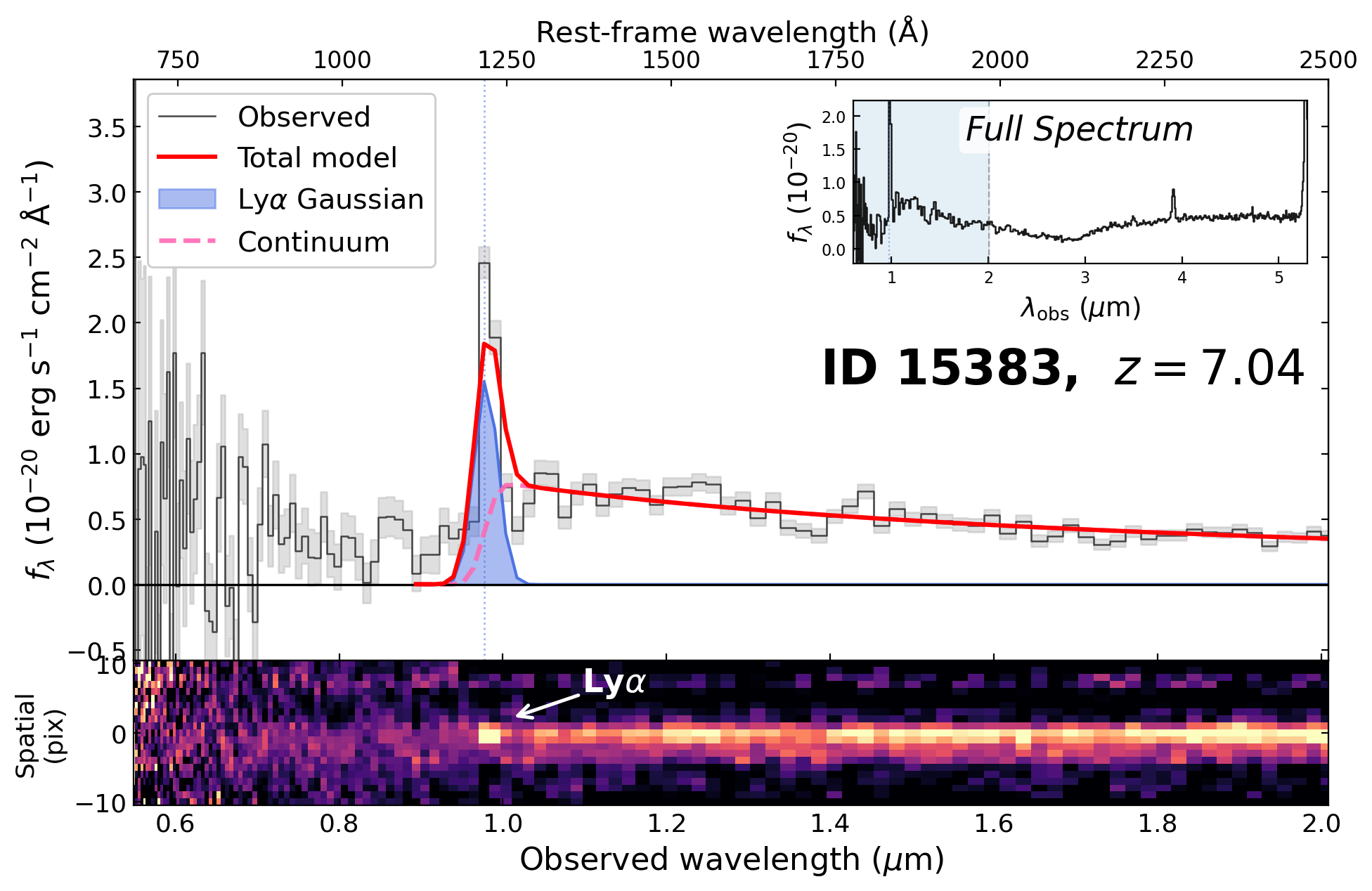}\\[1pt]
\includegraphics[width=0.32\textwidth]{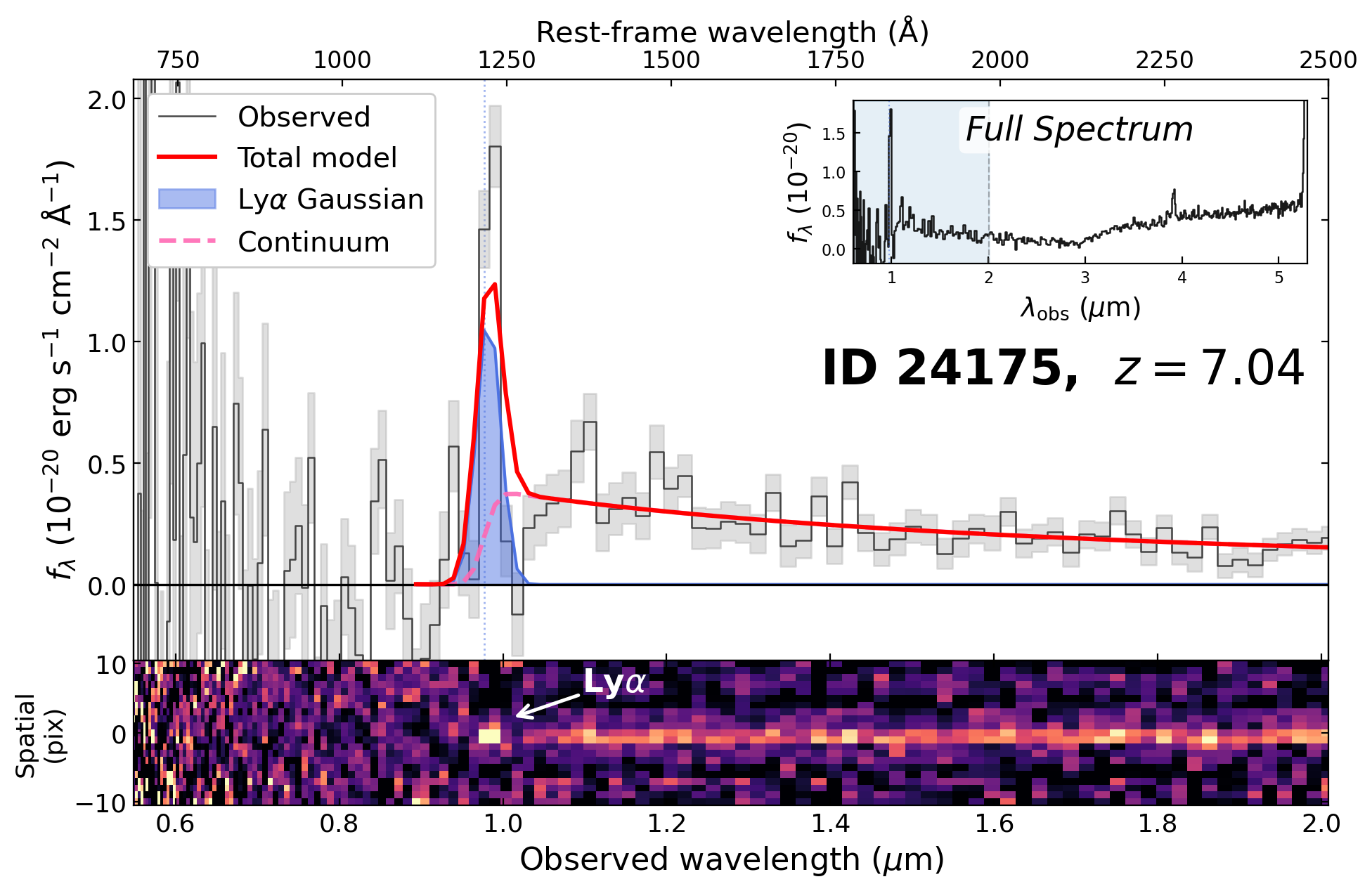}
\includegraphics[width=0.32\textwidth]{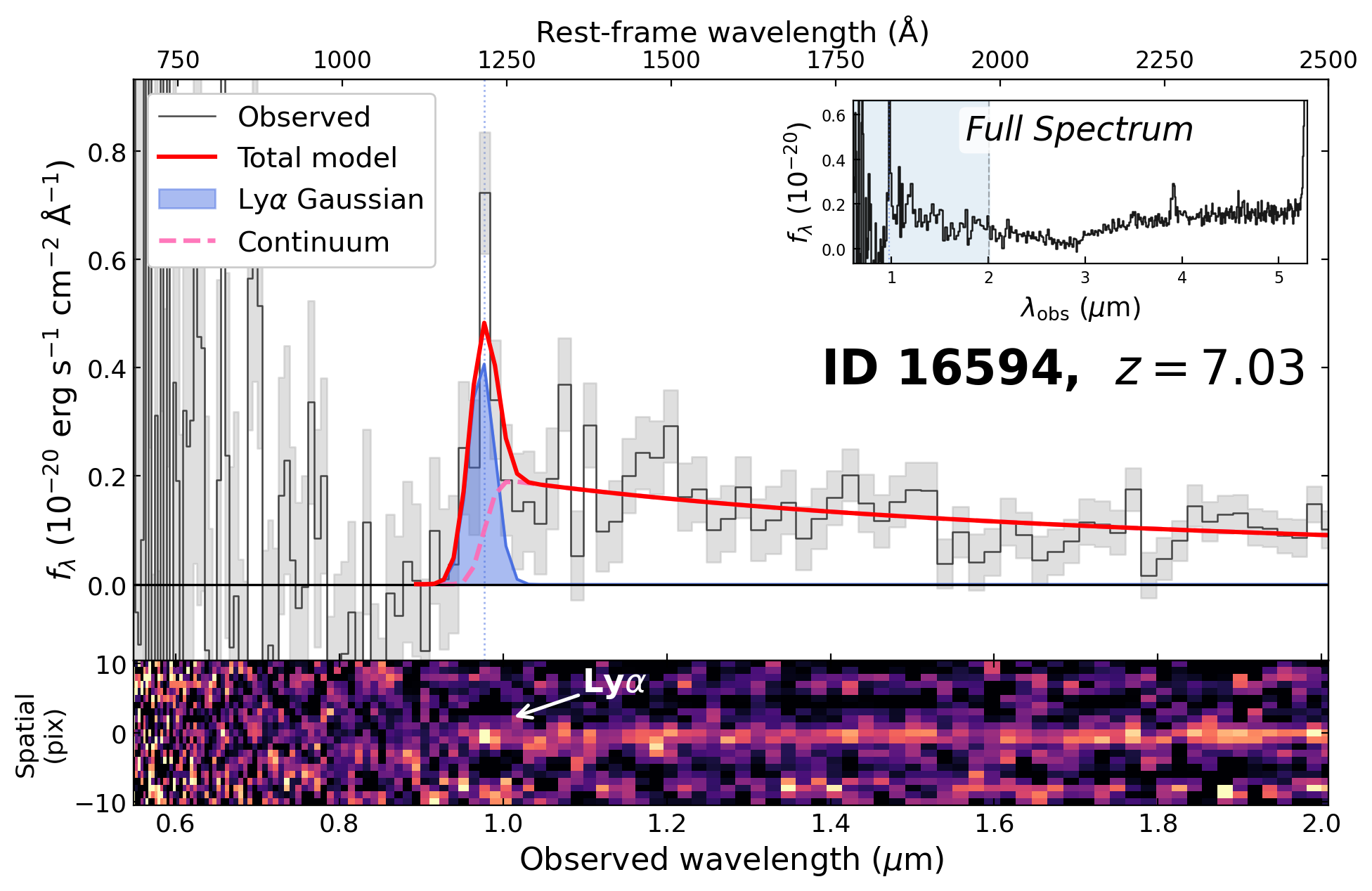}
\includegraphics[width=0.32\textwidth]{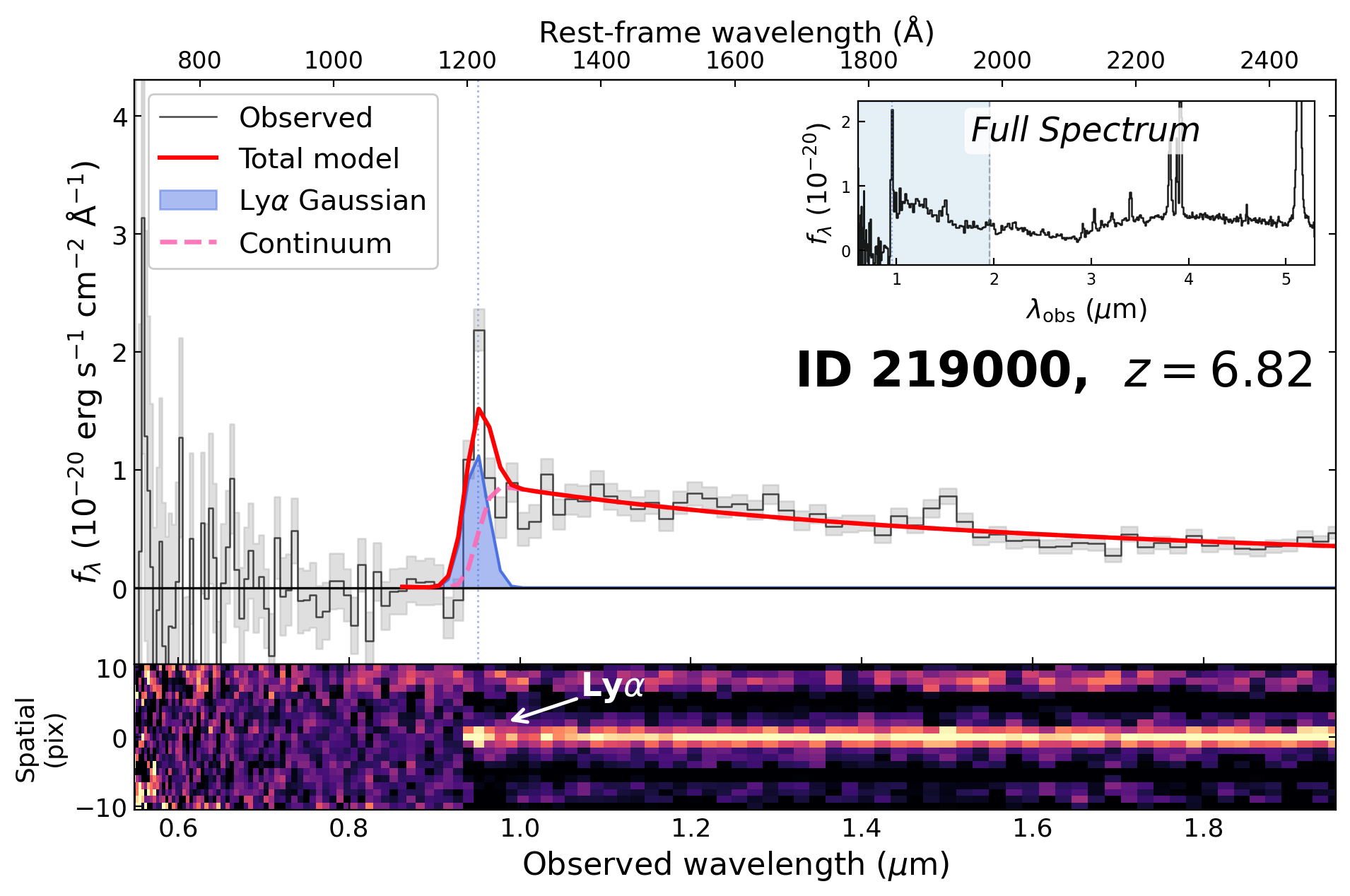}\\[1pt]
\includegraphics[width=0.32\textwidth]{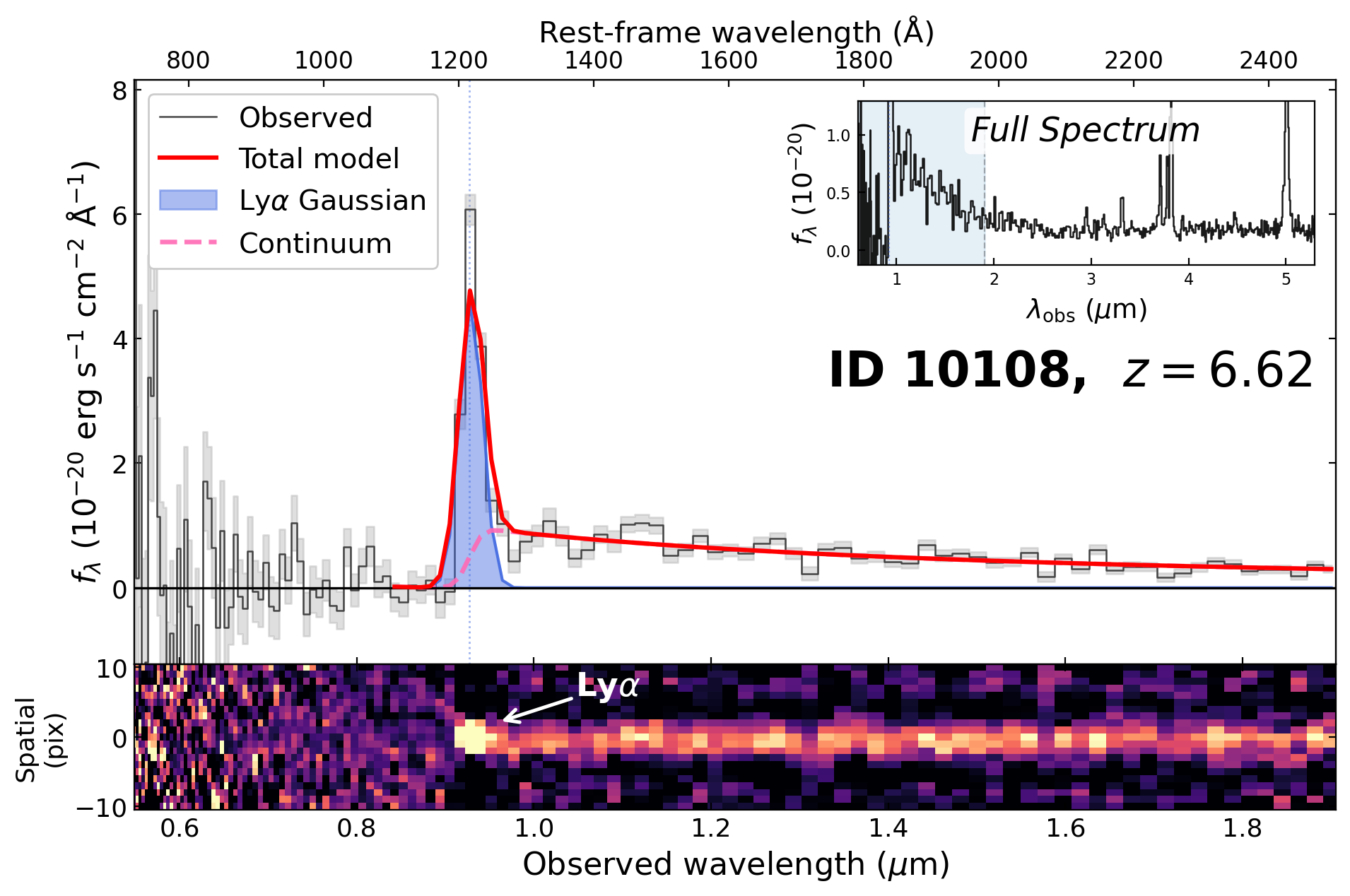}
\includegraphics[width=0.32\textwidth]{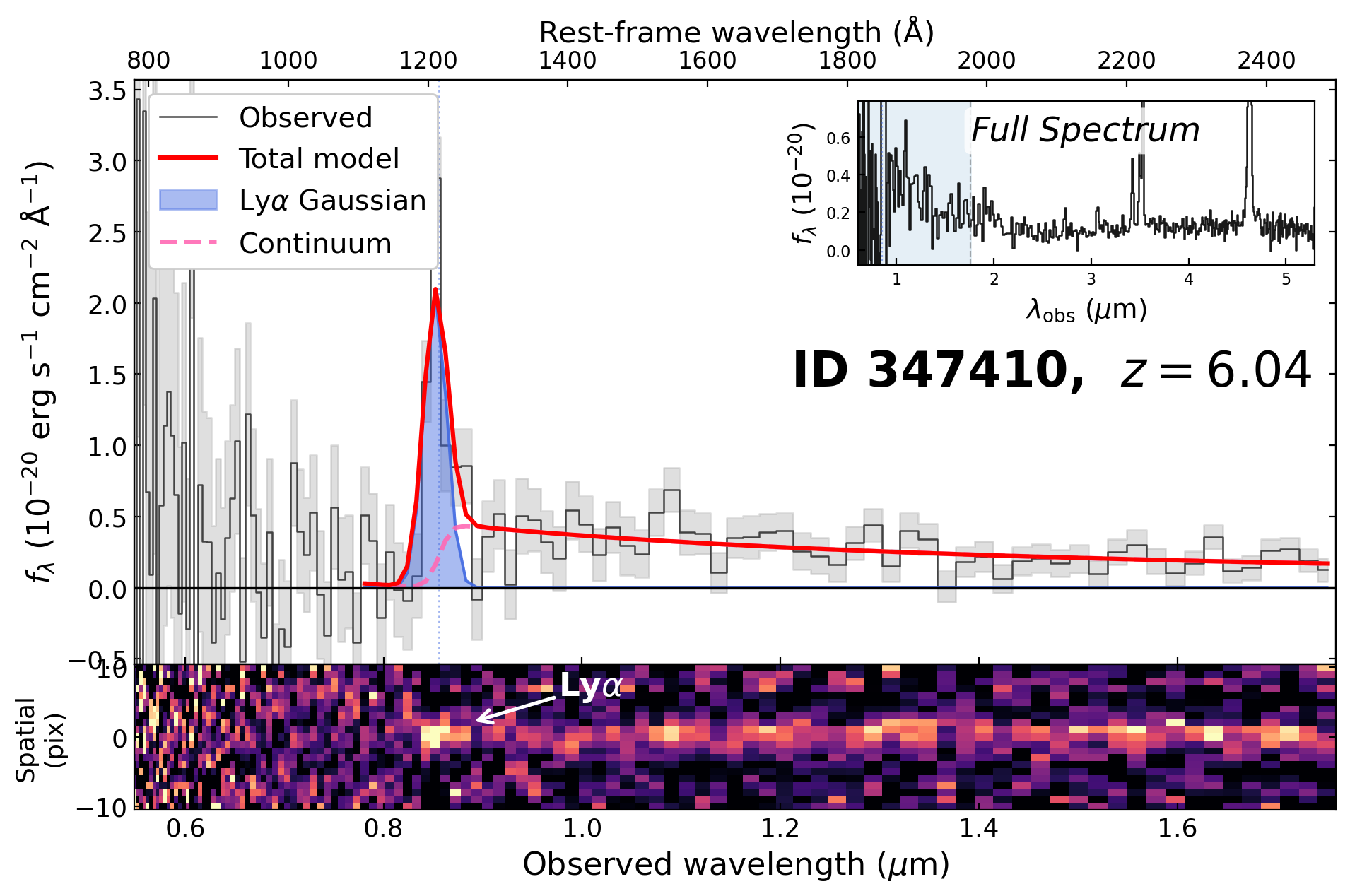}
\includegraphics[width=0.32\textwidth]{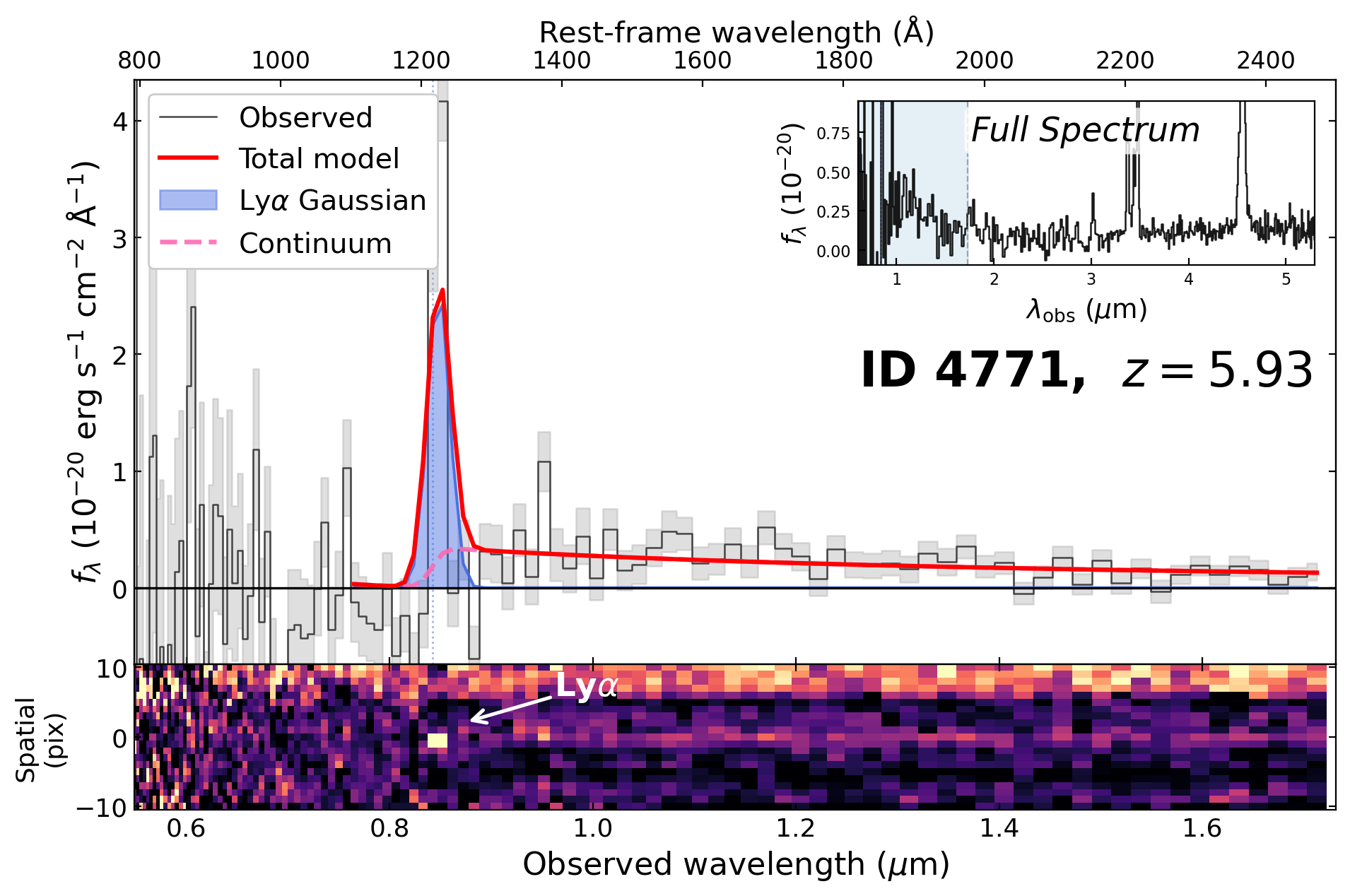}\\[1pt]
\includegraphics[width=0.32\textwidth]{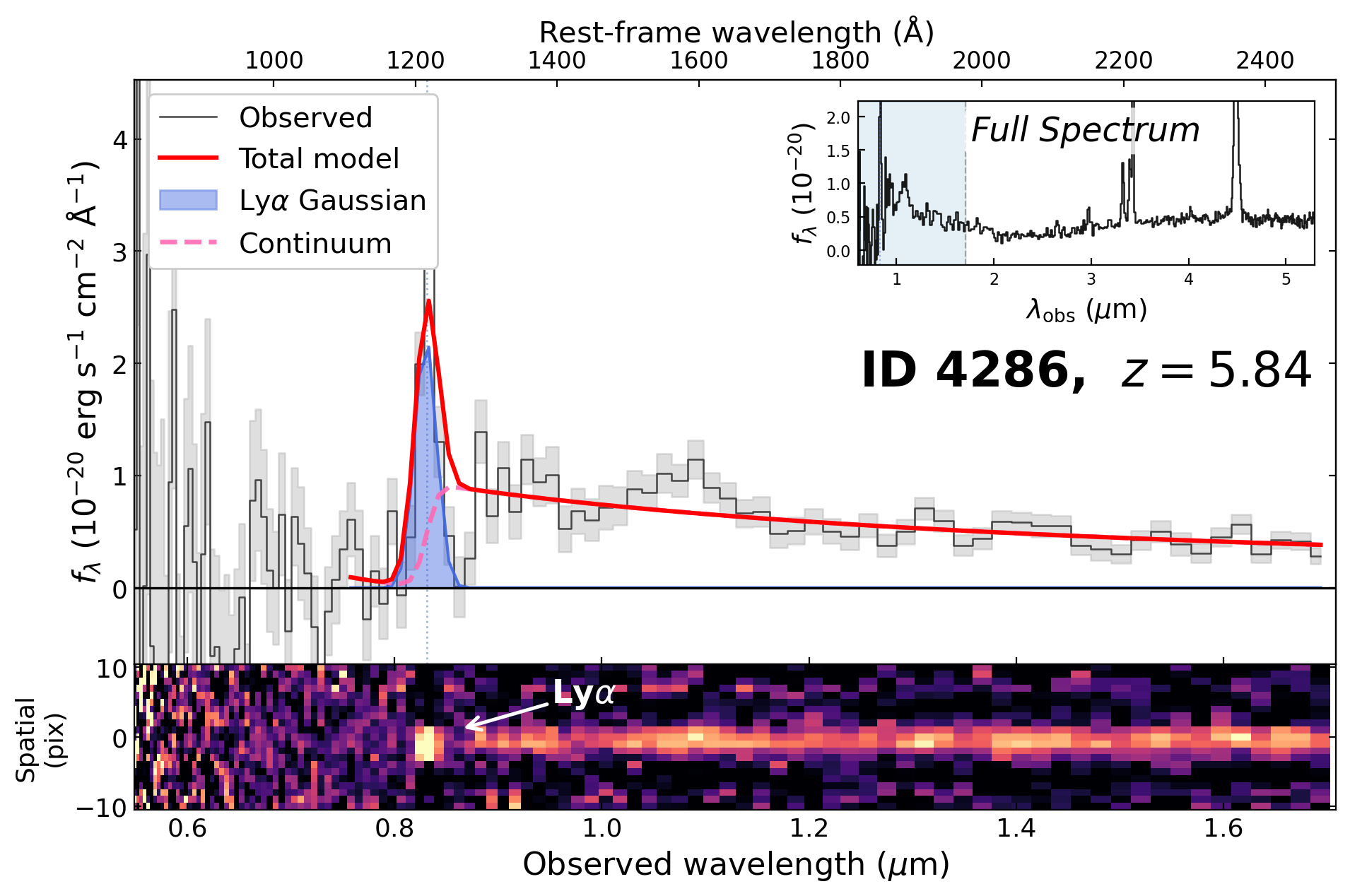}
\includegraphics[width=0.32\textwidth]{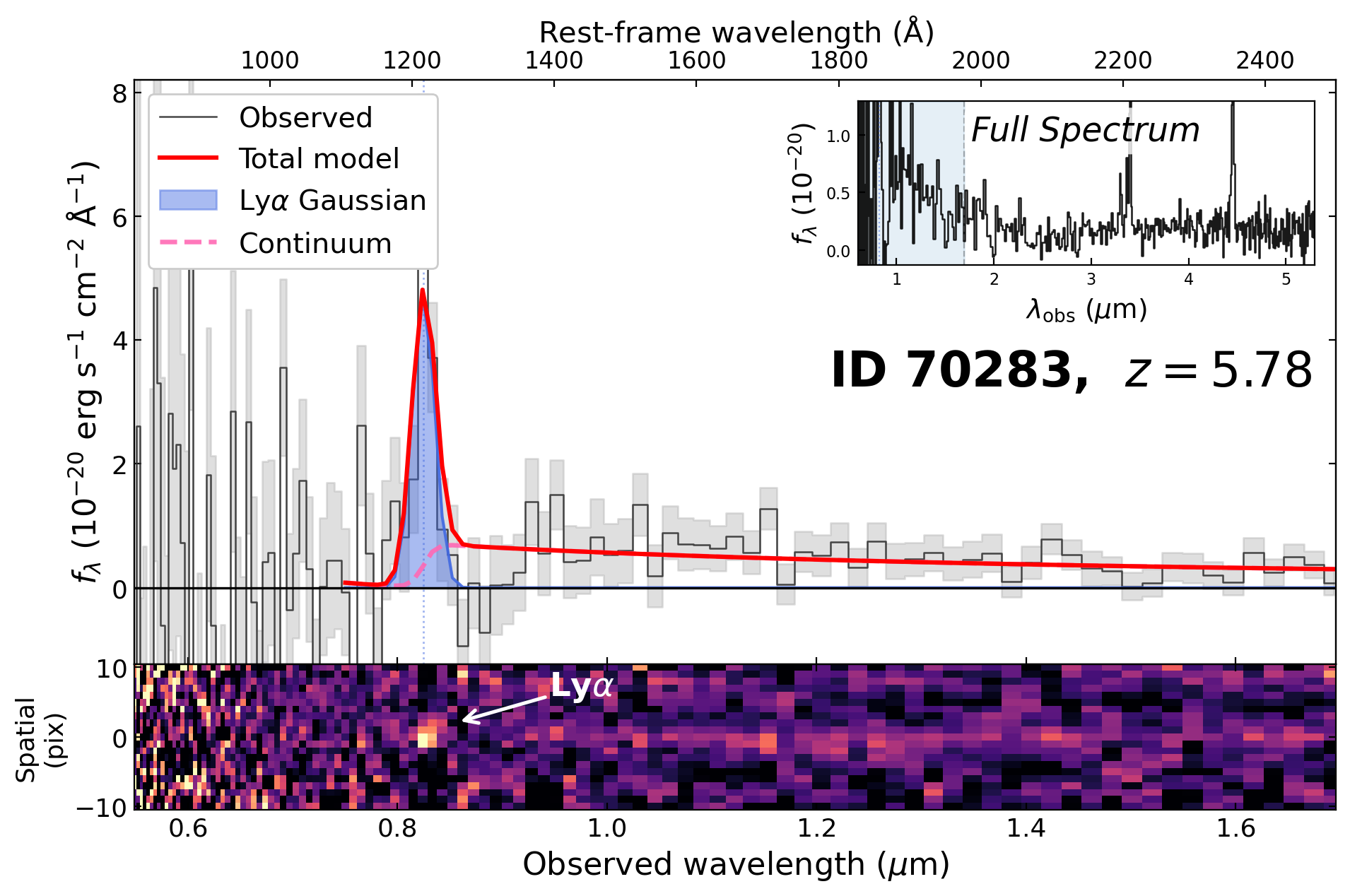}
\includegraphics[width=0.32\textwidth]{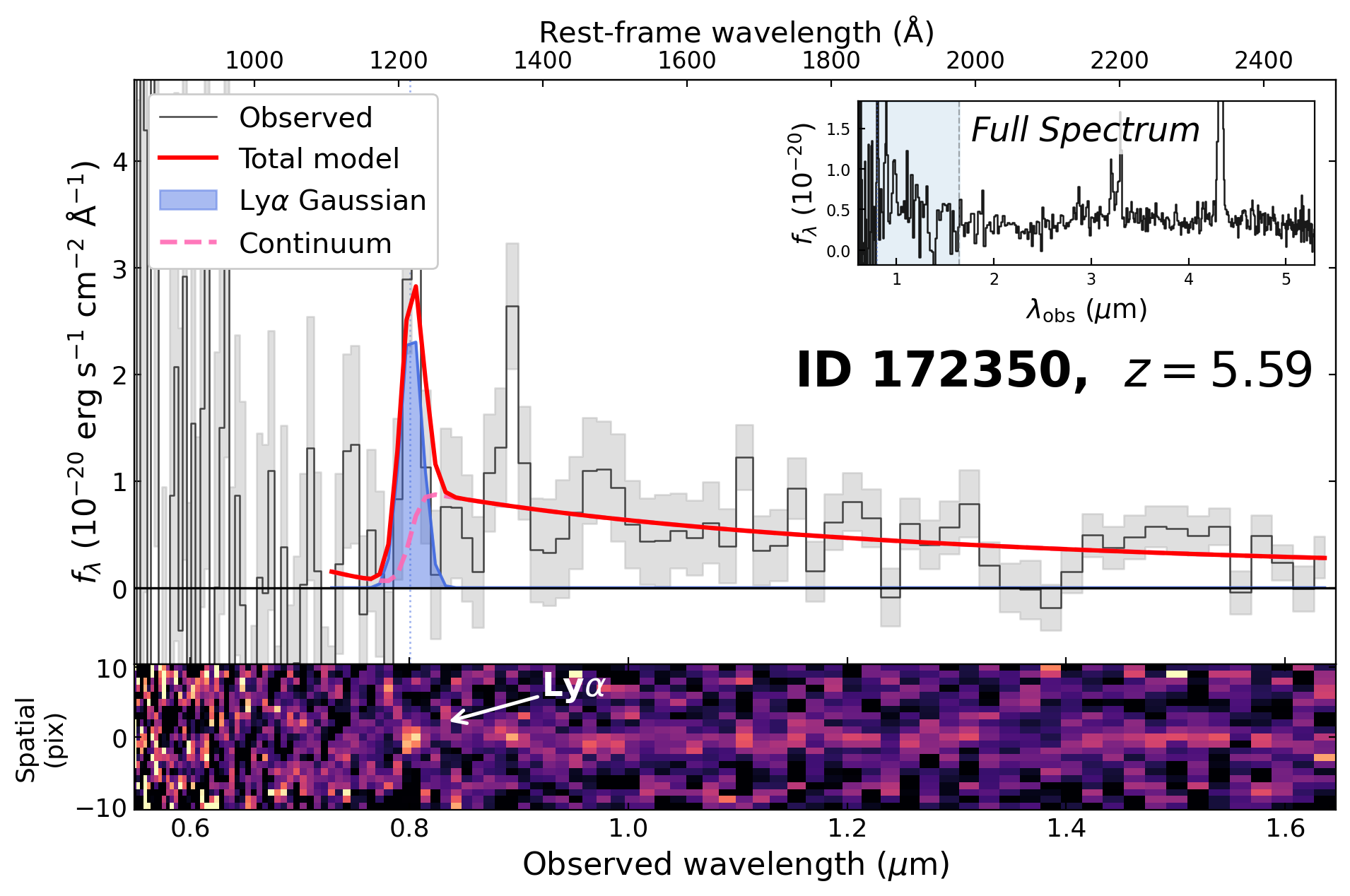}\\[1pt]
\includegraphics[width=0.32\textwidth]{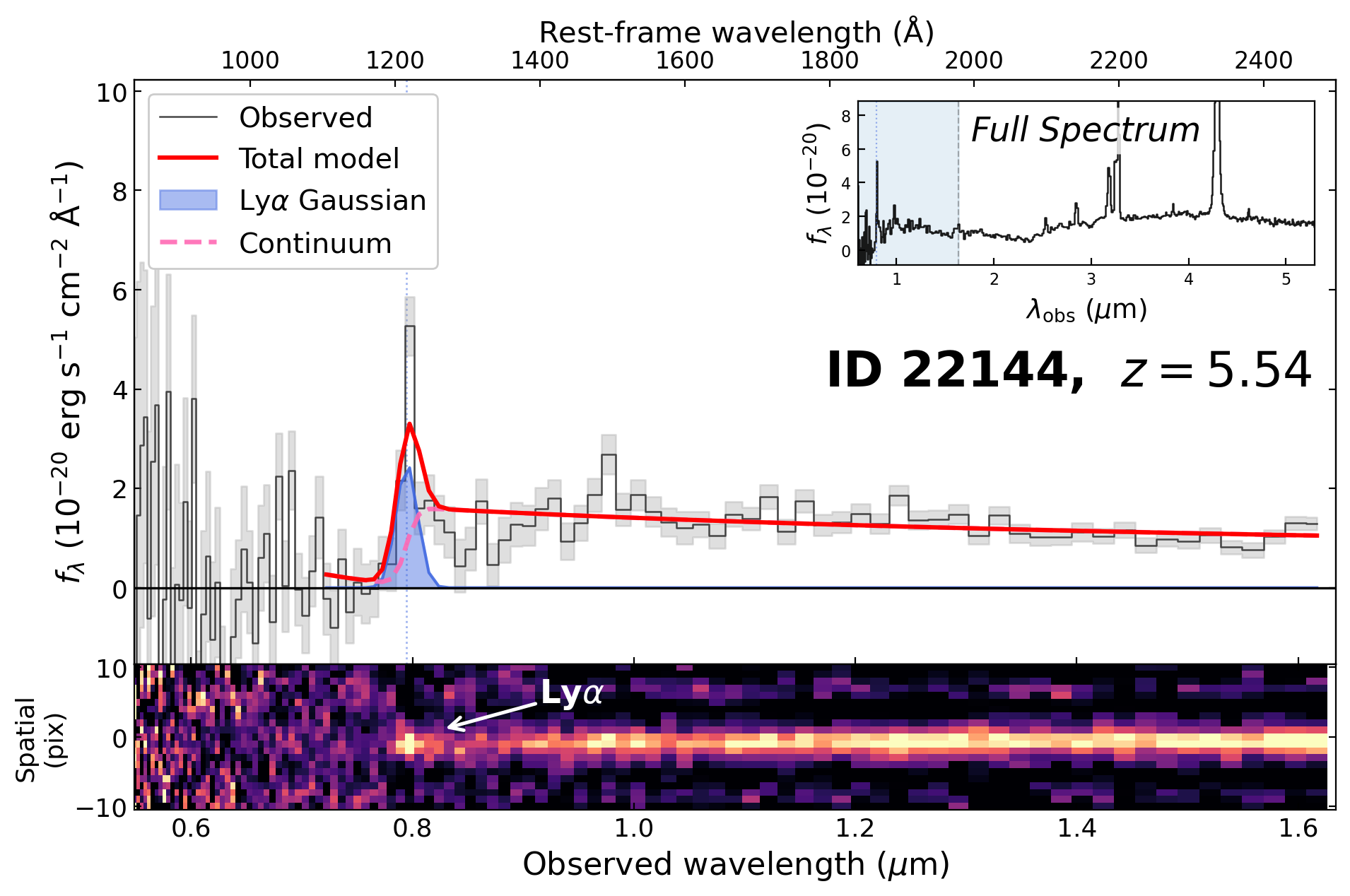}
\includegraphics[width=0.32\textwidth]{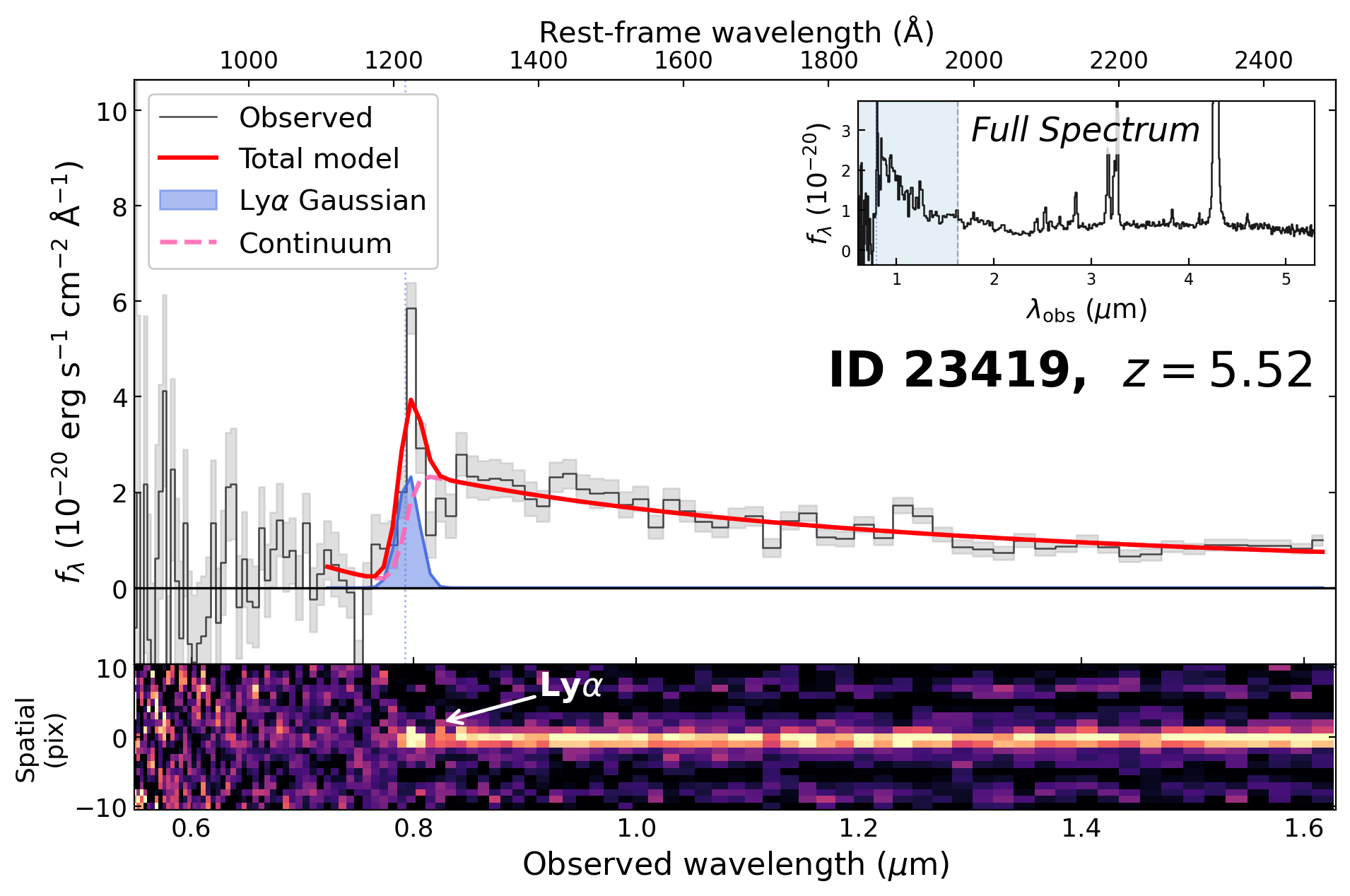}
\includegraphics[width=0.32\textwidth]{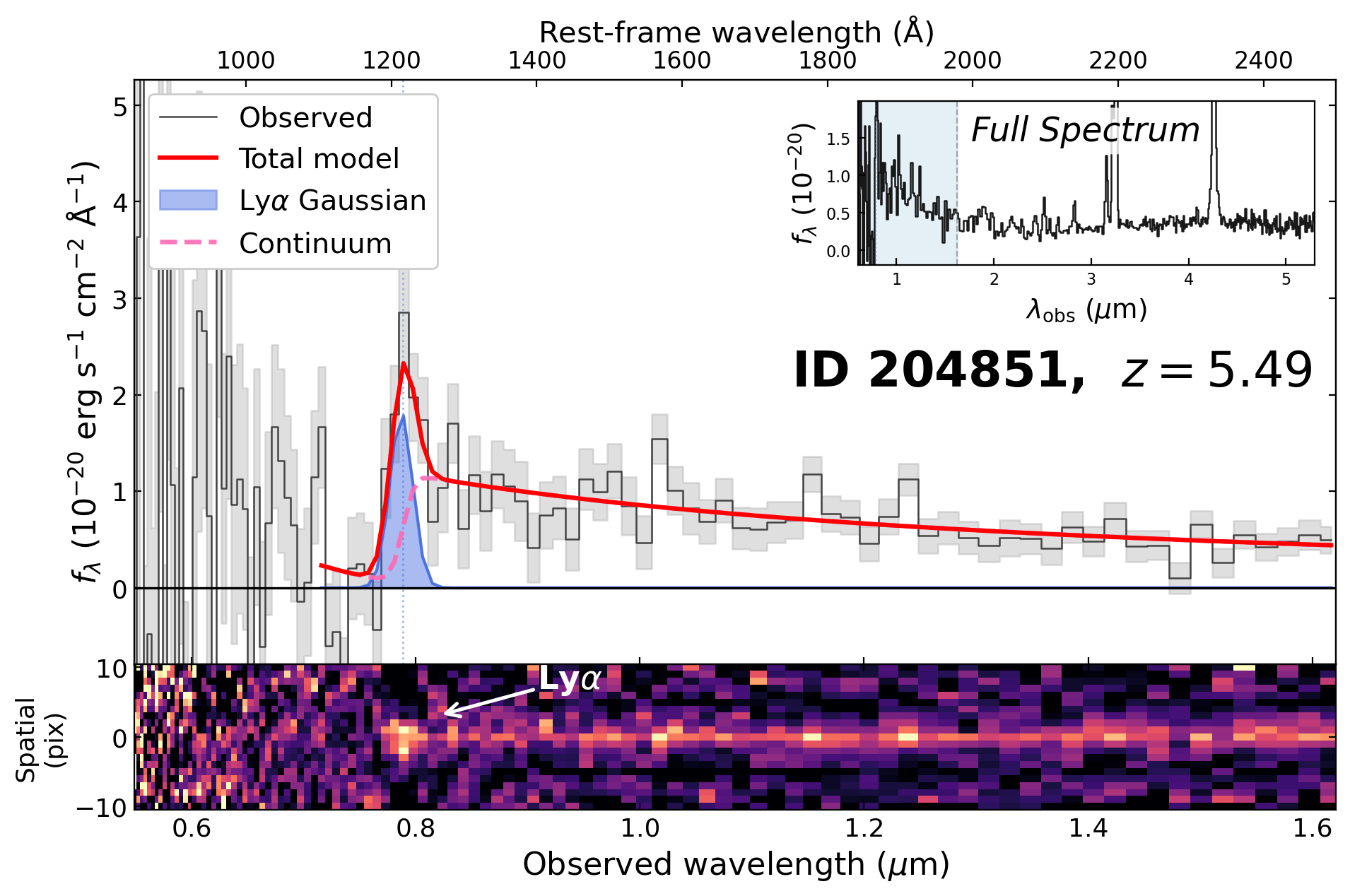}
\caption{1D and 2D spectra for LRDs with $z \geq 4$ and Ly$\alpha$ S/N $\geq 3$, sorted by decreasing redshift. For each source, the main panel shows the observed 1D spectrum (gray) with the best-fit total model (red), Ly$\alpha$ Gaussian component (blue shaded), and UV continuum (pink dashed). The bottom sub-panel shows the corresponding 2D spectrum. The inset displays the full-wavelength 1D spectrum, with the blue-shaded region marking the wavelength range shown in the main panel.}
\label{fig:spec_all_1}
\end{figure}

\begin{figure}
\centering
\includegraphics[width=0.32\textwidth]{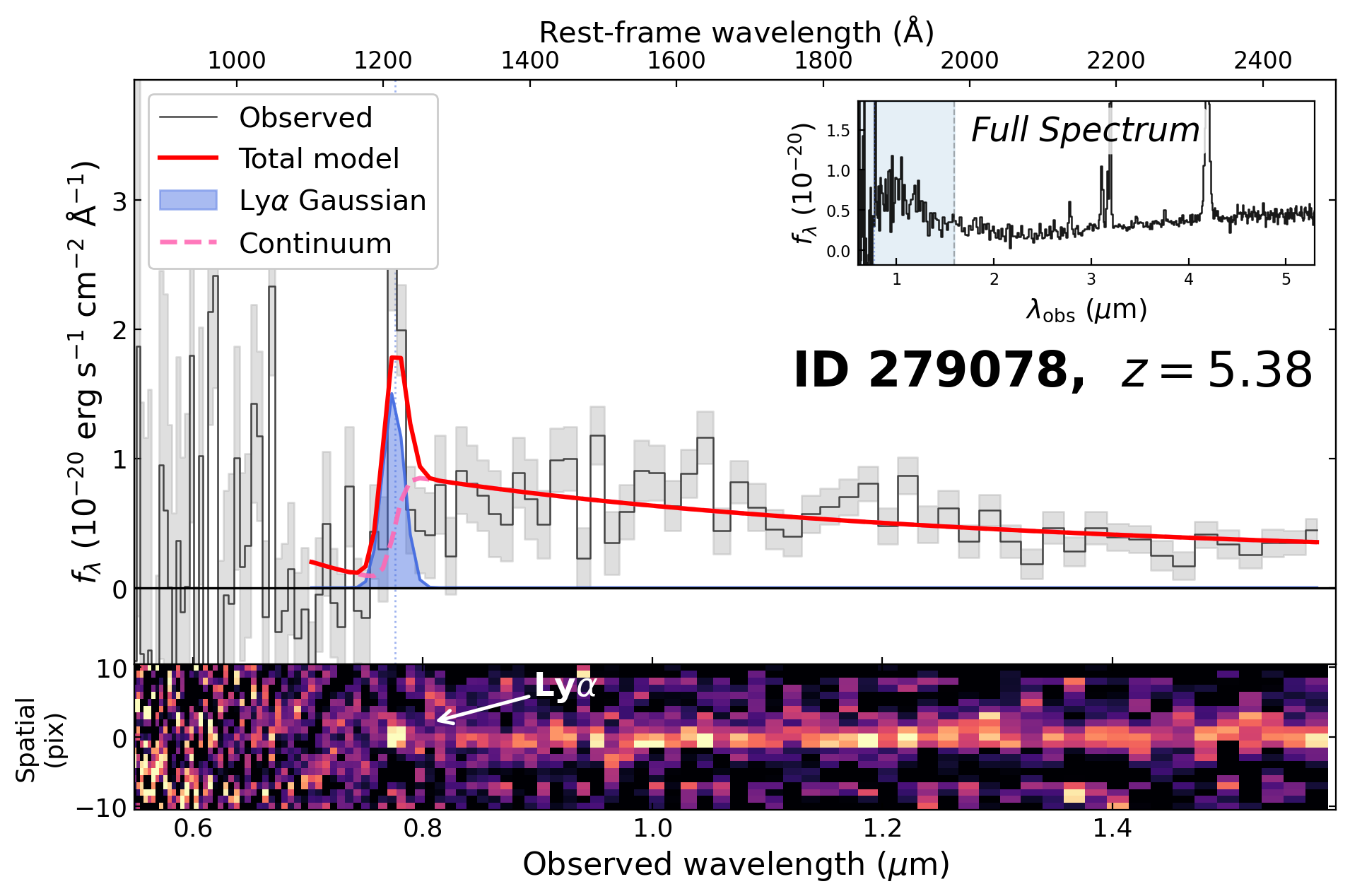}
\includegraphics[width=0.32\textwidth]{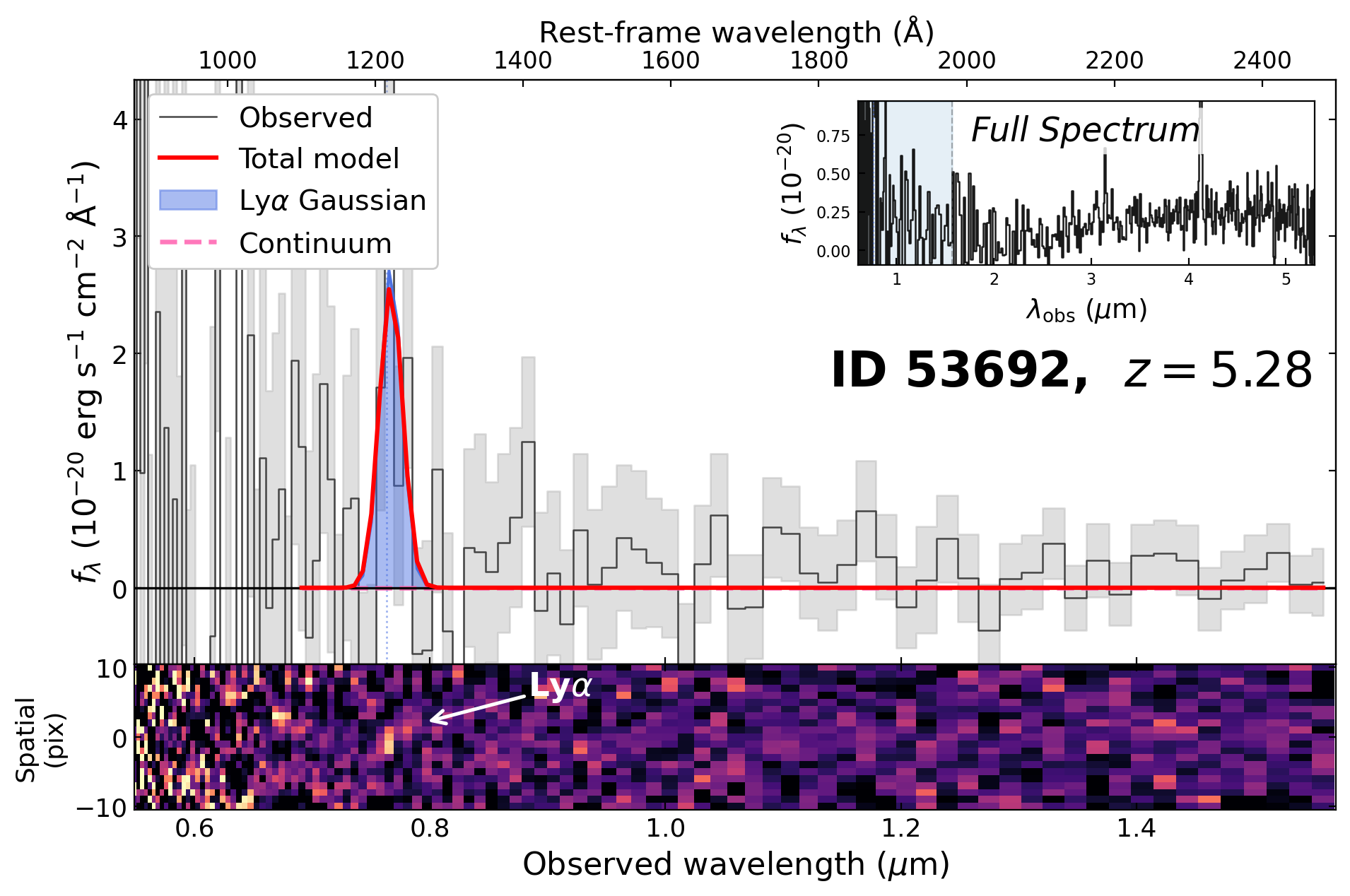}
\includegraphics[width=0.32\textwidth]{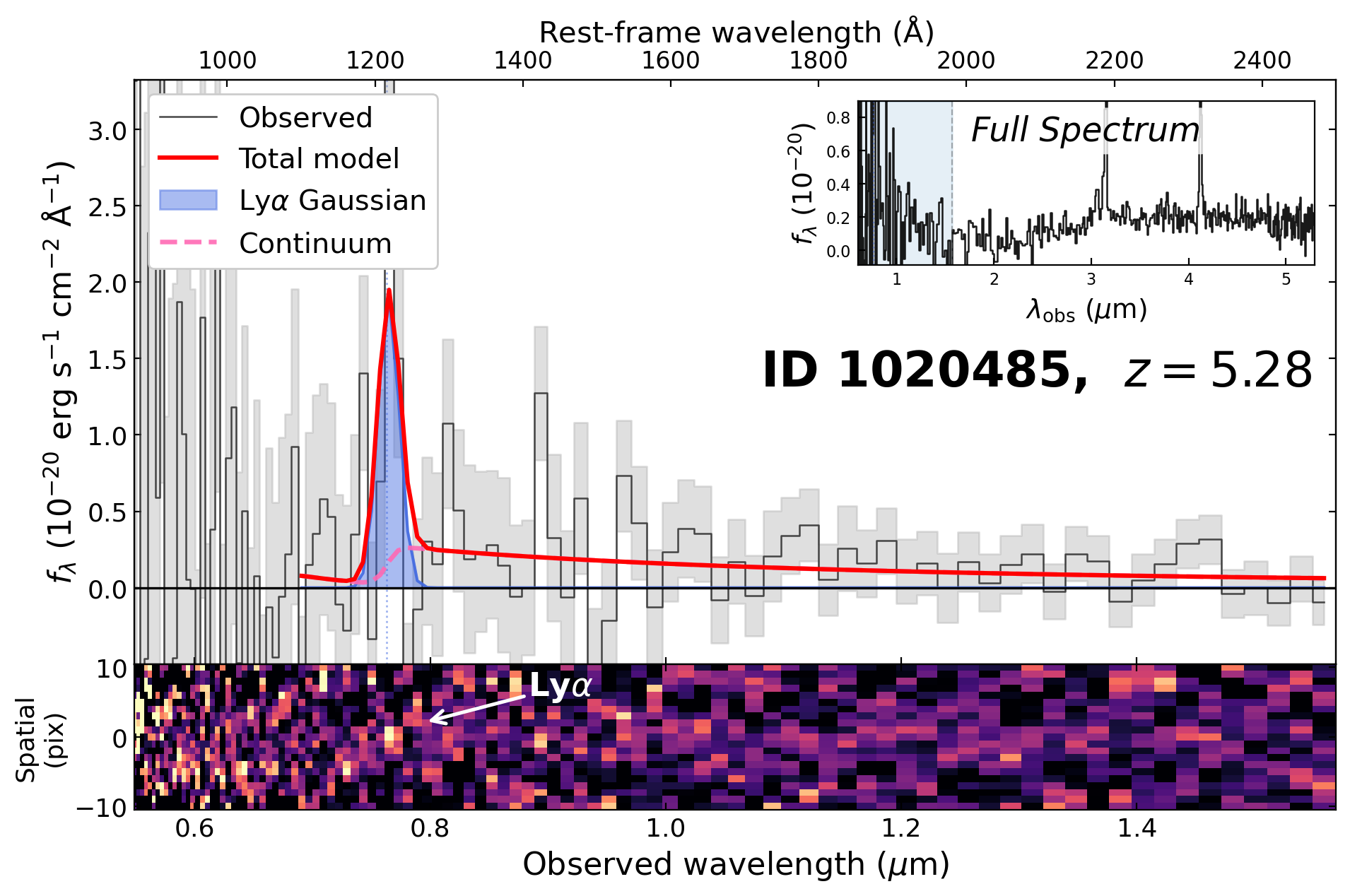}\\[1pt]
\includegraphics[width=0.32\textwidth]{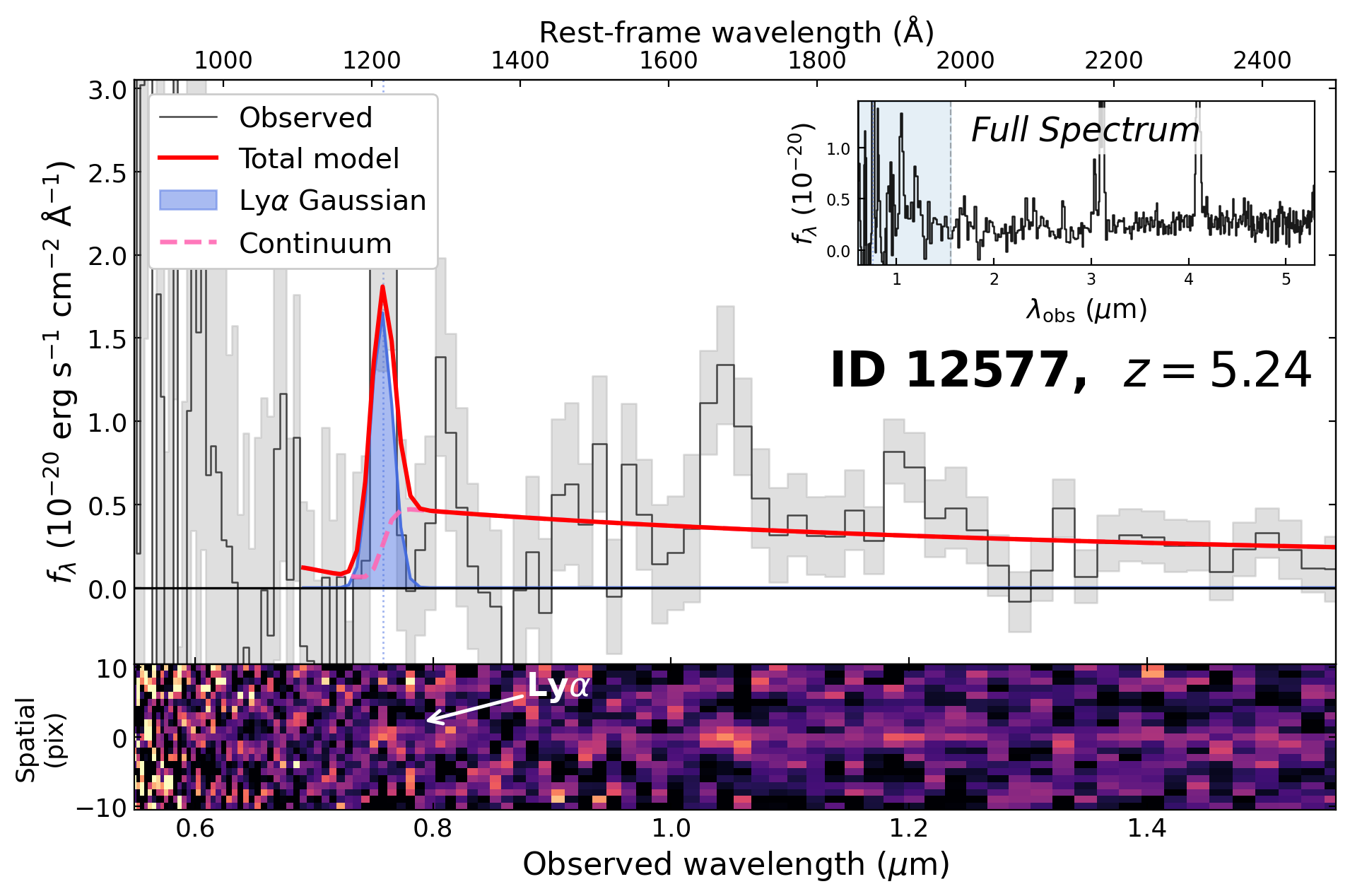}
\includegraphics[width=0.32\textwidth]{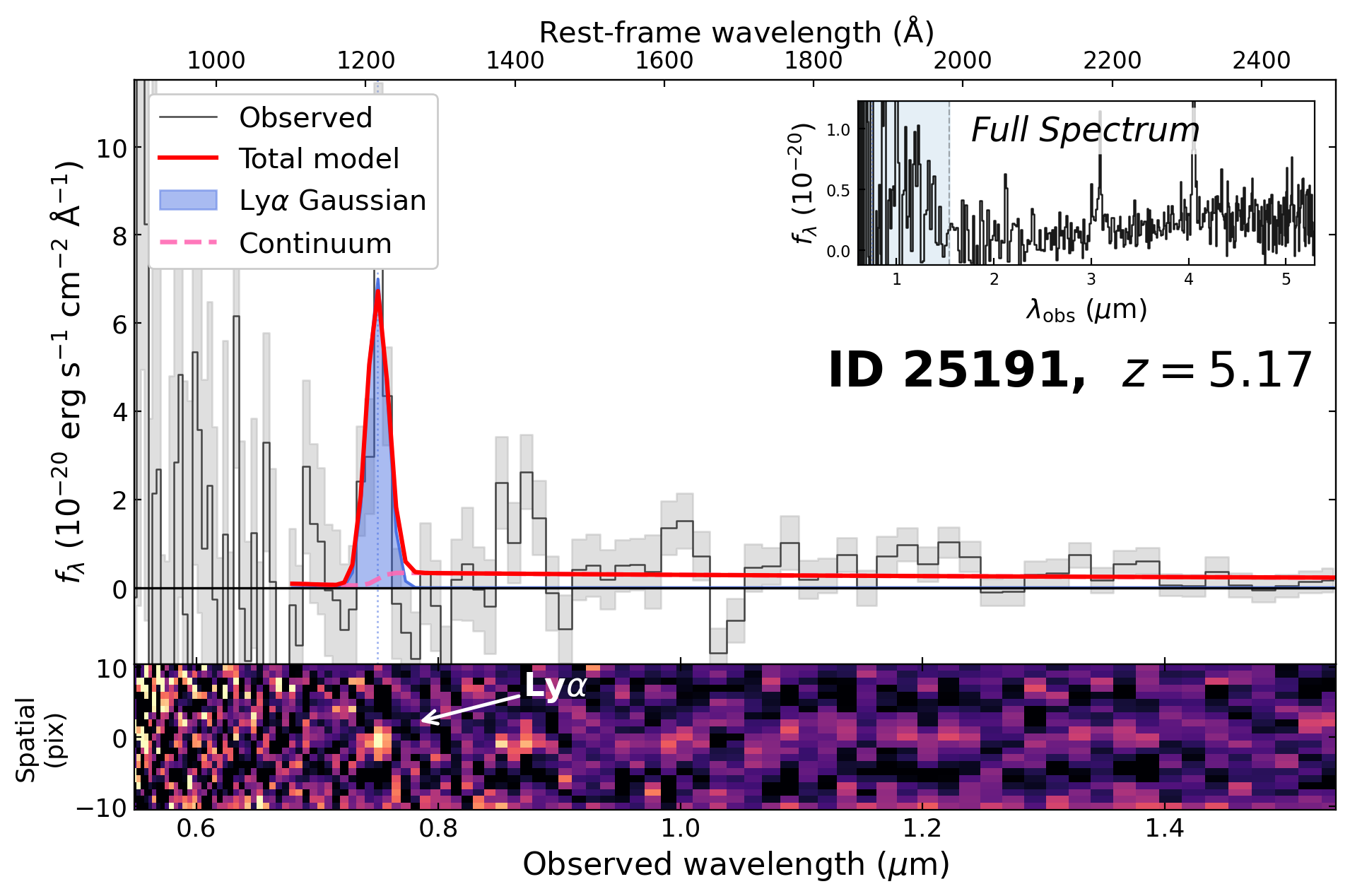}
\includegraphics[width=0.32\textwidth]{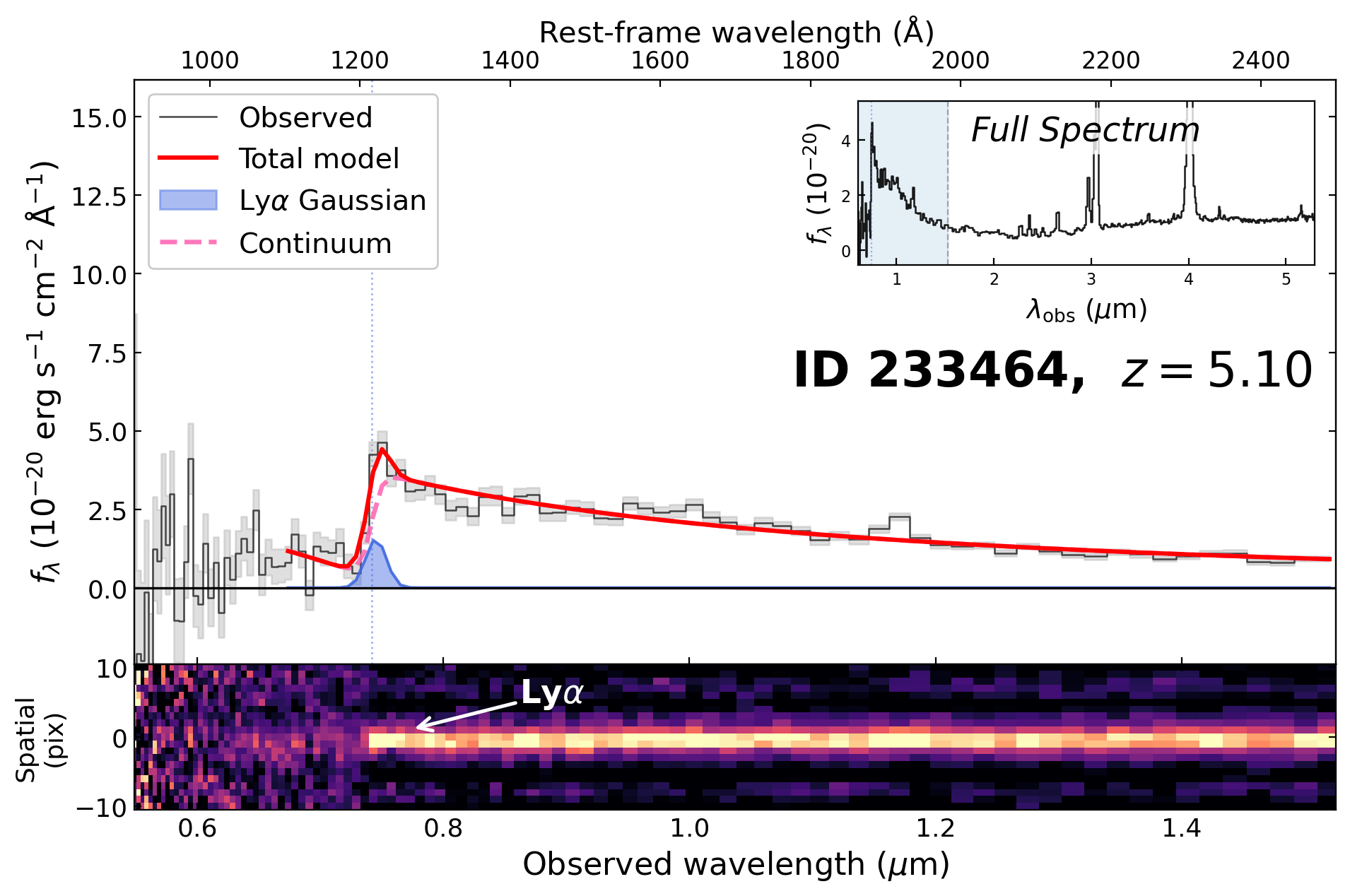}\\[1pt]
\includegraphics[width=0.32\textwidth]{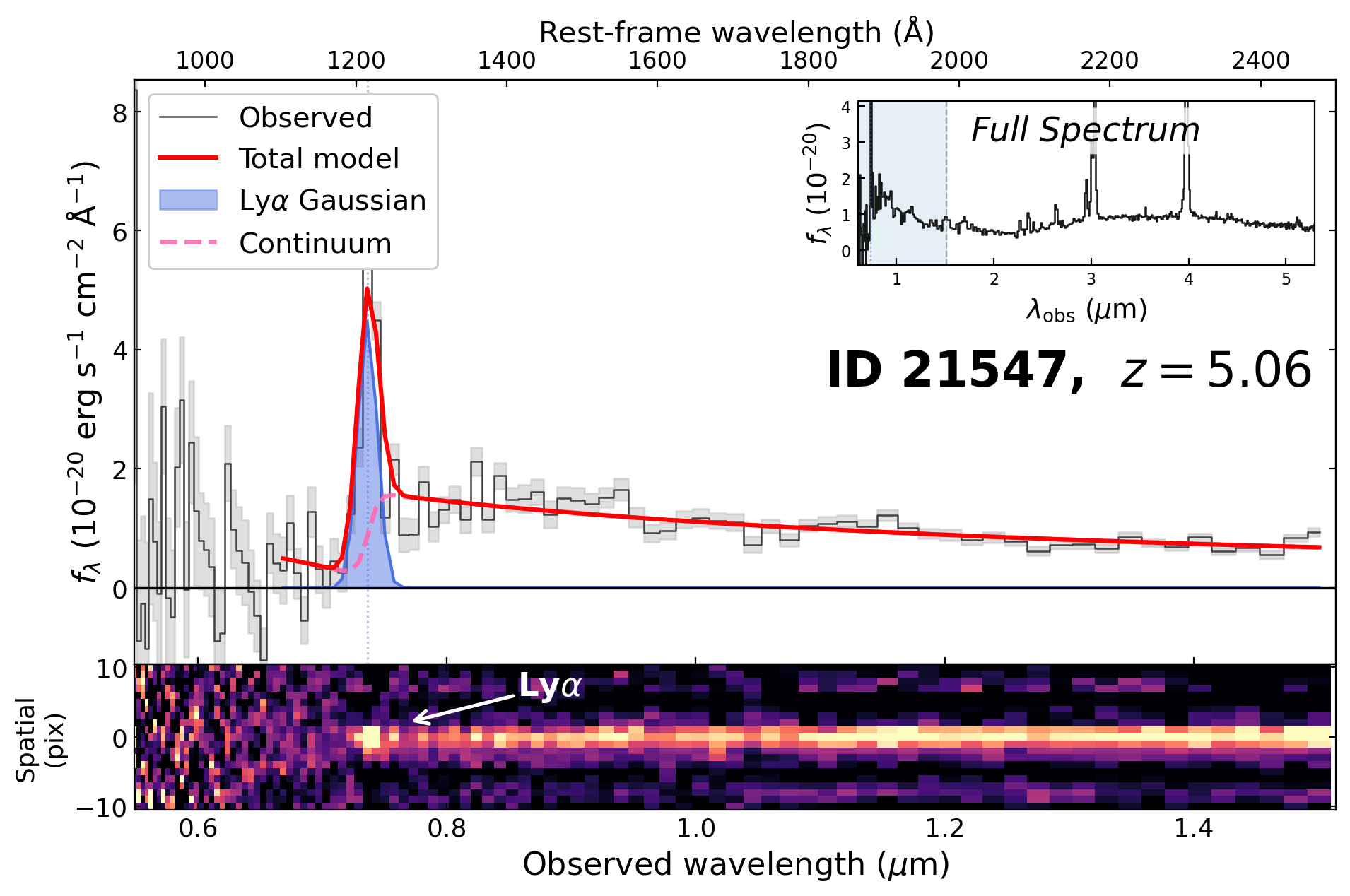}
\includegraphics[width=0.32\textwidth]{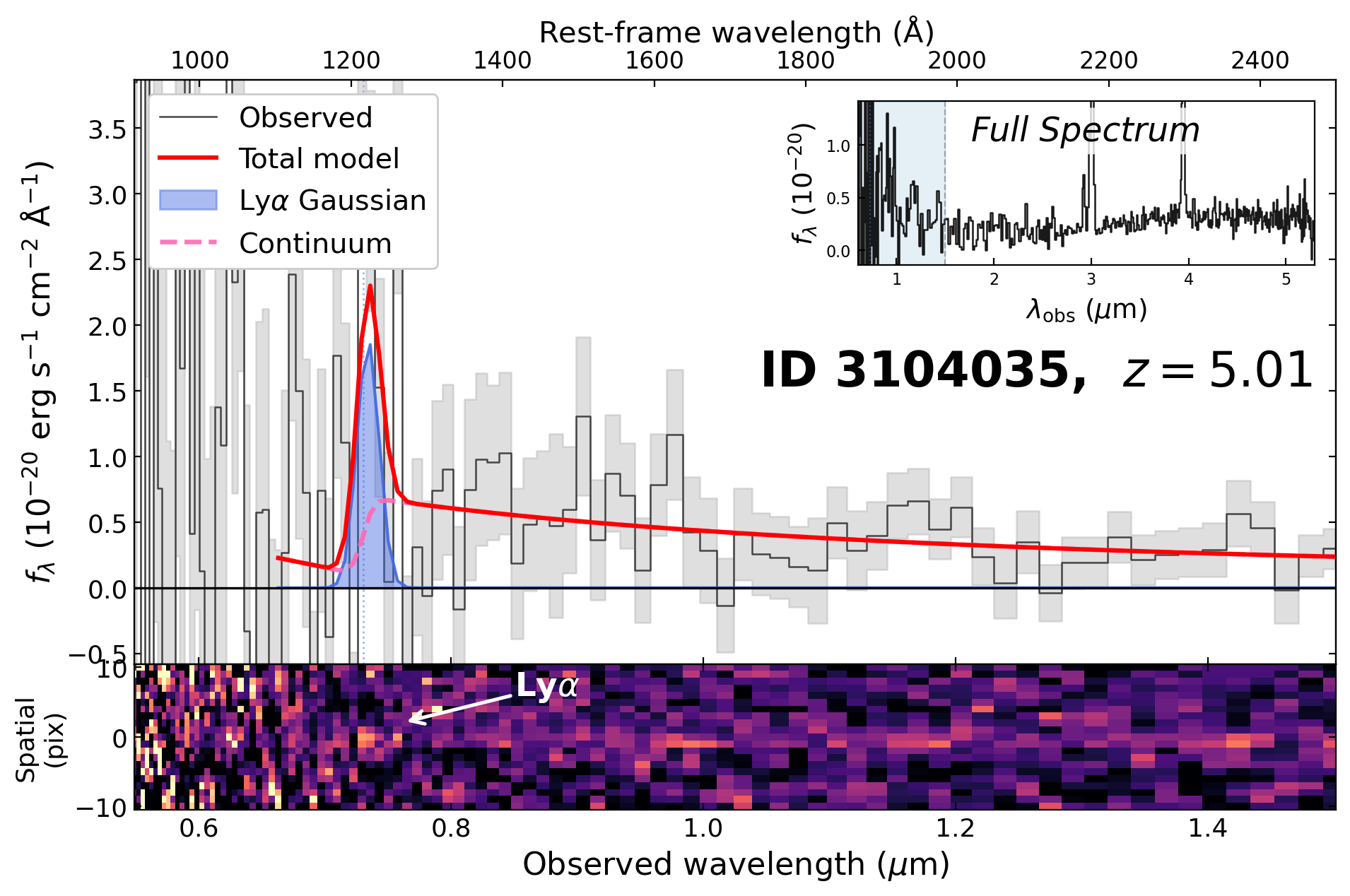}
\includegraphics[width=0.32\textwidth]{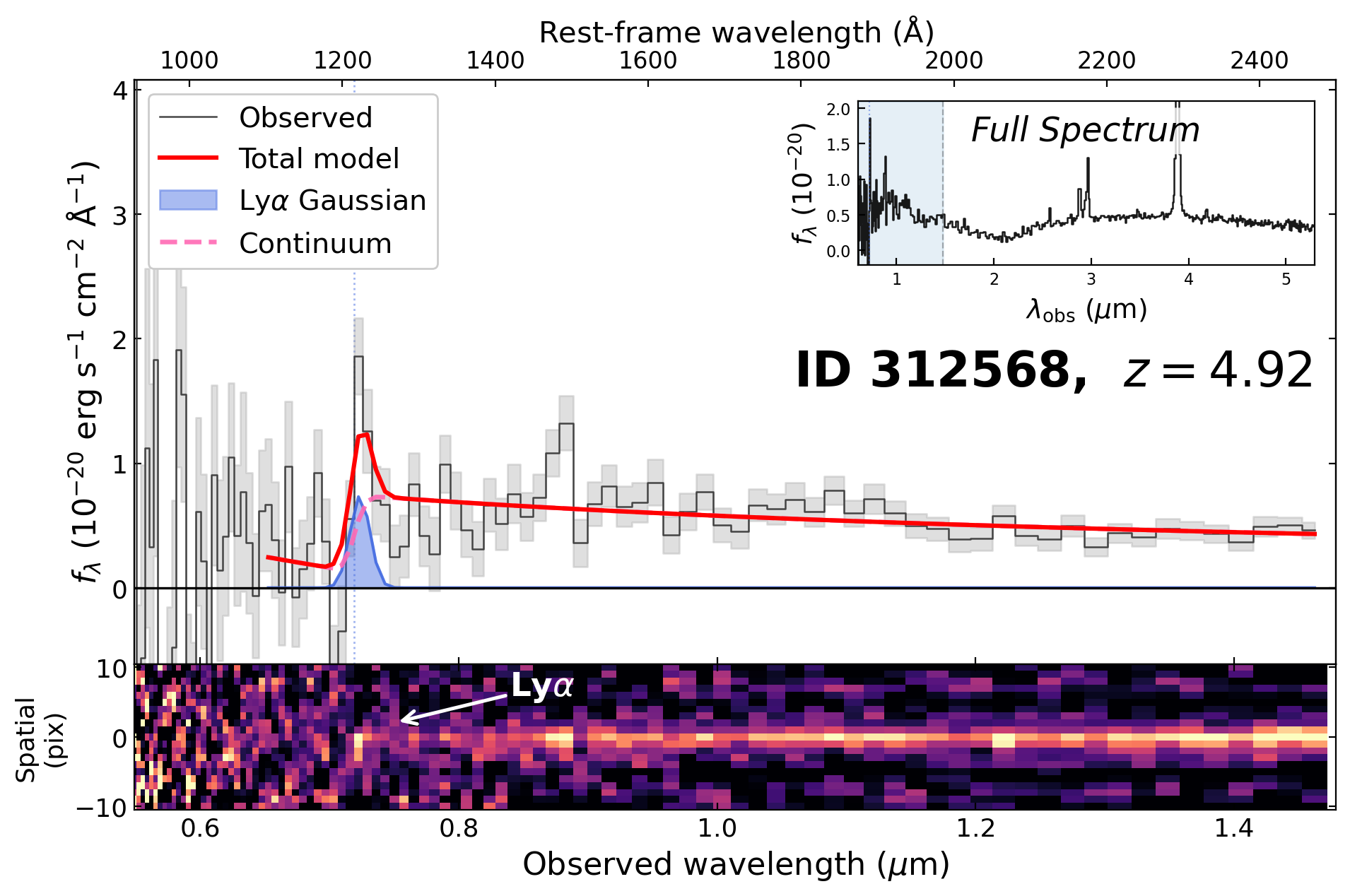}\\[1pt]
\includegraphics[width=0.32\textwidth]{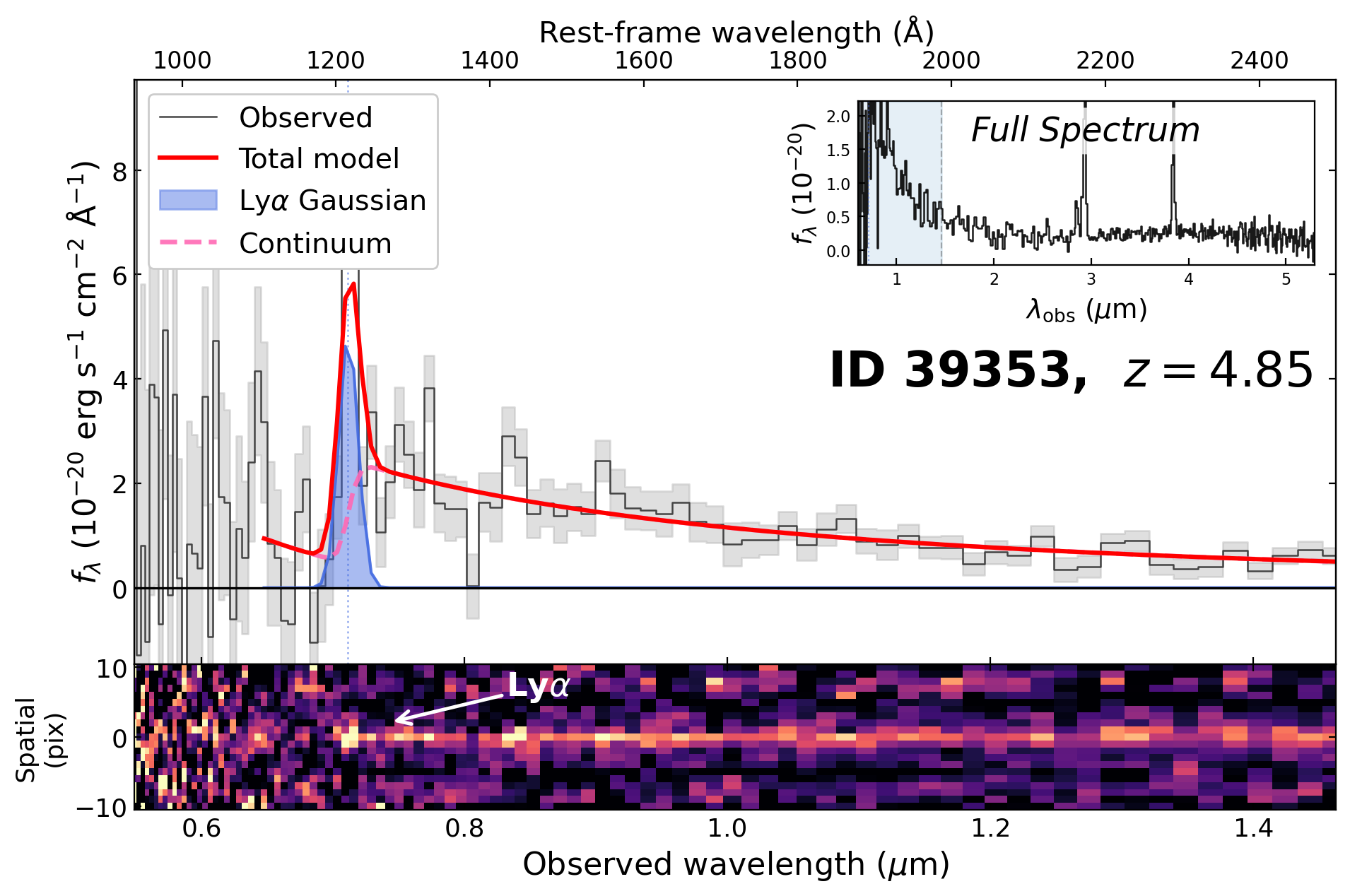}
\includegraphics[width=0.32\textwidth]{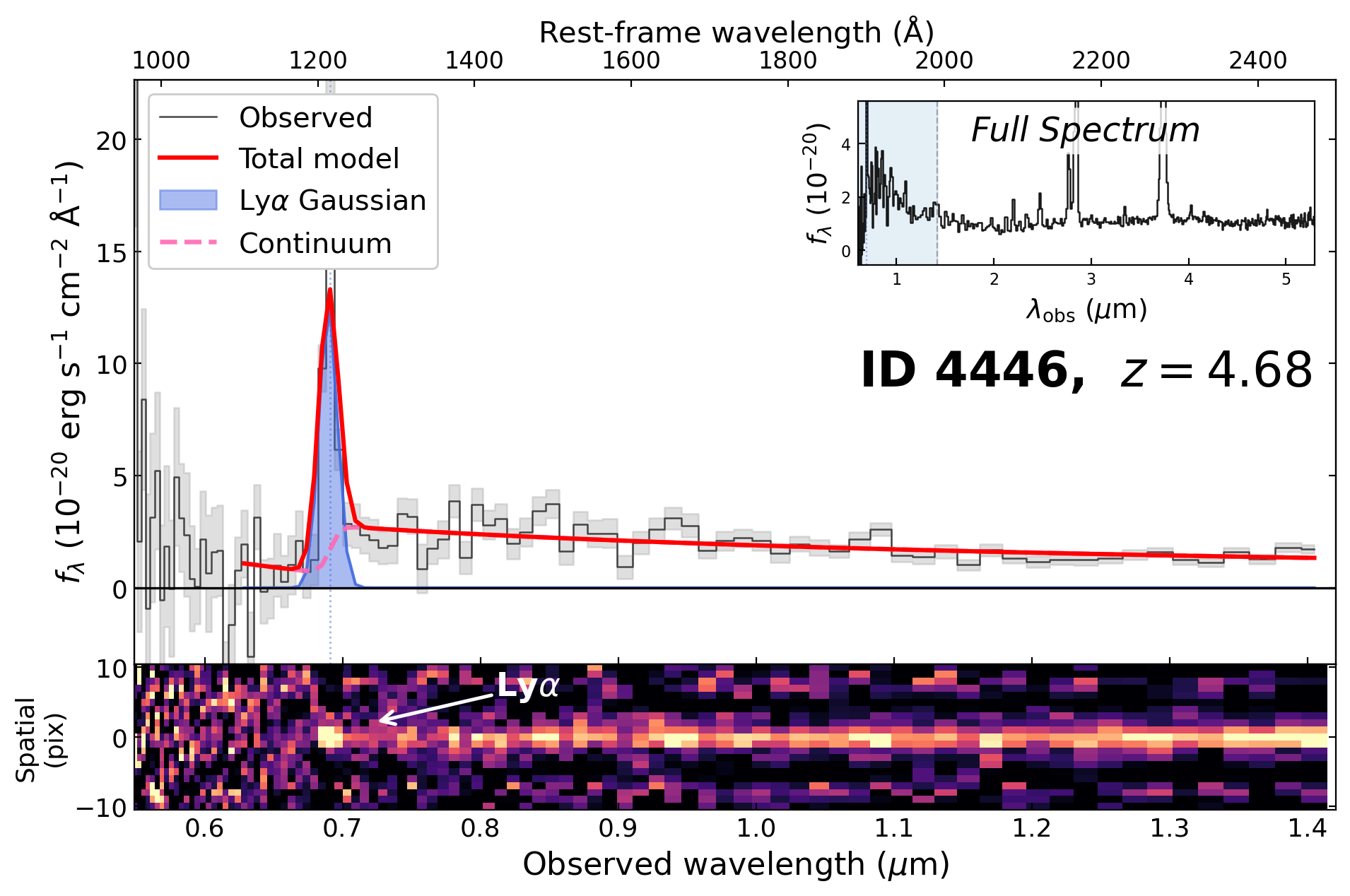}
\includegraphics[width=0.32\textwidth]{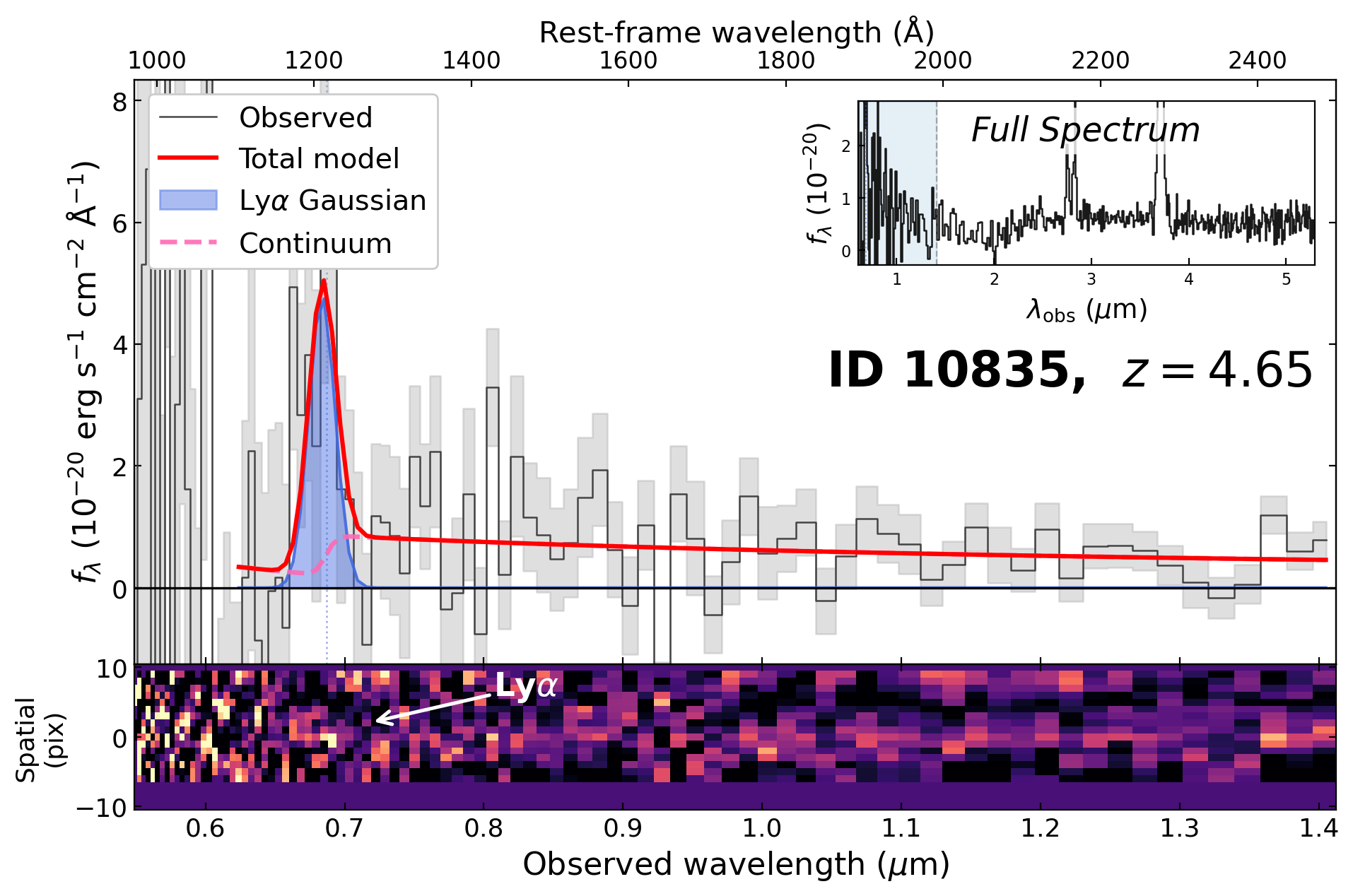}\\[1pt]
\includegraphics[width=0.32\textwidth]{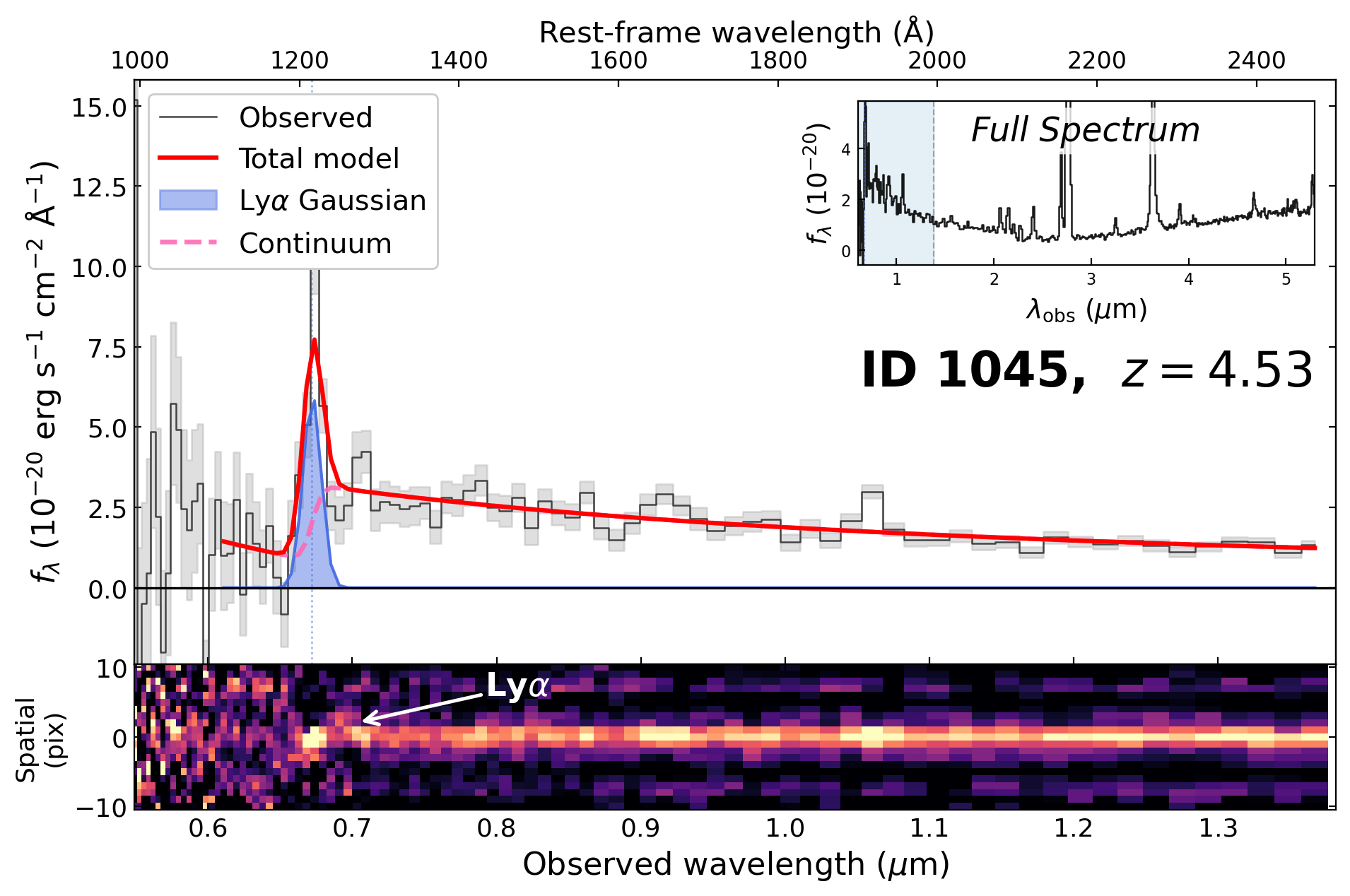}
\includegraphics[width=0.32\textwidth]{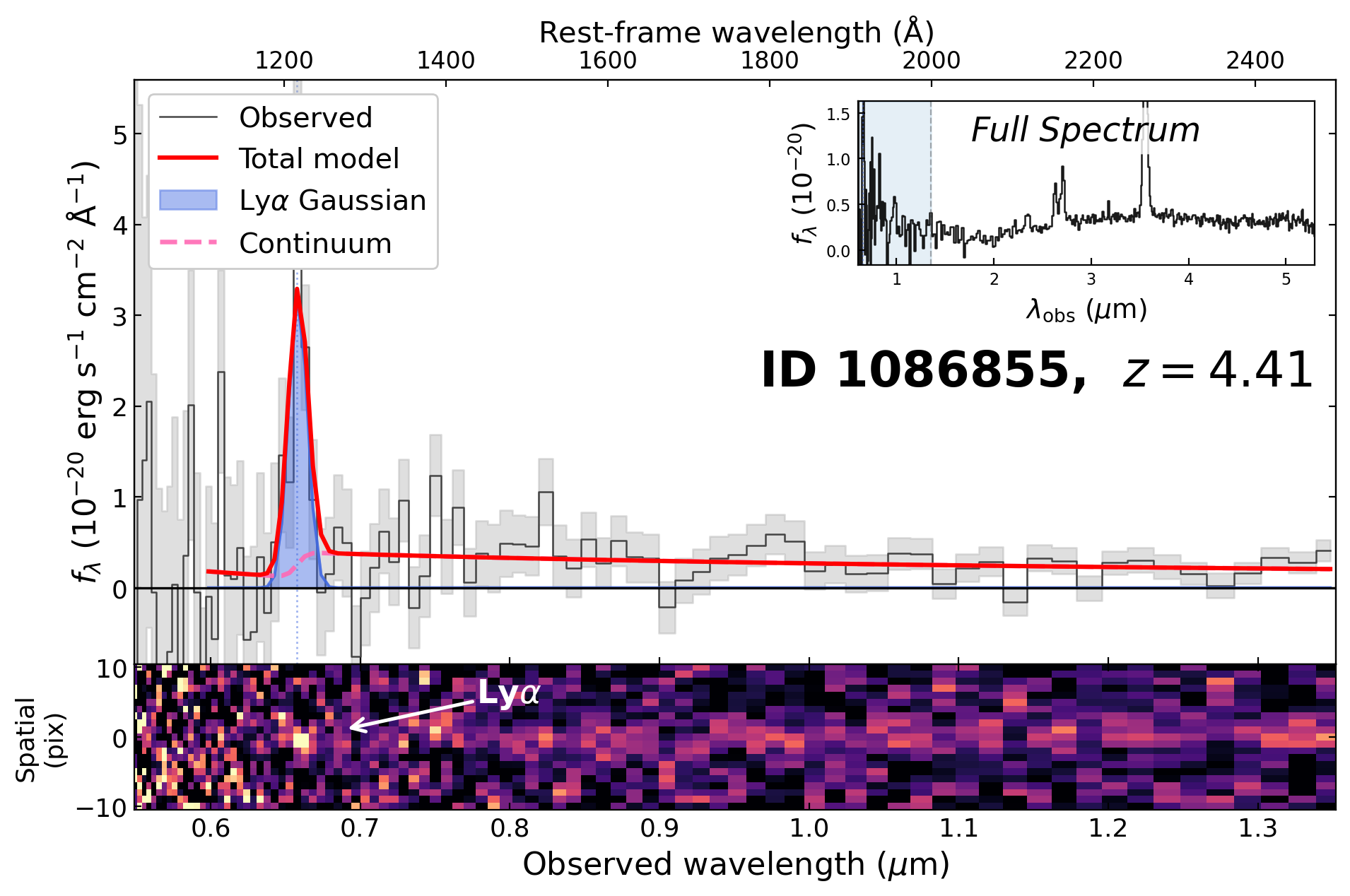}
\includegraphics[width=0.32\textwidth]{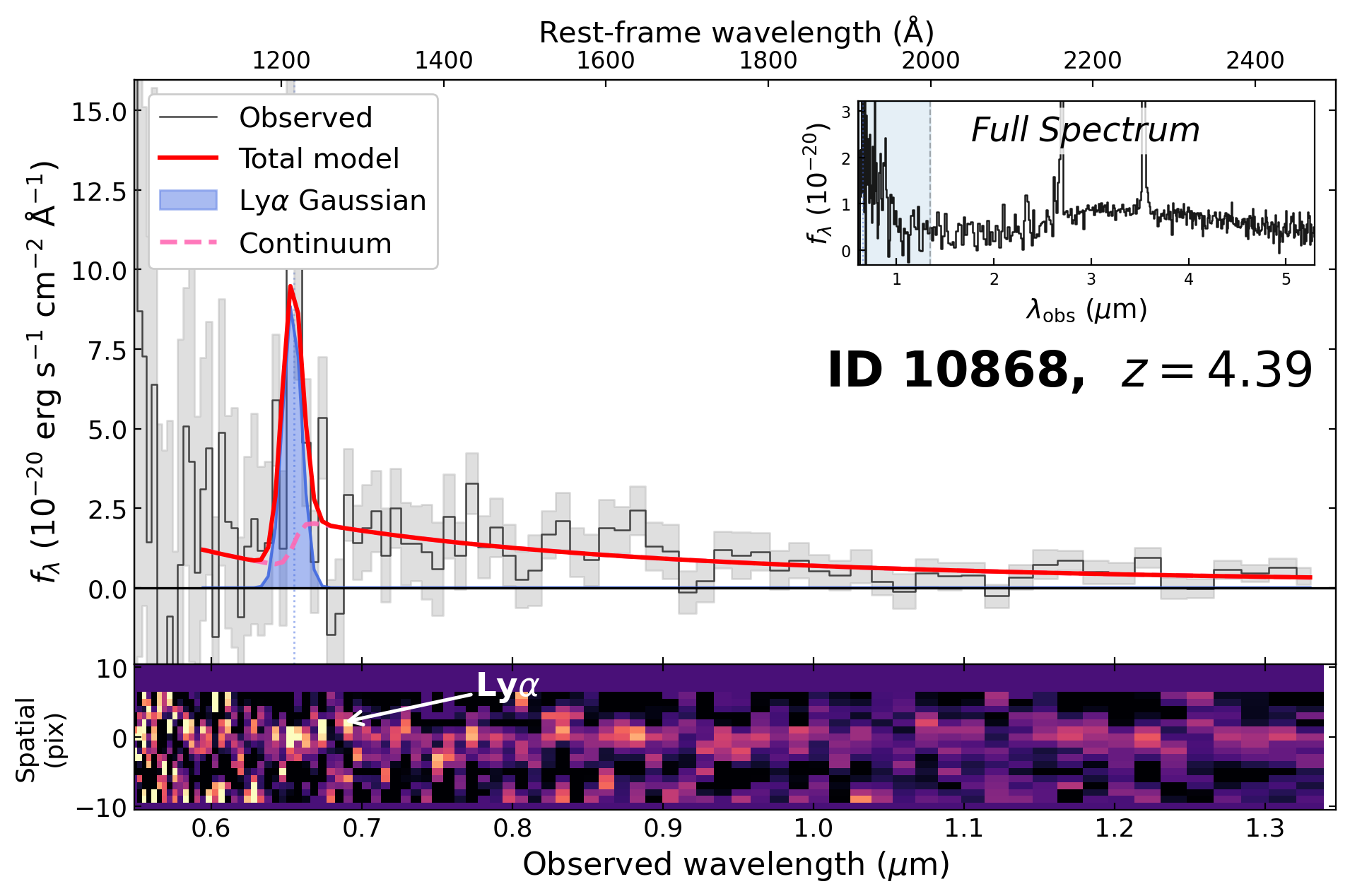}
\caption{Same as Figure~\ref{fig:spec_all_1}, continued.}
\label{fig:spec_all_2}
\end{figure}

\begin{figure}
\centering
\includegraphics[width=0.32\textwidth]{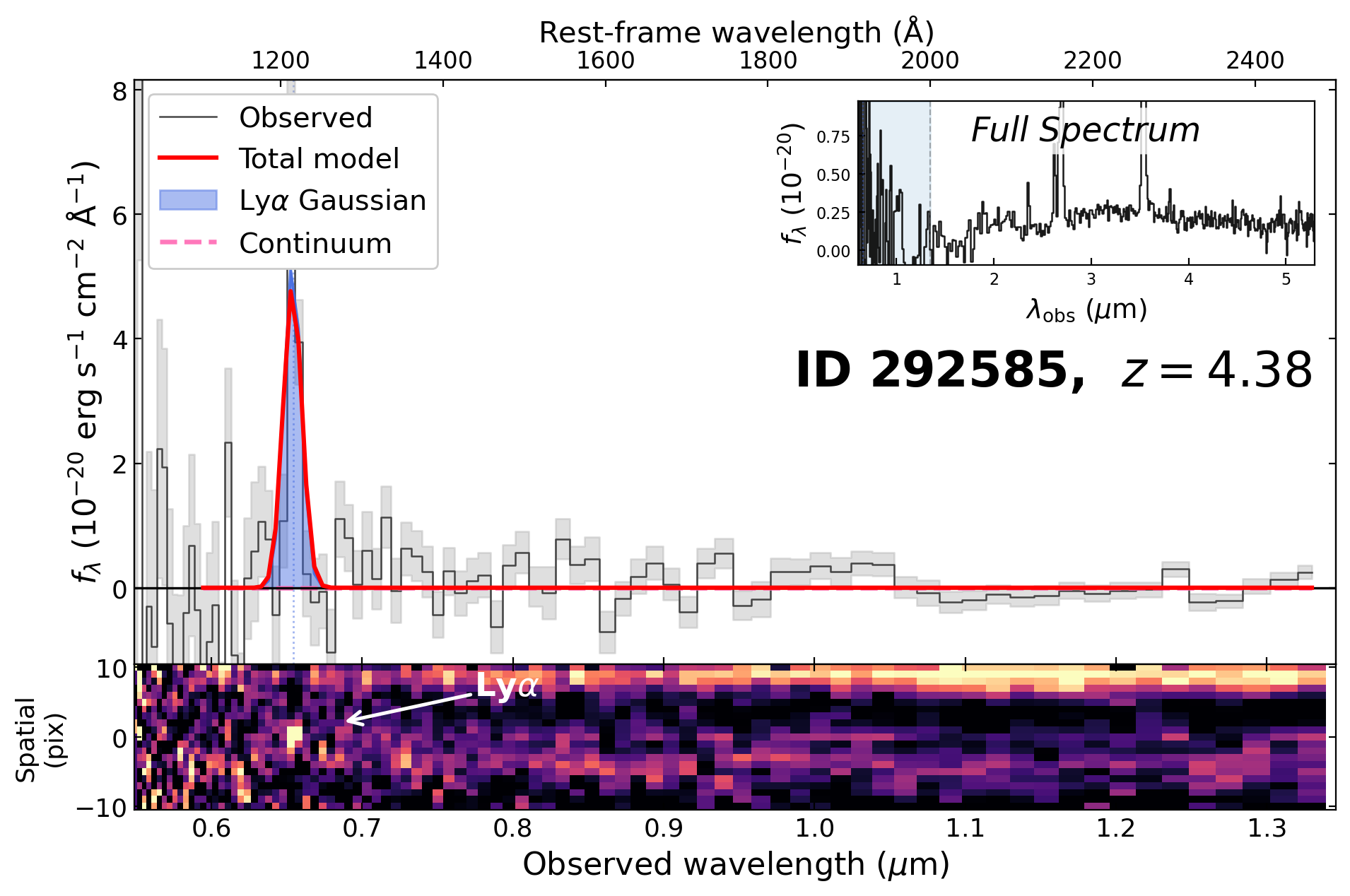}
\includegraphics[width=0.32\textwidth]{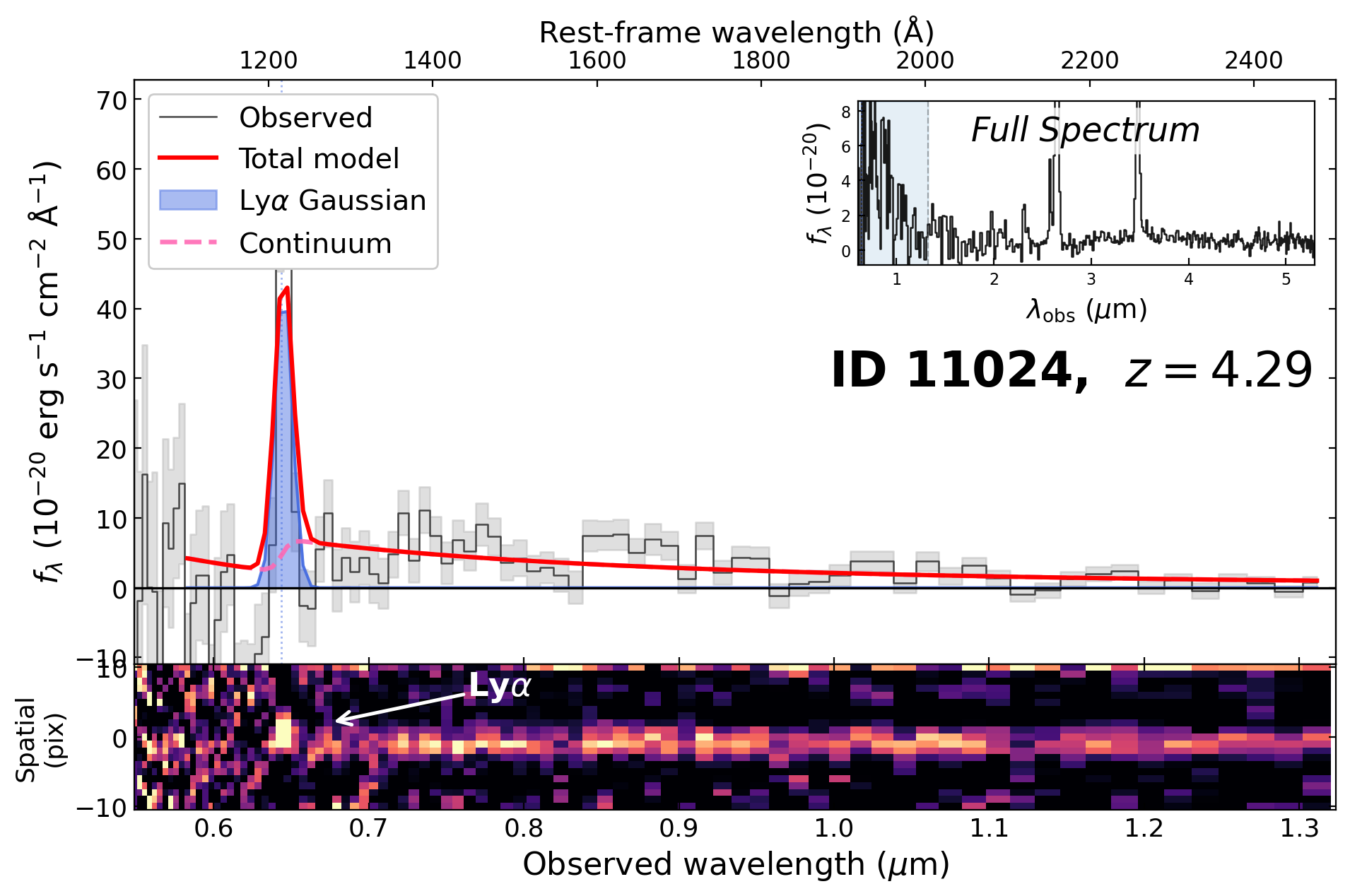}
\includegraphics[width=0.32\textwidth]{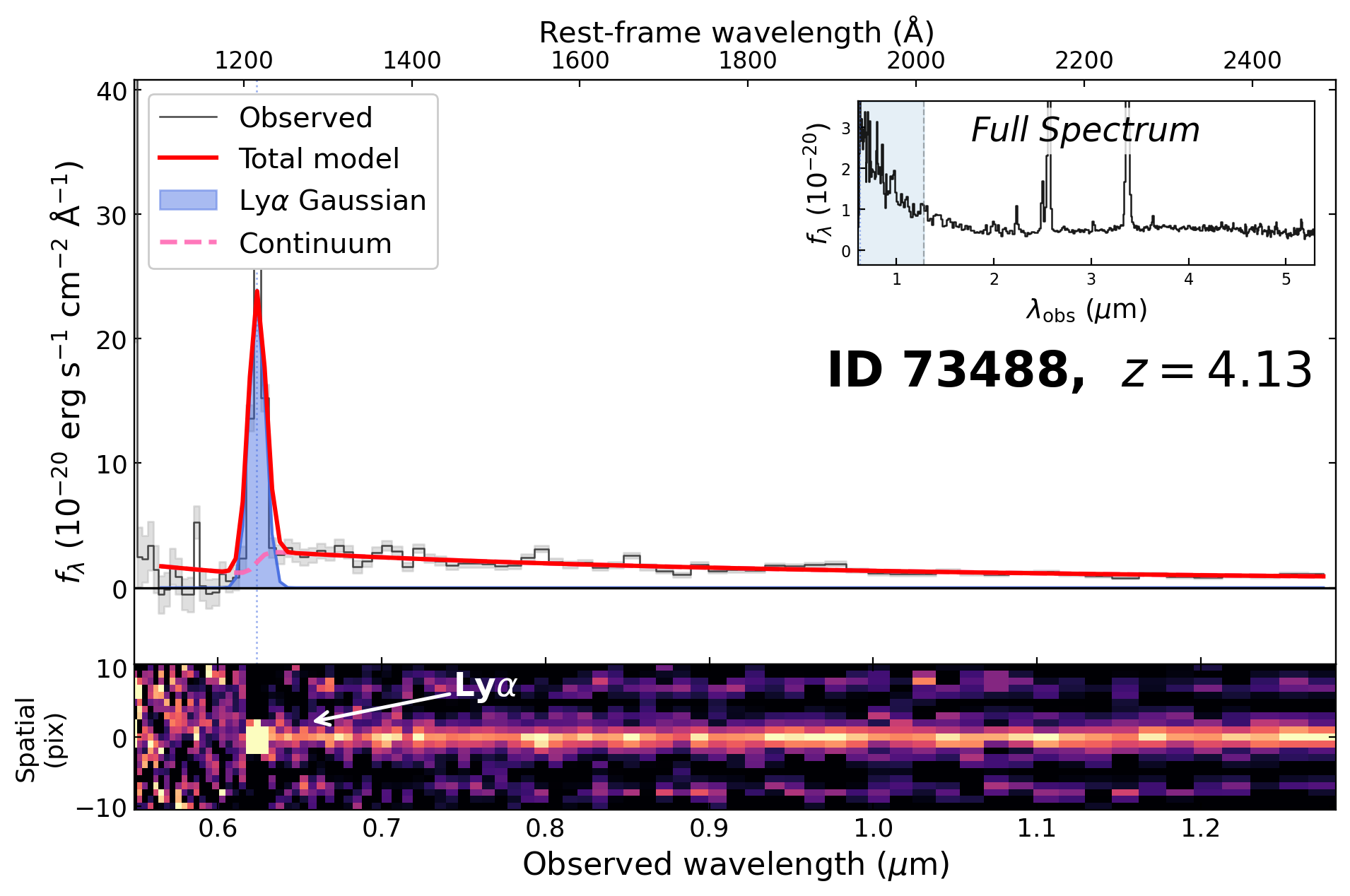}\\[1pt]
\includegraphics[width=0.32\textwidth]{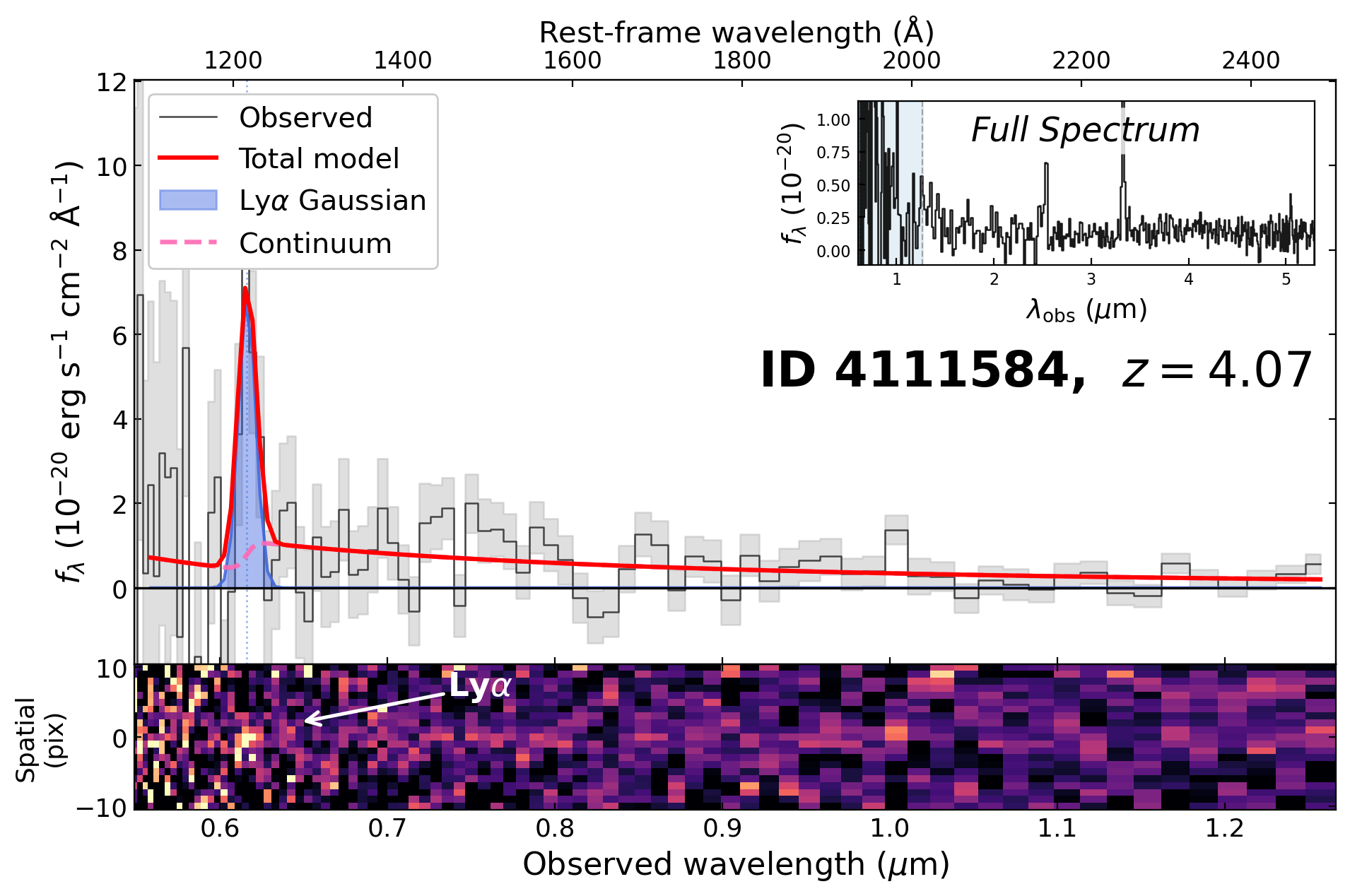}
\caption{Same as Figure~\ref{fig:spec_all_1}, continued.}
\label{fig:spec_all_3}
\end{figure}

Table \ref{tab:spec_fit_det} presents our spectral fitting results for all 32 LRDs at $z \geq 4$ from \citet{deGraaff2026} that have Ly$\alpha$ detections with S/N $\geq 3$. Figures \ref{fig:spec_all_1}, \ref{fig:spec_all_2}, and \ref{fig:spec_all_3}  show the individual spectral fits for the LRDs with S/N(Ly$\alpha$) $\geq 3$.

\begin{deluxetable}{lcccccccccc}
\tablecaption{UV spectral measurements of the LRDs with Ly$\alpha$ detections (S/N $\geq 3$)\label{tab:spec_fit_det}}
\tablewidth{0pt}
\tablehead{
\colhead{Source ID} & \colhead{PID} & \colhead{R.A.} & \colhead{Decl.} & \colhead{$z_{\rm spec}$} & \colhead{$\mu$} & \colhead{$F_{\rm Ly\alpha}$} & \colhead{EW$^{0}_{\rm Ly\alpha}$} & \colhead{S/N} & \colhead{$\beta_{\rm UV}$} & \colhead{$M_{\rm UV}$} \\
& & \colhead{(deg)} & \colhead{(deg)} & & & \colhead{($10^{-18}$ erg\,s$^{-1}$\,cm$^{-2}$)} & \colhead{(\AA)} & & & \colhead{(mag)}
}
\startdata
20466 & 2561 & 3.6404084 & -30.3864376 & 8.5095 & $1.35$ & $4.51_{-0.59}^{+0.60}$ & $219.0_{-52.3}^{+71.8}$ & 7.5 & $-1.52_{-0.47}^{+0.48}$ & $-18.45 \pm 0.15$ \\
920396 & 5224 & 34.2478277 & -5.1519963 & 7.0976 & $1.00$ & $1.90_{-0.48}^{+0.49}$ & $38.8_{-10.3}^{+11.2}$ & 3.9 & $-1.01_{-0.13}^{+0.13}$ & $-19.05 \pm 0.04$ \\
15383\tablenotemark{$^\dagger$} & 2561 & 3.5835346 & -30.3966786 & 7.0370 & $6.75$ & $5.75_{-0.33}^{+0.32}$ & $89.7_{-6.4}^{+6.5}$ & 17.6 & $-1.14_{-0.07}^{+0.07}$ & $-19.30 \pm 0.02$ \\
24175\tablenotemark{$^\dagger$} & 2561 & 3.5798316 & -30.4015692 & 7.0357 & $6.88$ & $4.14_{-0.43}^{+0.44}$ & $130.6_{-18.3}^{+21.9}$ & 9.6 & $-1.31_{-0.20}^{+0.20}$ & $-18.49 \pm 0.06$ \\
16594\tablenotemark{$^\dagger$} & 2561 & 3.5972027 & -30.3943287 & 7.0346 & $3.60$ & $1.70_{-0.34}^{+0.34}$ & $106.7_{-25.9}^{+29.2}$ & 5.0 & $-1.08_{-0.27}^{+0.27}$ & $-17.80 \pm 0.08$ \\
219000 & 8060 & 53.1613700 & -27.7376600 & 6.8215 & $1.00$ & $4.06_{-0.44}^{+0.43}$ & $58.0_{-6.8}^{+7.2}$ & 9.3 & $-1.29_{-0.10}^{+0.09}$ & $-19.28 \pm 0.03$ \\
10108 & 6368 & 214.7975318 & 52.8187520 & 6.6231 & $1.00$ & $15.21_{-0.56}^{+0.57}$ & $204.0_{-14.2}^{+15.2}$ & 26.9 & $-1.65_{-0.12}^{+0.12}$ & $-19.19 \pm 0.03$ \\
347410 & 5224 & 150.0729041 & 2.3236212 & 6.0405 & $1.00$ & $6.03_{-0.59}^{+0.62}$ & $187.6_{-32.0}^{+40.2}$ & 10.0 & $-1.38_{-0.32}^{+0.33}$ & $-18.11 \pm 0.09$ \\
4771 & 6368 & 150.1610287 & 2.4658044 & 5.9273 & $1.00$ & $7.10_{-0.64}^{+0.61}$ & $293.9_{-58.2}^{+75.2}$ & 11.3 & $-1.40_{-0.40}^{+0.40}$ & $-17.74 \pm 0.11$ \\
4286 & 2561 & 3.6192010 & -30.4232703 & 5.8352 & $1.61$ & $6.02_{-0.53}^{+0.54}$ & $94.2_{-10.9}^{+12.2}$ & 11.3 & $-1.25_{-0.15}^{+0.15}$ & $-18.80 \pm 0.04$ \\
70283 & 5545 & 150.0783330 & 2.3726864 & 5.7834 & $1.00$ & $13.18_{-1.70}^{+1.68}$ & $271.0_{-68.6}^{+107.9}$ & 7.8 & $-1.21_{-0.49}^{+0.52}$ & $-18.49 \pm 0.15$ \\
172350 & 4233 & 34.3689512 & -5.1039415 & 5.5865 & $1.00$ & $6.53_{-1.30}^{+1.29}$ & $107.3_{-29.9}^{+38.0}$ & 5.0 & $-1.67_{-0.43}^{+0.45}$ & $-18.53 \pm 0.12$ \\
22144 & 5545 & 150.0637500 & 2.2467978 & 5.5370 & $1.00$ & $6.27_{-1.00}^{+1.05}$ & $59.1_{-10.6}^{+12.0}$ & 6.1 & $-0.61_{-0.13}^{+0.13}$ & $-19.36 \pm 0.04$ \\
23419 & 6368 & 34.4711074 & -5.1904554 & 5.5160 & $1.00$ & $5.93_{-0.94}^{+0.96}$ & $37.2_{-6.5}^{+7.0}$ & 6.2 & $-1.66_{-0.11}^{+0.11}$ & $-19.55 \pm 0.03$ \\
204851 & 1286 & 53.1385932 & -27.7902534 & 5.4855 & $1.00$ & $4.83_{-0.98}^{+0.98}$ & $62.3_{-14.4}^{+16.6}$ & 4.9 & $-1.38_{-0.22}^{+0.22}$ & $-18.82 \pm 0.05$ \\
279078 & 5224 & 150.0828163 & 2.2777263 & 5.3801 & $1.00$ & $3.59_{-0.63}^{+0.62}$ & $63.7_{-13.1}^{+14.0}$ & 5.8 & $-1.29_{-0.20}^{+0.20}$ & $-18.45 \pm 0.05$ \\
53692 & 4233 & 34.4553763 & -5.2318140 & 5.2793 & $1.00$ & $6.84_{-1.78}^{+1.75}$ & -- & 3.9 & -- & -- \\
1020485 & 8018 & 189.1130858 & 62.2923908 & 5.2769 & $1.00$ & $4.49_{-1.10}^{+1.04}$ & $245.0_{-114.4}^{+332.7}$ & 4.2 & $-2.09_{-1.40}^{+1.35}$ & $-16.94 \pm 0.43$ \\
12577 & 2198 & 53.0484548 & -27.8151409 & 5.2363 & $1.00$ & $3.94_{-1.24}^{+1.22}$ & $128.6_{-48.8}^{+67.4}$ & 3.2 & $-0.96_{-0.47}^{+0.50}$ & $-17.78 \pm 0.15$ \\
25191 & 5105 & 268.3795708 & 65.2023333 & 5.1674 & $1.00$ & $15.12_{-1.88}^{+1.89}$ & $719.7_{-291.9}^{+873.6}$ & 8.0 & $-0.56_{-0.97}^{+1.06}$ & $-17.45 \pm 0.38$ \\
233464 & 5224 & 150.0716706 & 2.2004085 & 5.1024 & $1.00$ & $3.37_{-0.55}^{+0.59}$ & $14.9_{-2.6}^{+2.8}$ & 5.9 & $-1.96_{-0.06}^{+0.06}$ & $-19.67 \pm 0.01$ \\
21547 & 2561 & 3.5508378 & -30.4065978 & 5.0579 & $1.61$ & $8.86_{-0.44}^{+0.45}$ & $91.1_{-6.3}^{+6.7}$ & 19.8 & $-1.21_{-0.08}^{+0.08}$ & $-18.90 \pm 0.02$ \\
3104035 & 1208 & 64.0562377 & -24.1135383 & 5.0068 & $1.17$ & $4.45_{-1.41}^{+1.46}$ & $105.6_{-44.4}^{+67.6}$ & 3.1 & $-1.50_{-0.61}^{+0.63}$ & $-17.89 \pm 0.16$ \\
312568 & 5224 & 150.1058493 & 2.4468234 & 4.9161 & $1.00$ & $1.51_{-0.52}^{+0.47}$ & $34.0_{-12.4}^{+12.2}$ & 3.0 & $-0.77_{-0.15}^{+0.16}$ & $-18.07 \pm 0.05$ \\
39353 & 1181 & 189.2939522 & 62.1530911 & 4.8520 & $1.00$ & $9.10_{-1.23}^{+1.21}$ & $63.7_{-10.8}^{+11.5}$ & 7.4 & $-2.20_{-0.18}^{+0.19}$ & $-18.99 \pm 0.05$ \\
4446 & 1215 & 34.2707118 & -5.2176705 & 4.6835 & $1.00$ & $21.33_{-1.35}^{+1.37}$ & $135.3_{-12.7}^{+15.2}$ & 15.6 & $-1.03_{-0.14}^{+0.14}$ & $-19.27 \pm 0.04$ \\
10835 & 5105 & 268.4376747 & 65.1674849 & 4.6501 & $1.00$ & $11.22_{-3.64}^{+3.81}$ & $227.0_{-96.1}^{+165.7}$ & 3.0 & $-0.89_{-0.70}^{+0.73}$ & $-18.01 \pm 0.21$ \\
1045 & 1433 & 101.9334060 & 70.1982680 & 4.5276 & $1.00$ & $9.77_{-1.10}^{+1.08}$ & $55.0_{-7.2}^{+7.0}$ & 9.0 & $-1.35_{-0.10}^{+0.10}$ & $-19.24 \pm 0.03$ \\
1086855 & 8018 & 189.2865122 & 62.2381380 & 4.4096 & $1.00$ & $5.25_{-0.86}^{+0.83}$ & $247.3_{-71.5}^{+106.5}$ & 6.2 & $-0.90_{-0.53}^{+0.58}$ & $-16.97 \pm 0.17$ \\
10868 & 5105 & 268.4410301 & 65.1676270 & 4.3886 & $1.00$ & $14.65_{-3.23}^{+3.26}$ & $125.7_{-34.6}^{+44.9}$ & 4.5 & $-2.67_{-0.51}^{+0.49}$ & $-18.39 \pm 0.12$ \\
292585 & 5224 & 150.0719633 & 2.2972652 & 4.3837 & $1.00$ & $7.66_{-0.71}^{+0.77}$ & -- & 10.3 & -- & -- \\
11024 & 2767 & 322.4163338 & 0.0981286 & 4.2949 & $1.00$ & $61.30_{-6.44}^{+6.33}$ & $163.0_{-30.1}^{+37.2}$ & 9.6 & $-2.73_{-0.39}^{+0.38}$ & $-19.60 \pm 0.10$ \\
73488 & 1181 & 189.1973959 & 62.1772331 & 4.1309 & $1.00$ & $29.40_{-1.00}^{+1.04}$ & $194.2_{-10.5}^{+11.2}$ & 28.8 & $-1.65_{-0.08}^{+0.08}$ & $-18.77 \pm 0.02$ \\
4111584 & 1208 & 215.9296283 & 24.1083515 & 4.0657 & $1.20$ & $9.22_{-1.86}^{+1.74}$ & $162.7_{-52.8}^{+81.5}$ & 5.1 & $-2.42_{-0.70}^{+0.71}$ & $-17.47 \pm 0.18$ \\
\enddata
\tablenotetext{$$^\dagger$$}{The multiply imaged source Abell2744 QSO1}
\tablecomments{Columns: (1) source ID in \citet{deGraaff2026}; (2) program ID; (3--4) coordinates; (5) spectroscopic redshift; (6) gravitational magnification factor based on \citet{Furtak2023lensing} and \citet{Sarrouh2026}, compiled by \citet{deGraaff2026}; (7) Ly$\alpha$ flux, the median and 16th/84th percentile uncertainties from the posterior; (8) rest-frame Ly$\alpha$ equivalent width; (9) Ly$\alpha$ signal-to-noise ratio; (10) UV continuum slope; (11) absolute UV magnitude. Sources are sorted by decreasing redshift within each subsample.}
\end{deluxetable}

\section{The impact of DLA on Ly$\alpha$ measurements} \label{app:dla}

In our fiducial analysis (Section~\ref{sec:int_lya_ana}) we do not include a DLA component, because the limited
spectral resolution of the NIRSpec/PRISM data ($R\sim100$) smears out the
characteristic damping wings and makes a DLA difficult to detect and to
disentangle from the IGM attenuation. To verify that this choice does not bias
our measurements, here we repeat the fits with a DLA component included. 

We model the DLA absorption following the same prescription as \citet[][their Section~5.2]{Ion12025}. The neutral hydrogen along the line of
sight is treated as an ensemble of \ion{H}{1} atoms, and the Ly$\alpha$ optical
depth is computed by summing the corresponding Voigt profiles
\citep{Bolton2007}. We fix the Doppler parameter to $b = 10~{\rm km\,s^{-1}}$, a
value commonly adopted in the literature, and place the absorbing gas at the
systemic redshift of each source. Because this gas is local to the host galaxy,
its transmission multiplies the intrinsic spectrum before IGM attenuation. We
then refit every Ly$\alpha$-detected LRD with this DLA-augmented model, retaining the same \texttt{emcee} set-up as in the fiducial analysis
(Section~\ref{sec:int_lya_ana}). Because the DLA model includes one additional free parameter, we assess whether
it is actually favored by the data using the Bayesian information criterion
\citep[BIC;][]{Schwarz1978}, computing
$\Delta{\rm BIC}$ for each source.

\begin{figure}
\centering
\includegraphics[width=0.7\linewidth]{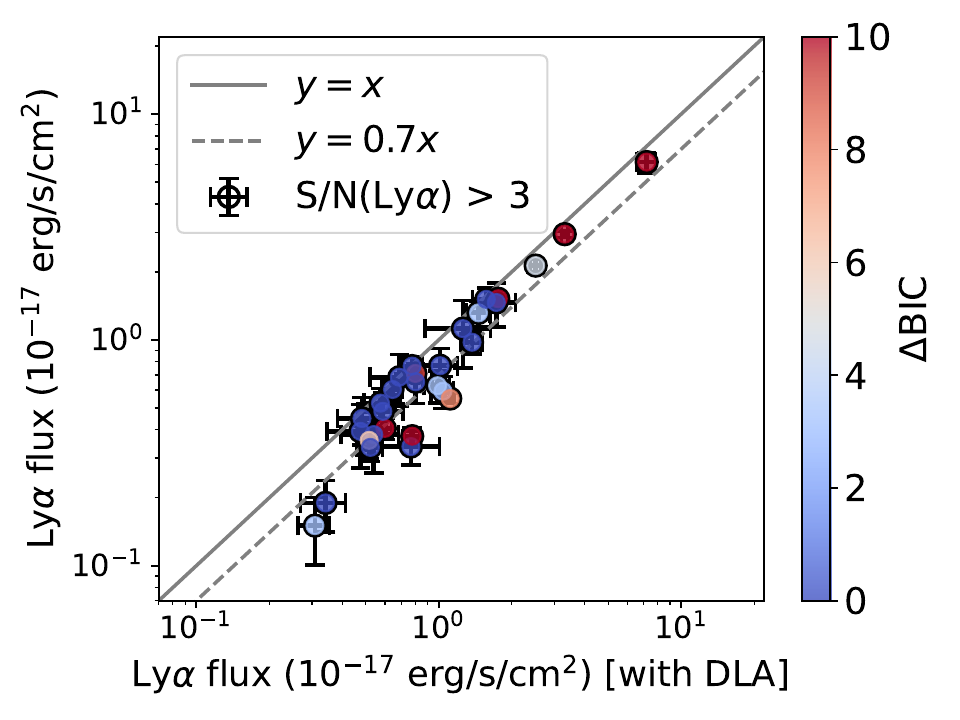}
\caption{Comparison of the integrated Ly$\alpha$ fluxes measured with the
DLA-augmented model against the fiducial (DLA-free, Section \ref{sec:int_lya_ana}) values for the
Ly$\alpha$-detected LRDs. The solid line marks the 1:1 relation. The two
measurements agree at the median level of $\sim20\%$, comparable to or smaller than the
typical measurement uncertainties.}
\label{fig:dla_flux}
\end{figure}

Figure~\ref{fig:dla_flux} compares the Ly$\alpha$ fluxes recovered with and
without the DLA component. For the vast majority of the LRDs the DLA model is not preferred ($\Delta{\rm BIC}<5$). The two sets of measurements track the 1:1 relation
closely: including the DLA only fills in part of the absorbed line and
continuum, so the fluxes are systematically equal to or slightly larger than
the fiducial values, with a median change of just $\sim20\%$. This is
comparable to or smaller than the typical measurement uncertainties on
$F_{\rm Ly\alpha}$. We therefore conclude that omitting a DLA component from our fiducial model introduces at most a $\sim20\%$ systematic in the integrated Ly$\alpha$ fluxes and does not affect any of the conclusions of this work.

\section{Comparison of PSF models} \label{app:psf}

As mentioned in the main text (Section~\ref{sec:map_analysis}), to retain the finer pixel scale of the DJA mosaics in Abell 2744, we construct our own ePSFs for the four NIRCam/SW filters (F090W, F115W, F150W, and F200W) using isolated, unsaturated stars identified in the A2744 mosaic. The ePSFs are built at the native $0\farcs02$ pixel scale of the DJA mosaics using 22--27 stars per filter, following the iterative ePSF-fitting procedure of \citet{Anderson2000}, with a quartic oversampling kernel and careful sky subtraction around each stellar cutout.

\begin{figure}
    \centering
    \includegraphics[width=1\linewidth]{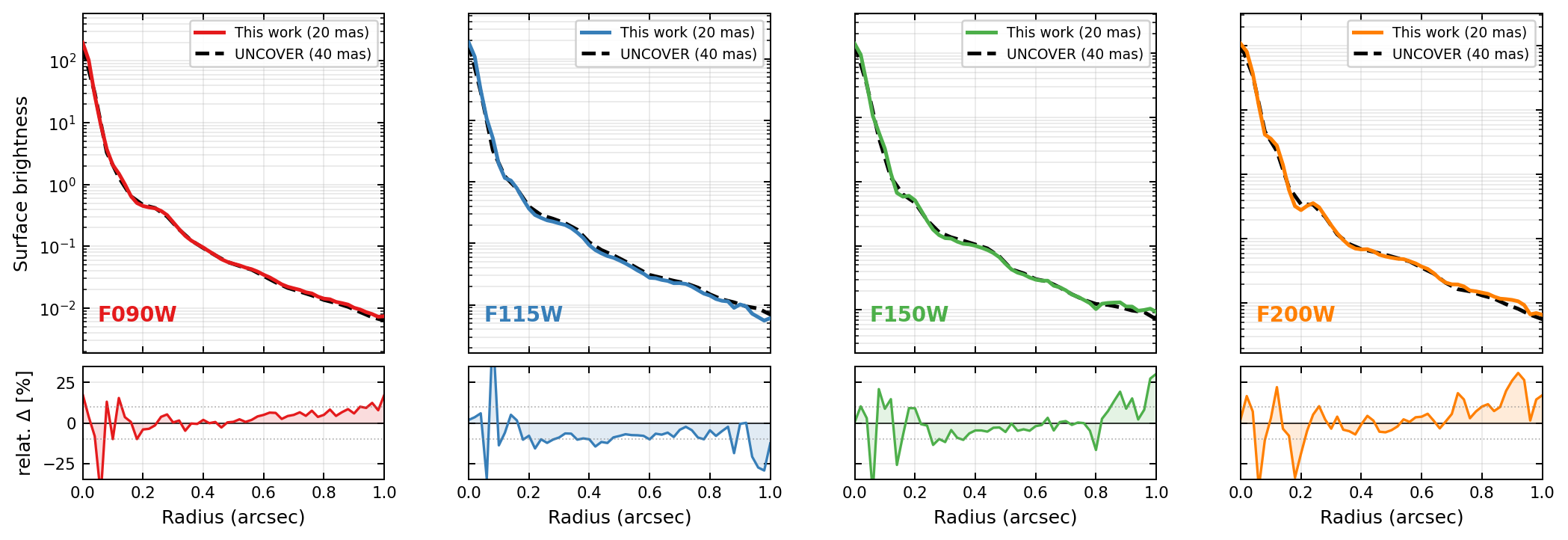}
    \caption{Radial surface brightness profiles of our ePSFs constructed from isolated stars in the A2744 mosaic (solid colored lines) compared with the UNCOVER ePSFs (black dashed lines; \citealt{Weaver2024}) for the four NIRCam/SW filters used in this work. Our ePSFs are built at the native $0\farcs02$ pixel scale using the DJA mosaics, while the UNCOVER ePSFs are sampled at $0\farcs04$. Lower panels show the relative difference with respect to the UNCOVER profiles.}
    \label{fig:epsf_comp}
\end{figure}

Figure~\ref{fig:epsf_comp} compares our ePSFs with those provided by the UNCOVER team \citep{Weaver2024} for all four filters. Overall, the two sets of profiles are highly consistent, agreeing to within $\sim 3-10\%$. A systematic trend is visible in the residuals at very small radii ($\lesssim 0.2''$), where our ePSFs exhibit slightly higher central surface brightness, indicating a sharper core. This behavior is expected and reflects the advantage of constructing ePSFs from mosaics with finer pixel sampling, when the observational setup permits drizzling onto a finer grid. At a $0\farcs02$ pixel scale, the NIRCam/SW PSF core, with FWHM $\sim 0.03$--$0.06''$ depending on wavelength, is sampled by 2--3 pixels across, preserving spatial information that is partially lost when the same data are drizzled to $0\farcs04$ pixels. The finer sampling allows the ePSF construction algorithm to recover a more faithful representation of the intrinsic PSF structure, particularly within the innermost resolution element, where the surface-brightness gradient is steepest. At larger radii ($\gtrsim 0.5''$), where the PSF profile varies more slowly and is adequately sampled at either pixel scale, the two ePSFs converge.

\section{Astrometric Precision of the DJA mosaics}
\label{app:wcs}
 
Here, we quantify the relative astrometric precision between the F444W and Ly$\alpha$-filter mosaics in the vicinity of each LRD. In particular, we select bright, compact sources within a $1\arcmin$ radius of the target LRD by requiring $\mathrm{S/N}_{\mathrm{F444W}} > 20$ and a half-light radius of $< 5$~pixels in F444W. For each selected source, we then measure its centroid position in both the F444W and Ly$\alpha$-filter mosaics using \texttt{photutils.centroid}, and finally calculate the median and standard deviation of the astrometric offsets.

Table~\ref{tab:wcs_offset} summarizes the median WCS offset and $1\sigma$ scatter for each LRD. The JWST--JWST pairs (LRDs in A2744, CEERS, GOODS, and PRIMER-COSMOS, all using NIRCam filters as the Ly$\alpha$ band) show offsets consistent with zero ($|\Delta\alpha| \lesssim 3$~mas, $|\Delta\delta| \lesssim 2$~mas). The JWST--HST pairs (LRDs in UDS with F814W and GOODS-S with F775W) show slightly larger systematic offsets of $5$--$11$~mas. All of these offsets and standard deviations are $\lesssim$ half a pixel. We thus conclude that the WCS alignment of the DJA mosaics is sufficiently precise for the spatially resolved Ly$\alpha$ analysis presented in the main text.

\section{Ly$\alpha$ maps in COSMOS and UDS}
\label{app:map_cosmos_uds}

\begin{figure}
   \centering
  \includegraphics[width=0.87\textwidth]{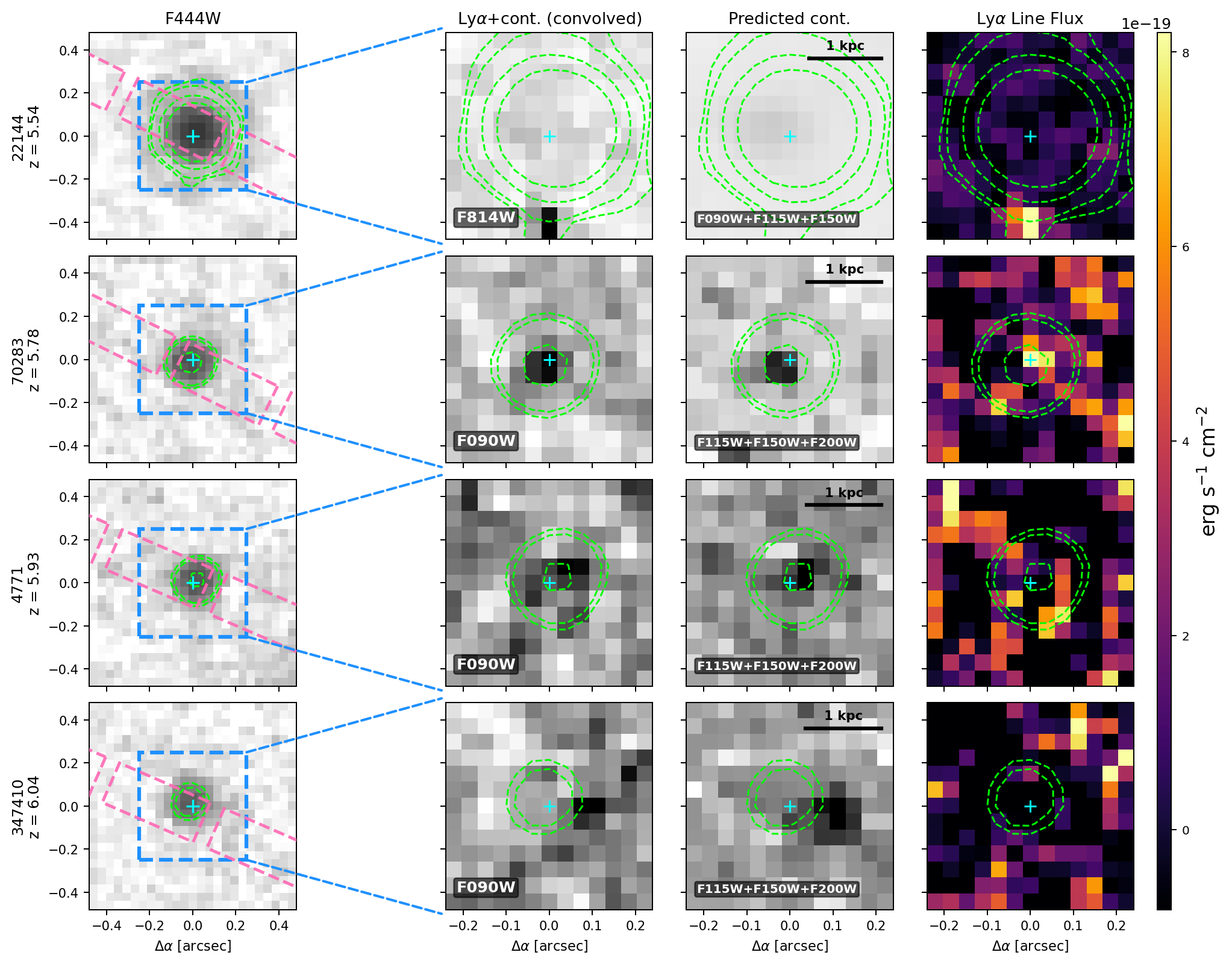}
    \includegraphics[width=0.87\textwidth]{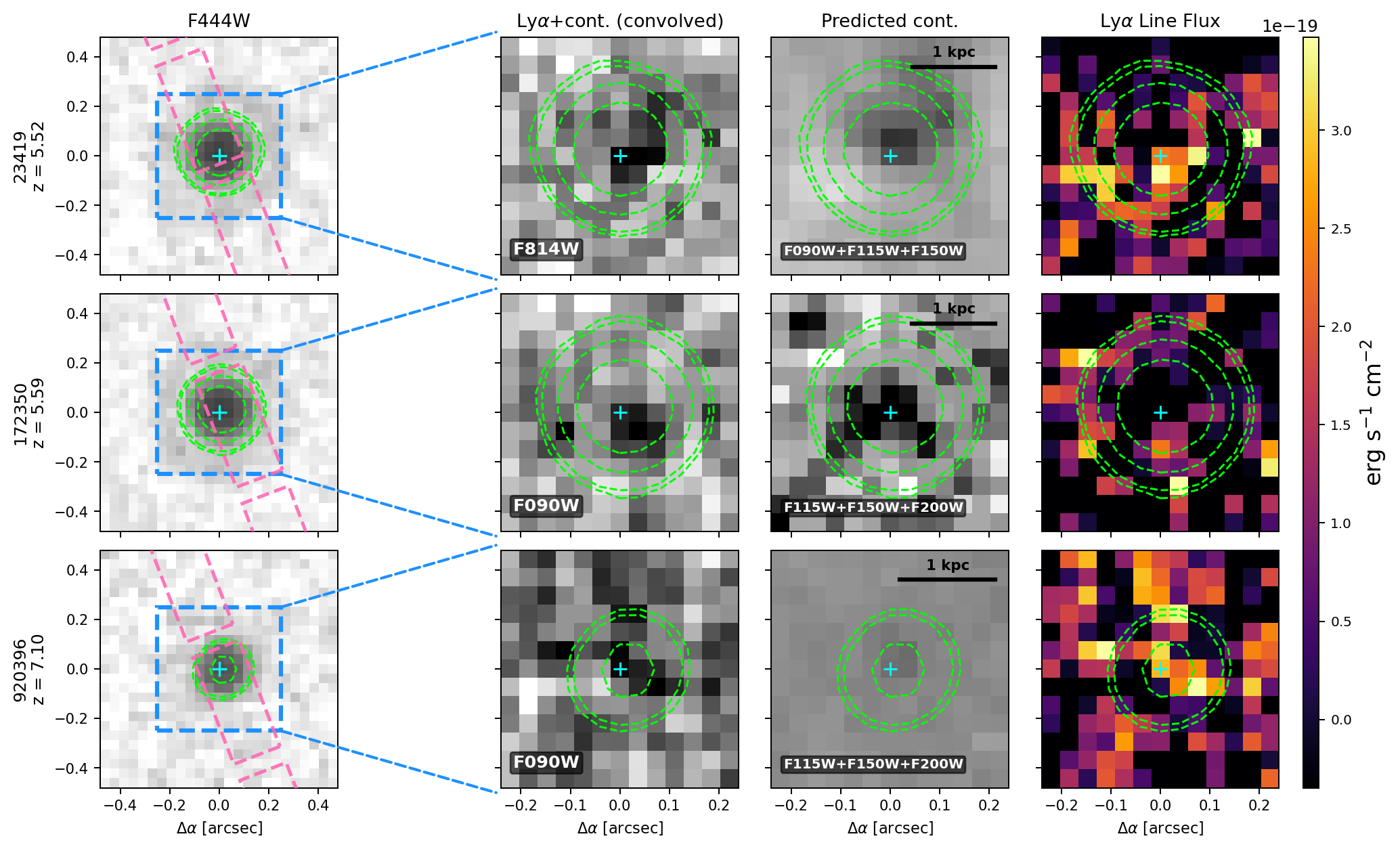}
    \caption{Similar to Figure~\ref{fig:map_1}, but for the LRDs in COSMOS (first two rows) and UDS (remaining three rows). We note that the F090W imaging used here is from PRIMER and is relatively shallow, with a 5$\sigma$ depth of 27.6 AB mag \citep{Donnan2024}, about 2 magnitudes shallower than JADES and UNCOVER.}
    \label{fig:map_3}
\end{figure}

Here we present the Ly$\alpha$ maps for five LRDs in COSMOS and UDS (Figure \ref{fig:map_3}), four of which are derived from NIRCam/F090W imaging obtained through the PRIMER program. These F090W images are 2 mag shallower than those in GOODS-S and Abell~2744, resulting in noisier Ly$\alpha$ maps.

\section{NIRSpec/PRISM spectrum of the southern clump of LRD 204851}
\label{app:204851}

\begin{figure}
    \centering
    \includegraphics[width=1\linewidth]{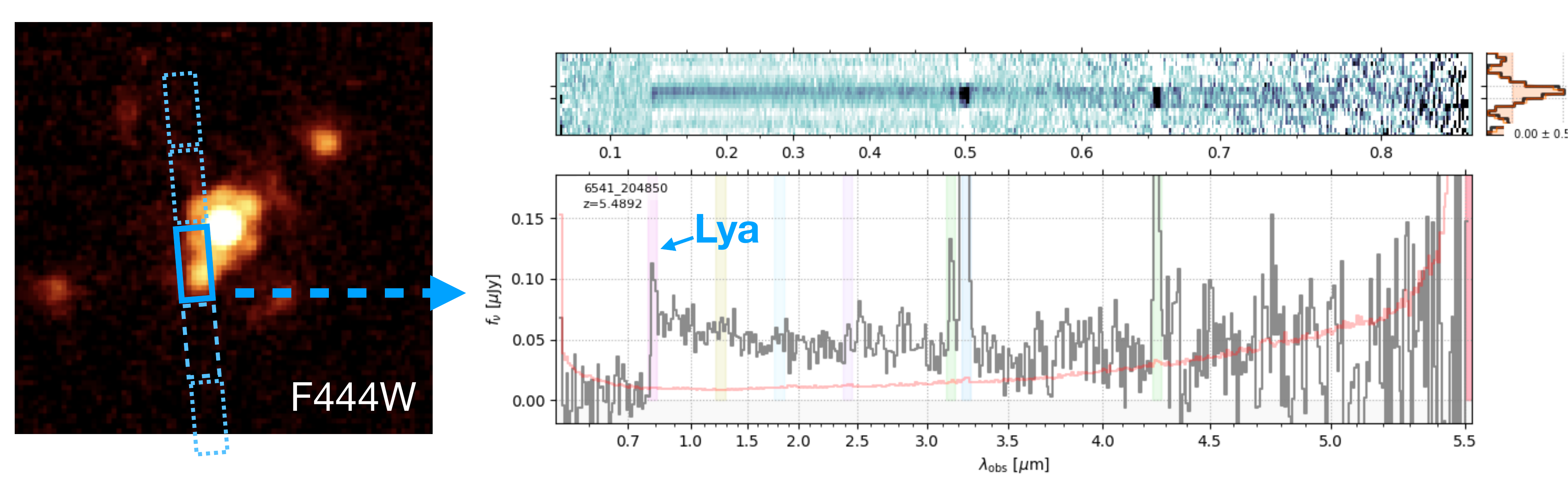}
    \caption{The NIRSpec/PRISM spectrum for the southern clump in LRD 204851 in GOODS-S (second row of Figure \ref{fig:map_2}).}
    \label{fig:24580}
\end{figure}

Our Ly$\alpha$ map reveals a remarkable spatial distribution of Ly$\alpha$ emission in LRD 204851 in GOODS-S. The central region of this galaxy is identified as an LRD based on a PRISM spectrum centered on it (second row of Figure \ref{fig:map_2}). The lower, southern clump of this LRD was observed in JWST/DDT program PID 6541 (PI: Egami). As shown in Figure \ref{fig:24580}, this clump is at the same redshift as the central LRD and exhibits strong Ly$\alpha$ emission, in excellent agreement with our Ly$\alpha$ map. A detailed analysis of this source will be presented in Ji et al. 2026 (in prep.).

\begin{deluxetable}{lccccccc}
\tablecaption{WCS Offset Between F444W and Ly$\alpha$-Filter Mosaics
\label{tab:wcs_offset}}
\tablewidth{0pt}
\tablehead{
    \colhead{Field} &
    \colhead{Source ID} &
    \colhead{Ly$\alpha$ Filter} &
    \colhead{Pixel scale\tablenotemark{$^\dagger$}} &
    \colhead{$\Delta\alpha$} &
    \colhead{$\sigma_{\Delta\alpha}$} &
    \colhead{$\Delta\delta$} &
    \colhead{$\sigma_{\Delta\delta}$} \\
    \colhead{} & \colhead{} & \colhead{} &
    \colhead{(mas)} & \colhead{(mas)} & \colhead{(mas)} & \colhead{(mas)} & \colhead{(mas)}
}
\startdata
A2744          & 4286   & F090W  & 20 & $+2.7$  & $13.0$ & $-1.6$ & $9.1$  \\
A2744          & 24175  & F090W  & 20 & $+1.4$  & $8.1$  & $-0.9$ & $7.0$  \\
A2744          & 20466  & F115W  & 20 & $+1.4$  & $6.4$  & $+0.6$ & $10.3$ \\
COSMOS  & 22144  & HST/F814W & 40 & $-7.0$  & $22.5$ & $+2.2$ & $20.4$ \\
COSMOS  & 4771   & F090W  & 40 & $+0.3$  & $17.2$ & $-1.2$ & $16.3$ \\
COSMOS & 347410 & F090W & 40 & $+0.2$ & $8.6$ & $-2.2$ & $7.3$ \\
COSMOS         & 70283  & F090W      & 40 & $-1.7$  & $7.3$  & $+0.6$ & $11.6$ \\
EGS          & 10108  & F090W & 40  & $-1.5$  & $8.9$  & $+0.2$ & $9.8$  \\
UDS            & 23419  & HST/F814W & 40  & $+5.4$  & $18.0$ & $+6.1$ & $14.2$ \\
UDS & 172350 & F090W & 40 & $+1.0$ & $11.2$ & $-1.2$ & $11.0$ \\
UDS & 920396 & F090W & 40 & $+0.0$ & $11.4$ & $-1.2$ & $10.9$ \\
GOODS-S        & 204851 & HST/F775W  & 40 & $+10.6$ & $9.7$  & $-6.9$ & $9.6$  \\
GOODS-S        & 219000 & F090W      & 20 & $-2.9$  & $12.1$ & $+0.5$ & $10.3$ \\
\enddata
\tablenotetext{$$^\dagger$$}{Pixel scale of the Ly$\alpha$ map.}
\tablecomments{Median offset ($\Delta\alpha$, $\Delta\delta$) and
standard deviation ($\sigma$) of the centroid position difference
between F444W and the Ly$\alpha$-filter mosaic, measured from bright
compact sources within $1\arcmin$ of each LRD. Positive $\Delta\alpha$
indicates the F444W centroid is east of the Ly$\alpha$-filter centroid.
Outliers are rejected via iterative $3\sigma$ clipping using the MAD.}
\end{deluxetable}

\bibliography{sample701}{}
\bibliographystyle{aasjournalv7}

\end{document}